\documentclass[a4paper,11pt]{article}

\usepackage[left = 0.8in, right = 0.8in, top = 1in, bottom = 1in]{geometry}
\usepackage{amsfonts,amsmath,amsthm,amssymb}
\usepackage{xcolor}
\usepackage{tikz, tcolorbox}
\usepackage{graphicx, subcaption}
\usepackage{booktabs}
\usepackage[linktocpage=true]{hyperref}
\hypersetup{
    linkcolor = blue,
    citecolor = blue,
    colorlinks = true
}
\usepackage[ruled,vlined]{algorithm2e}

\theoremstyle{definition}
\newtheorem{definition}{Definition}
\newtheorem{theorem}{Theorem}

\newtheorem{example}{Example}
\newtheorem{remark}{Remark}

\newcommand{\R}{\mathbb{R}}

\newcommand{\var}{\text{Var}}
\newcommand{\cov}{\text{Cov}}
\newcommand{\E}{\mathbb{E}}
\newcommand{\transpose}{^\intercal}
\newcommand{\indep}{\perp \!\!\! \perp }

\newcommand{\tr}{\text{trace}}
\newcommand{\scalbb}[1]{\boldsymbol{#1}}
\newcommand{\rv}[1]{\mathsf{#1}}
\newcommand{\rvbb}[1]{\boldsymbol{\mathsf{#1}}}

\begin{document}

\title{Trustworthy Dimensionality Reduction \\[1ex] \normalsize A dissertation report presented for the completion of\\
the degree of Master of Statistics (M. Stat.) from Indian Statistical Institute}
\author{Subhrajyoty Roy}
\date{June 25, 2021}

\maketitle

\begin{abstract}
Different unsupervised models for dimensionality reduction like PCA, LLE, Shannon’s mapping, tSNE, UMAP, etc. work on different principles, hence, they are difficult to compare on the same ground. Although they are usually good for visualisation purposes, they can produce spurious patterns that are not present in the original data, losing its trustability (or credibility). On the other hand, information about some response variable (or knowledge of class labels) allows us to do supervised dimensionality reduction such as SIR, SAVE, etc. which work to reduce the data dimension without hampering its ability to explain the particular response at hand. Therefore, the reduced dataset cannot be used to further analyze its relationship with some other kind of responses, i.e., it loses its generalizability. 

To make a better dimensionality reduction algorithm with a better balance between these two, we shall formally describe the mathematical model used by dimensionality reduction algorithms and provide two indices to measure these intuitive concepts such as trustability and generalizability. Then, we propose Localized Skeletonization and Dimensionality Reduction (LSDR) algorithm which approximately achieves optimality in both these indices to some extent. The proposed algorithm has been compared with the state of the art algorithms such as tSNE and UMAP and is found to be better overall in preserving global structure while retaining useful local information as well. We also propose some of the possible extensions of LSDR which could make this algorithm universally applicable for various types of data similar to tSNE and UMAP.
\end{abstract}

\section*{Notations}

\begin{table}[h]
    \centering
    \begin{tabular}{rl}
        $x$ & A fixed scalar variable \\
        $\scalbb{x}$ & A fixed scalar vector \\
        $\scalbb{X}$ & A fixed scalar matrix \\
        $\rv{x}$ & A scalar random variable\\
        $\rvbb{x}$ & A random vector \\
        $\rvbb{X}$ & A random matrix \\
        $\R$ & Set of real numbers \\
        $\rvbb{1}_n$ & The $(n\times 1)$ column vector of all ones\\
        $\E(\cdot)$ & Expectation operator\\
        $\var(\cdot)$ & Variance operator\\
        $\cov(\cdot)$ & Covariance operator
    \end{tabular}
\end{table}

\pagebreak

\tableofcontents

\pagebreak

\section{Introduction}\label{ch:intro}

During the current age of big data, data sources have become easily accessible, leading to an abundance of available real-world data to solve any practical problems. This abundance is mostly related to the sheer increase in the number of features or variables about a particular sample, rather than the increase in the number of sample points. However, analysing these high-dimensional datasets in their entirety can pose various challenges. Because of the curse of dimensionality, the samples become sparse in high dimensions leading to the overfitting of many models. Also, most of these high-dimensional features are not linked to the response variable under study, and, one cannot use visualisation techniques to see the common patterns to obtain insight into the variables. Therefore, it is natural to have a reduced representation of these high dimensional datasets which should not compromise the subsequent analysis. \textbf{Dimensionality Reduction} is such a transformation of high-dimensional data into a meaningful representation of reduced dimensionality~\cite{van2009dimensionality}. Ideally, this reduced dimension should be equal to the intrinsic dimension of the data, which is described as the minimum number of parameters required to account for the variability presented in the observed data~\cite{fukunaga2013introduction}. For example, it is well known that only $(k-1)$ variables are enough to describe the segregation between $k$ well-separated classes of any dimension~\cite{fukunaga2013introduction}.

There are various applications of dimensionality reduction in different fields of science. Compression of digital signals from audio, video and images~\cite{jayant1997signal}, analysis of fMRI scans, gene expression, protein-coding datasets in biostatistics~\cite{tarca2007machine}, finding word embeddings in Natural Language Processing~\cite{jagarlamudirevisiting}, stock market analysis in Finance~\cite{chalup2008kernel,zhong2017forecasting} are some of the real-life use-cases for dimensionality reduction techniques.

Dimensionality reduction algorithms can broadly be classified into three groups based on their focus and approach.

\begin{enumerate}
    \item \textbf{Metric preserving Algorithms} focus on reducing any high dimensional data by preserving local distances or increasing global variability. PCA, Kernel PCA, Maximum variance unfolding, Hessian Map, Multidimensional scaling, tSNE etc. are examples of such algorithms. Neural network models such as autoencoders try to minimize the reconstruction error which is also a metric in high dimensional space.
    \item \textbf{Graphical Algorithms} focus on reducing the dimensionality by approximating the geodesic distance based on the creation of neighbourhood graphs. Isomap, UMAP, Locally linear embedding, Laplacian eigenmaps etc. are examples of such algorithms.
    \item \textbf{Supervised Algorithms} focus on reducing the dimensionality of predictor variables so that the ability to explain the response variable is not reduced. Sliced Inverse Regression (SIR), Sliced Average Variance Estimation (SAVE), Mean Average Variance Estimation (MAVE) are some of the examples of this nature.
\end{enumerate}

\subsection{Mathematical Setup of Dimensionality Reduction}

The mathematical setup of dimensionality reduction starts with an $n \times p$ size dataset $\scalbb{X}$, where each row represents a sample $\scalbb{x}_i \in \R^p$, for $i = 1, 2, \dots n$. For the sake of simplicity, one assumes that these datapoints $\scalbb{x}_1, \dots \scalbb{x}_n$ are independent and identically distributed observations from a distribution $F$. To make dimensionality reduction meaningful, it is also assumed that the support of this distribution $F$ has an intrinsic dimension $d$ (with $d << p$), i.e. the datapoints $\scalbb{x}_i$ lie on or near a $d$-dimensional manifold in the $\R^p$ space. 

It is often easier to work if we restrict the attention of the Riemannian manifold only, but the general problem makes no assumption about the manifold and allows discontinuous clusters to be present. The goal is to obtain a low dimensional representation $\scalbb{y}_i \in \R^d$ corresponding to each $\scalbb{x}_i$ such that the geometry of the manifold is preserved in the low dimensional space as much as possible. Since the problem does not associate a metric to measure the distortion, it is an ill-posed problem. Almost all of the existing DR algorithms define objective functions (or a metric) which is a characterization of the dissimilarity between geometric properties of the high and low dimensional manifolds, and then it tries to minimize that metric to obtain the ``best" lower dimensional representation. Since there is no universal characterization, most of these objective function building process is ad-hoc in nature. Later, in Section~\ref{sec:framework}, we shall provide a unified framework to define the problem more concretely, and also formally define what is meant by a dimensionality reduction algorithm.

\subsection{Existing Dimensionality Reduction Algorithms}\label{sec:flaws}

\subsubsection{Metric Preserving Algorithms}

Principal Component Analysis (PCA) is the most popular metric-preserving DR algorithm. It aims to find the linear projections $\rvbb{y} = \scalbb{M}\rvbb{x}$ of the data $\rvbb{x}$ to preserve the highest variability. The objective becomes same as maximizing $\tr\left(\scalbb{M}\transpose \cov(\rvbb{x}) \scalbb{M}\right)$. An analytical solution of this problem can be obtained by finding the normalized eigenvectors associated with the $d$ largest eigenvalues of the empirical variance-covariance matrix $\scalbb{\Sigma}$ of the zero mean data (i.e. the centered data). Although it has been used in various applications~\cite{turk1991face,posadas1993spatial,iezzoni1991applications,jeffers1967two}, it suffers from many problems.

\begin{enumerate}
    \item PCA restricts its attention to only linear functions of the variables. However, if the underlying manifold is nonlinear, PCA cannot recover such structure. Localized PCA~\cite{kambhatla1997dimension,roweis1998algorithms} and Kernel PCA~\cite{scholkopf1998nonlinear,scholkopf1997kernel} are two extensions that considers functions in an RKHS (see Appendix~\ref{appendix:RKHS}) instead of only linear functions to counter this problem.
    \item It is not necessarily true that the information contained in the dataset must be explained by variance (not by other moments). An instance of this problem is described in detail in Example~\ref{example:pca-linear}.
    
    \begin{example}\label{example:pca-linear}
    Since PCA, Kernel PCA or probabilistic PCA uses the variance in the data as a measure of information content, it can fail when most of the information presented in the data is along the direction with minimum variance. For example, if the dataset contains two long linear clusters as shown in Figure~\ref{fig:pca-problem}, PCA fails to detect the shape or the structure of the manifold. A desirable dimensionality reduction algorithm should be able to detect these clusters, and create a one-dimensional mapping where the clusters are mapped to some disjoint intervals in $\R$. 

    \begin{figure}[ht]
        \centering
        \begin{subfigure}{0.8\textwidth}
            \includegraphics[width = \textwidth]{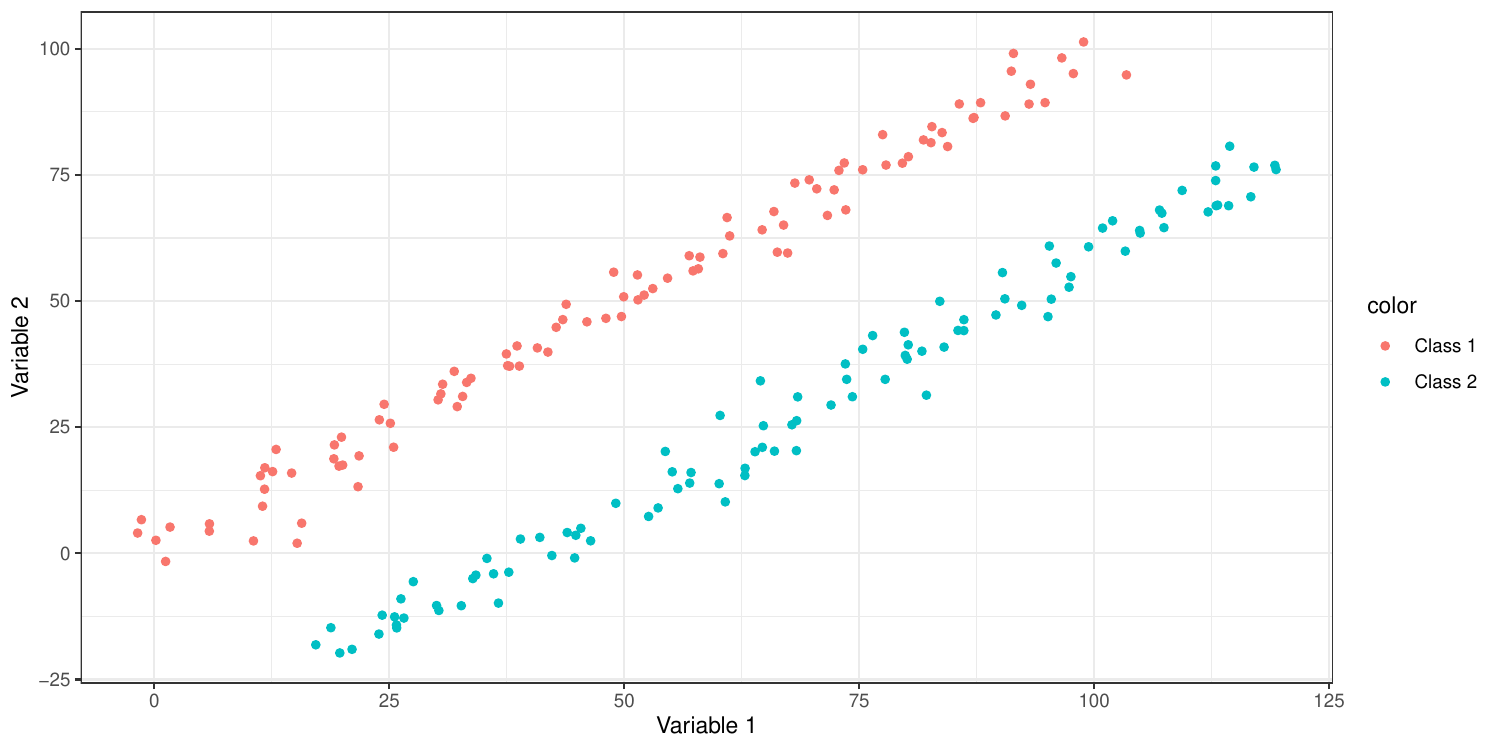}
            \caption{A 2-dimensional dataset with two linear clusters}
        \end{subfigure}
        \begin{subfigure}{0.8\textwidth}
            \includegraphics[width = \textwidth]{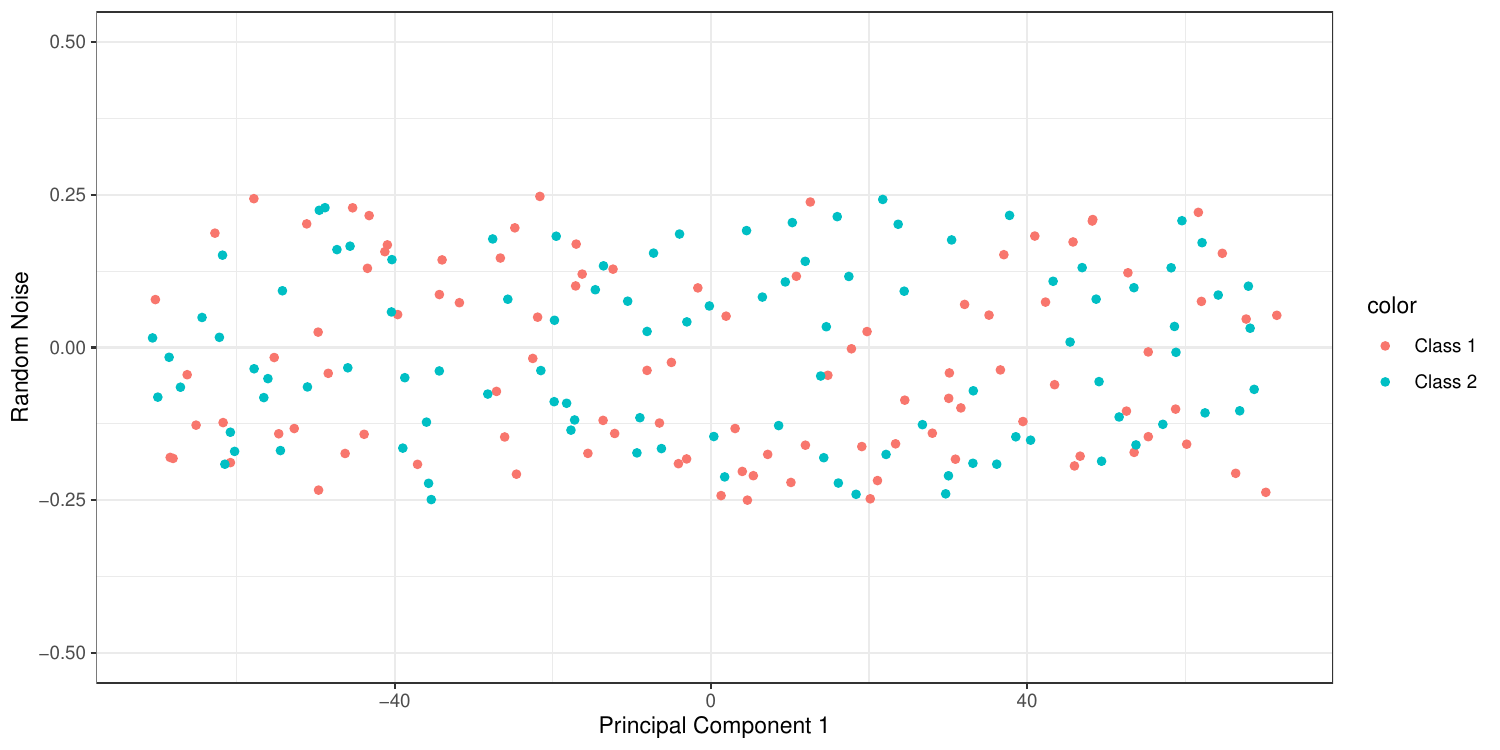}
            \caption{Output of 1 dimensional PCA with random noise in $y$-axis}
        \end{subfigure}
        \caption{PCA fails to capture the class difference by first principal component}
        \label{fig:pca-problem}
    \end{figure}
    
    \end{example}
    
    \item The third problem is the trustability issues. Since PCA takes the approach to maximize the variances, and the solution is given by the principal eigenvectors of the covariance matrix. These principal eigenvectors have an asymptotic normal distribution irrespective of the distribution of the original $\rvbb{x}$~\cite{marcinek2020high}, hence the top few principal components tend to look like Gaussian distributed random variables. A simulated experiment has been performed in Example~\ref{example:pca-trust} to visualize this issue. Although all the principal components taken together returns only a rotated version of the original dataset, which makes PCA very desirable, its lower dimensional embeddings tend to distort the distribution of the data, which is undesirable. Any statistical inference based on the PCA reduced form could then be misleading in nature.
\end{enumerate}

\begin{example}\label{example:pca-trust}
    As noted earlier, PCA has trustability issues as the principal eigenvectors are asymptotically normally distributed~\cite{marcinek2020high}, hence the low dimensional representation usually tend to look like normally distributed. A simulation was performed to demonstrate this effect.
    \begin{enumerate}
        \item In first simulation, $1000$ datapoints are generated from $N(\scalbb{0},\scalbb{I}_5)$. From the data, the first two principal component directions were estimated, and then the projection onto them is obtained. As shown in Figure~\ref{fig:normal-pca}, the algorithm outputs a 2-dimensional Gaussian data as best embedding of the original $5$-dimensional data.
        \item In the second simulation, $1000$ datapoints are generated from the unit $5$-dimensional hypercube, where each of the variables are i.i.d. uniform $(0, 1)$. Again, the first two principal components are extracted and plotted in Figure~\ref{fig:uniform-pca}. The output again looks similar to a Gaussian distribution. 
    \end{enumerate}

    \begin{figure}[ht]
        \centering
        \begin{subfigure}{0.33\linewidth}
            \includegraphics[width = \textwidth]{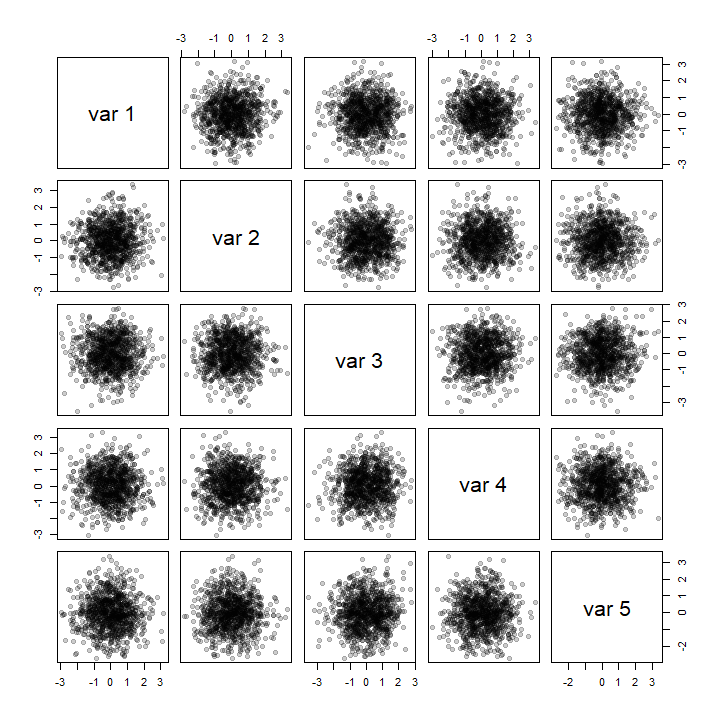}
        \end{subfigure}
        \begin{subfigure}{0.66\linewidth}
            \includegraphics[width = \textwidth]{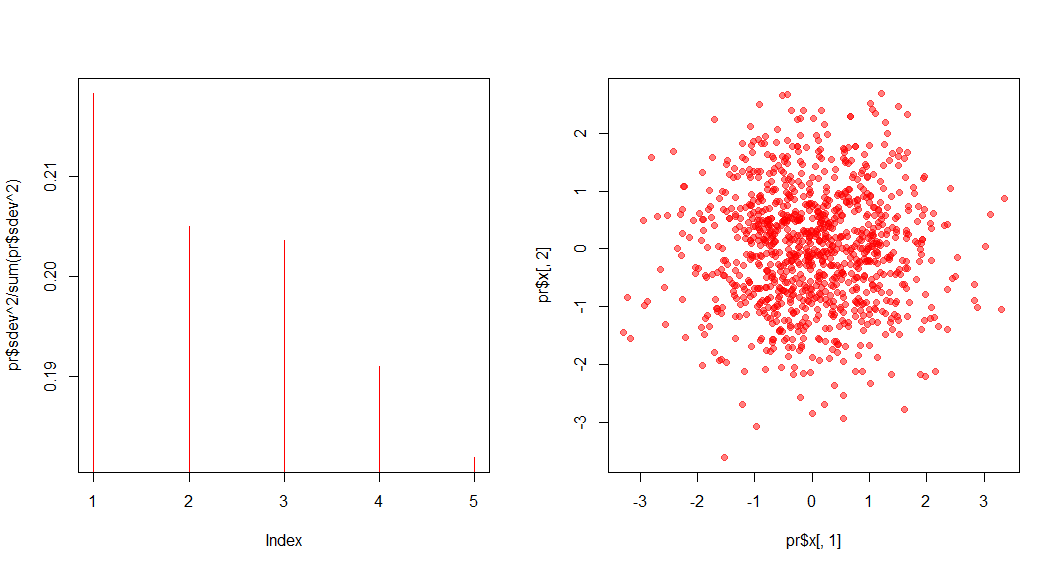}
        \end{subfigure}
        \caption{First two principal components as extracted by normally distributed noise data}
        \label{fig:normal-pca}
    \end{figure}

    \begin{figure}[ht]
        \centering
        \begin{subfigure}{0.33\linewidth}
            \includegraphics[width = \textwidth]{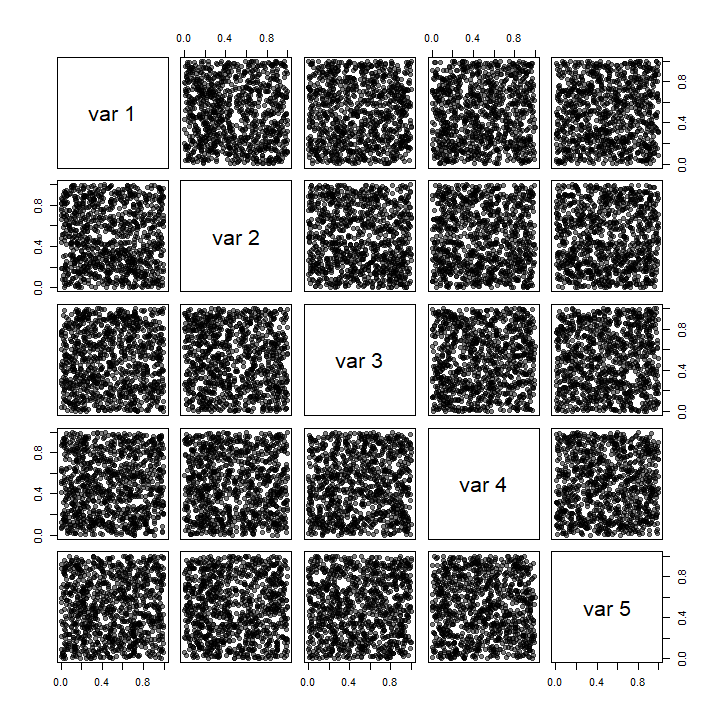}
        \end{subfigure}
        \begin{subfigure}{0.66\linewidth}
            \includegraphics[width = \textwidth]{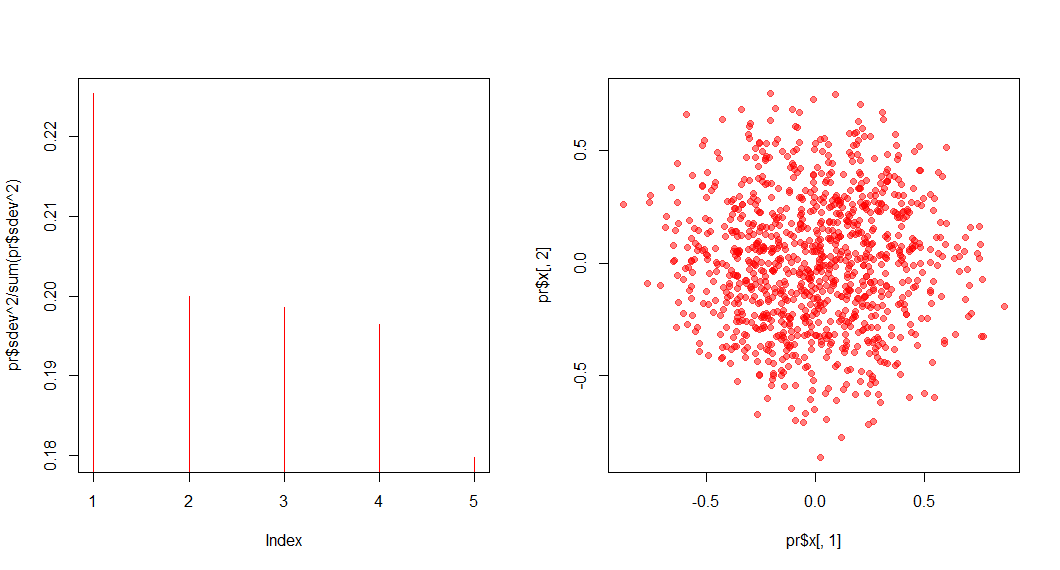}
        \end{subfigure}
        \caption{First two principal components as extracted by uniformly distributed noise data}
        \label{fig:uniform-pca}
    \end{figure}    

    Therefore, despite the original data being very different, the reduction outputted by PCA is similar, resulting in similar inference about the data (whenever the analysis first performs PCA as dimensionality reduction and analyzes the reduced data).
\end{example}

\subsubsection{Graphical Algorithms}

Another approach to dimensionality reduction is by approximating the geodesic distance through the curved manifold. These algorithms start by creating a neighbourhood graph $G$ where each datapoint $\scalbb{x}_i$ is connected to its $k$-nearest neighbours. The length of the shortest path between two points $\scalbb{x}_i$ and $\scalbb{x}_j$ through the graph $G$ serves as an approximation of the geodesic distance between $\scalbb{x}_i$ and $\scalbb{x}_j$. Isomap, a popular graph-based DR algorithm~\cite{lim2003planar,niskanen2003comparison} uses multidimensional scaling to find low dimensional points to approximate these distances through $G$. Uniform Manifold Approximation and Projection (UMAP)~\cite{lel2018umap}, one of the most recent DR algorithms, creates the $k$-neighbourhood graph $G$ and computes the simplicial sets~\cite{friedman2008elementary} generated by that neighbourhood graph. Then the high and low dimensional simplicial sets are matched against each other.

\begin{figure}[ht]
    \centering
    \begin{subfigure}{0.32\linewidth}
        \includegraphics[width = \textwidth]{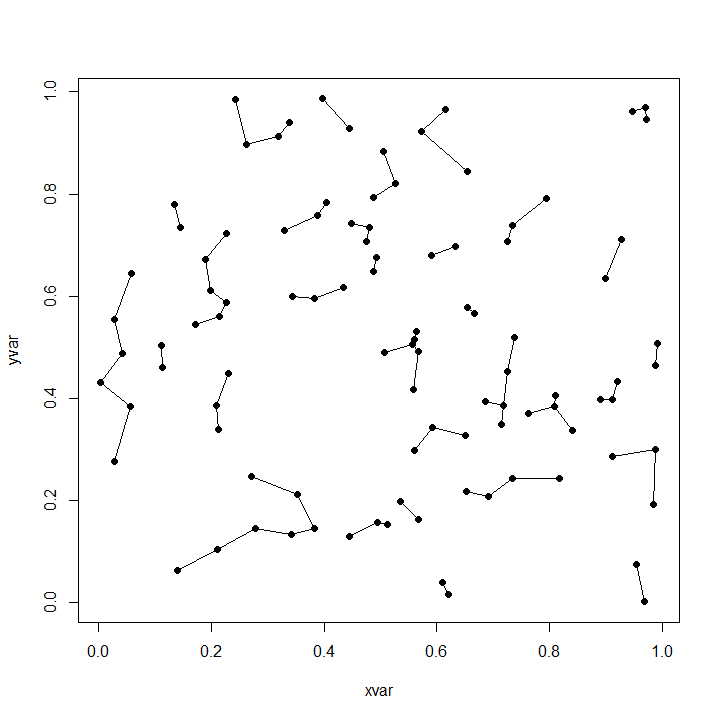}
    \end{subfigure}
    \begin{subfigure}{0.32\linewidth}
        \includegraphics[width = \textwidth]{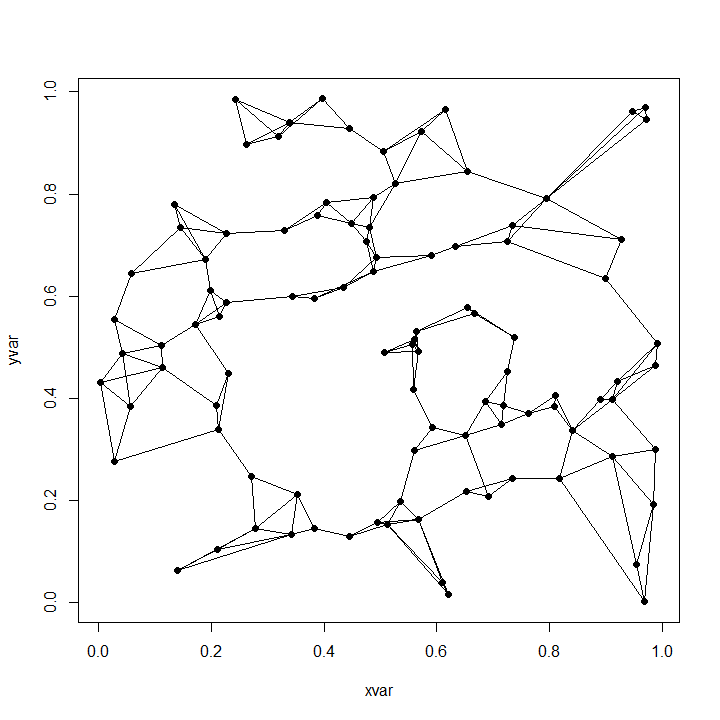}
    \end{subfigure}
    \begin{subfigure}{0.32\linewidth}
        \includegraphics[width = \textwidth]{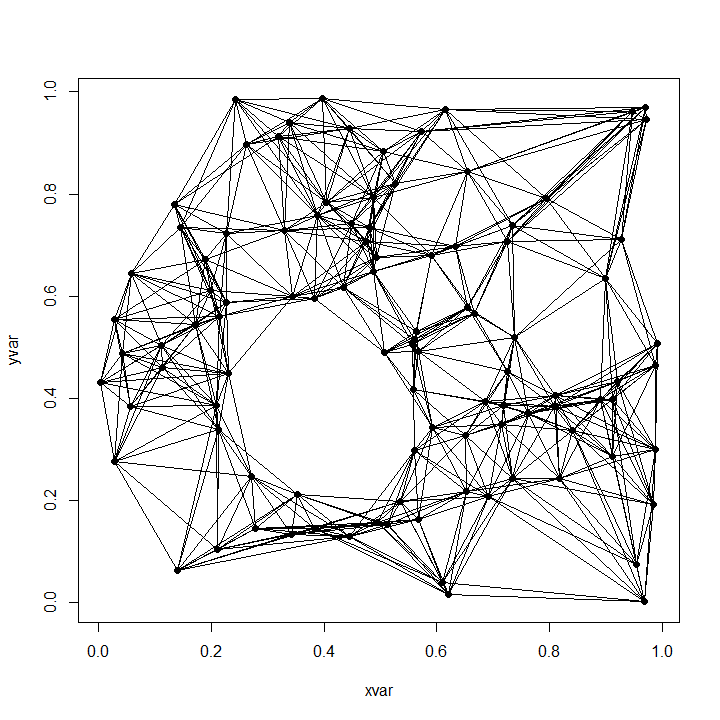}
    \end{subfigure}
    \caption{Neighbourhood graphs with $k = 1, 3, 10$, for $100$ points generated uniformly in the unit square}
    \label{fig:isomap-problem}
\end{figure}

These graph-based algorithms mainly suffer from two problems. The first one being the choice of the parameter $k$, the optimal number of neighbours, which if not properly chosen, may lead to artificial holes. For the purpose of demonstration, we generated $100$ datapoints uniformly from the unit square and obtained the neighbourhood graphs for $k = 1, 3$ and $10$. As shown in Figure~\ref{fig:isomap-problem}, for $k = 1$, the neighbourhood graph is highly disconnected, and even for $k = 10$, there remains an artificial hole in the dataset. There is no consensus about the choice of $k$ in a data-dependant way, since there is no established theory about what metric should one look at.

\begin{figure}[ht]
    \centering
    \begin{subfigure}{0.49\linewidth}
        \includegraphics[width = \textwidth]{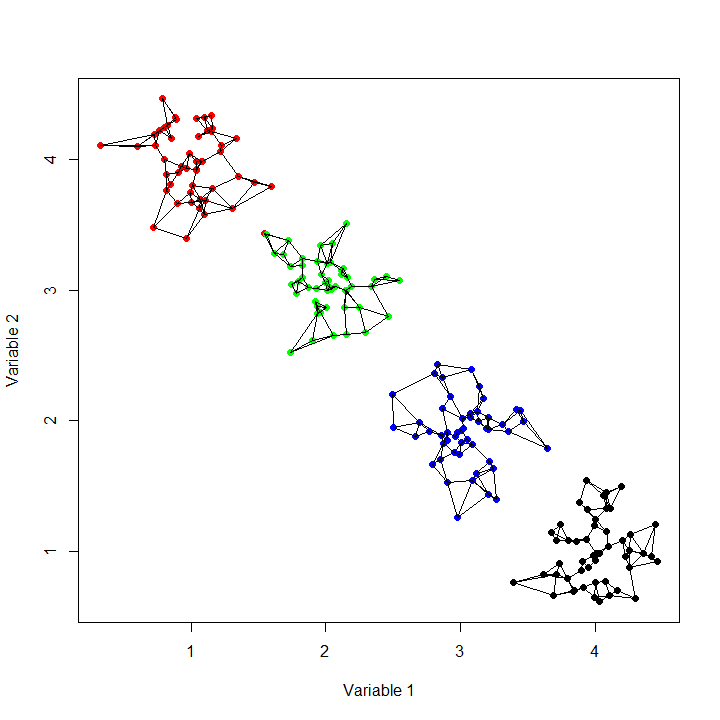}
        \caption{The Neighborhood graph}
    \end{subfigure}
    \begin{subfigure}{0.49\linewidth}
        \includegraphics[width = \textwidth]{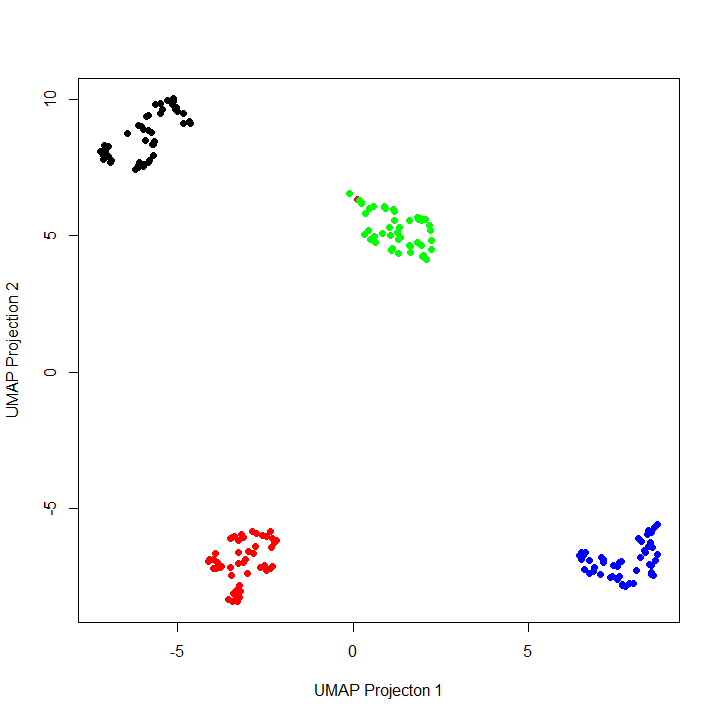}
        \caption{2 dimensional projection by UMAP}
    \end{subfigure}
    \caption{UMAP fails to preserve global structure because of disconnected graphs}
    \label{fig:umap-problem}
\end{figure}

On the other hand, another problem occurs when there are tightly clustered points having small intra-cluster distances and having high inter-cluster distances. In such a case, the neighbourhood graph often becomes disconnected. Therefore, the preservation of distances based on the neighbourhood graph leads to distortion of the global structure of the dataset. UMAP, the most recent graph-based DR algorithm, suffers from a similar problem~\cite{van2008visualizing,parametric-tsne,oskolov2020}. The algorithm assumes that the data is uniformly distributed on the manifold $\mathcal{M}$, and hence, away from any boundaries, any ball of fixed volume should contain approximately the same number of points of $\scalbb{X}$ regardless of where on the manifold it is centered. However, if there are more than one clusters present in the data, the neighbourhood graph $G$ becomes disconnected, i.e. for two clusters present in the data, the neighbourhood graph can be split into two disjoint neighbourhood graph $G = G_1 \cup G_2$, where $G_1, G_2$ are neighbourhood graphs for each of the clusters. In that case, the ordering or the position of the clusters have very little effect, since the disconnected graphs can be rearranged without any constraint. For this reason, UMAP fails to convey the ordering of the clusters (present in the original data), into the reduced dimensionality, as shown in Figure~\ref{fig:umap-problem}. 

A few other graph-based algorithms such as Locally Linear Embedding (LLE)~\cite{roweis2000nonlinear}, Laplacian eigenmaps~\cite{belkin2004semi} etc., all suffer from similar problems described above.

\subsubsection{Supervised Dimensional Reduction Algorithms}

Sufficient Dimensionality Reduction Framework describes the supervised framework for dimensionality reduction. Any kind of dimensionality reduction analysis is generally followed by clustering, classification, regression analysis using the reduced set of variables. If such classification labels or regression variables are known a-priori, one may use this information on response variable in formulation of the dimensionality reduction problem, so that the reduced set of variables is more meaningful. It starts with a framework given as $\rvbb{Y} \indep \rvbb{X} \mid \sigma\left(\left\{ \phi_1(\rvbb{X}), \dots \phi_d(X) \right\}\right)$, where $\phi_1, \dots \phi_d$ are unknown functions to be estimated. Specifically for linear embedding functions, one needs to estimate $\beta_1, \dots \beta_d$ such that $\rvbb{Y} \indep \rvbb{X} \mid (\beta_1\transpose \rvbb{X}, \beta_2\transpose \rvbb{X}, \dots \beta_d\transpose\rvbb{X})$. Bing Li~\cite{li2018sufficient} provides a useful discussion on the numerous methods that has been developed to solve this problem. Popular methods under sufficient dimensionality reduction framework includes,

\begin{enumerate}
    \item Sliced Inverse Regression~\cite{li1991sliced,wang2008sliced}. Here, the idea is that the difference between the conditional expectations $\E(\rvbb{X} \mid \rvbb{Y} = y_1) - \E(\rvbb{X} \mid \rvbb{Y} = y_2)$ contains only the effect due the principal SDR directions.
    \item Sliced Average Variance Estimation (SAVE)~\cite{dennis2000save}. The basic idea is that $\var(\rvbb{X} \mid \rvbb{Y} = y)$ contains only the variation due to the directions orthogonal to SDR. 
    \item Generalized SIR and Generalized SAVE attempts to solve the general problem by restricting the $\sigma$-field $\sigma\left(\left\{ \phi_1(\rvbb{X}), \dots \phi_d(X) \right\}\right)$ to a RKHS instead.
\end{enumerate}

The main problem with the SDR approach is that it performs poorly if the response variable is categorical. For example, if there are $c$ levels in the categorical variable $\rvbb{Y}$, then only $(c-1)$ SDR directions can be obtained. Also, since the reduction is performed only with a particular response in mind, the reduced set of variables cannot be reused in other problems. An example of how SIR may result in loss of information has been given as follows.

\begin{example}\label{appendix:sdr-problem}

We apply dimensionality reduction for Wisconsin breast cancer data based on two algorithms, tSNE and SIR. As shown in Figure~\ref{fig:wb-cancer}, they output a very different shape of the reduced data. While SIR only tries to best preserves the classification of benign and malignant cells, tSNE does it unsupervisedly, and outputs a one dimensional curvilinear manifold of the data. It was seen that the horizontal axis outputted by tSNE roughly aligns to the size of the tumour, with a strongly negative correlation of $-0.7013$. The tSNE output also says that it is much easier to classify the malignant and benign cells if the tumour is large, while it is much harder to segregate if it is small. SDR techniques will generally fail to consider this as it is optimized only to discriminate between the two classes.

\begin{figure}[ht]
    \begin{subfigure}{0.49\linewidth}
        \includegraphics[width = \textwidth]{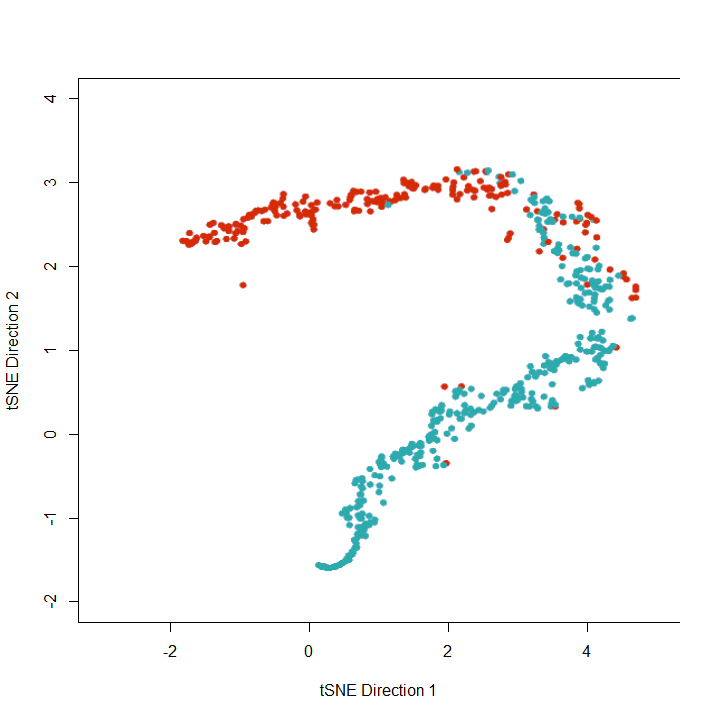}
        \caption{Dimensionality reduction via tSNE}
    \end{subfigure}
    \begin{subfigure}{0.49\linewidth}
        \includegraphics[width = \textwidth]{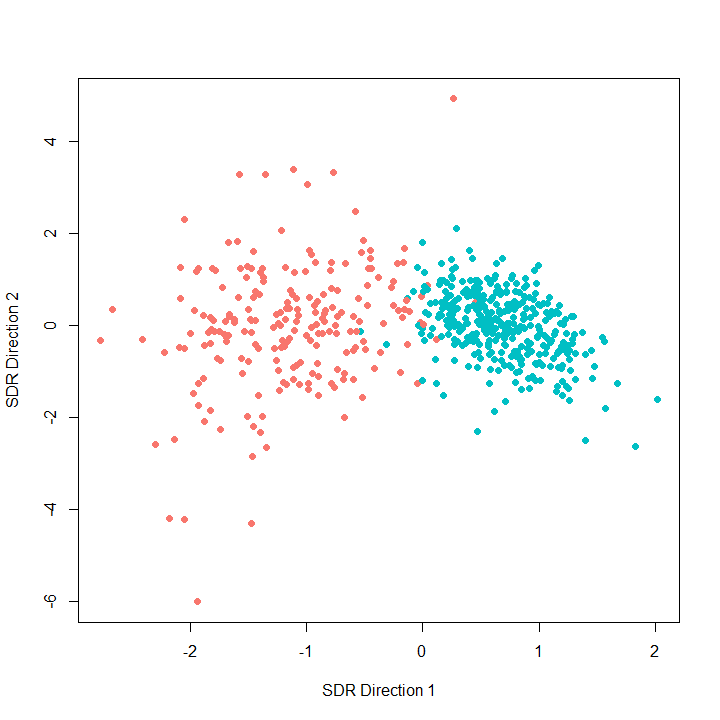}
        \caption{Dimensionality reduction via SIR}
    \end{subfigure}
    \caption{Dimensionality reduction via tSNE and SIR for Winconsin breast cancer data, Red means malignant and Turquoise means benign cancer cells.}
    \label{fig:wb-cancer}
\end{figure}
\end{example}

\subsection{Existing Metrics}\label{sec:metrics}

As described in Section~\ref{sec:flaws}, despite the existence of a large number of dimensionality reduction algorithms, most of these suffer from some kind of trustability issues. Also, there is little consensus about how to evaluate the performance of a dimensionality reduction algorithm since there exists a diverse range of frameworks under which the existing algorithms are derived and they use different metrics for defining their corresponding loss functions. Let us define $A_k(\scalbb{x}_i)$ (and $B_k(\scalbb{y}_i)$) as the set of indices $j$ such that the $j$-th datapoint is a $k$ nearest neighbour of the $i$-th datapoint in high (and low) and $r_{k, i}(\scalbb{x}_j)$ (and $\widetilde{r}_{k, i}(\scalbb{y}_j)$) denote the corresponding ranks according to the distances from $i$-th datapoint in high (and low) dimensional space, with the nearest neighbour having rank $1$. Almost all of the existing metrics to evaluate dimensionality reduction work on the idea of preserving local structures, as follows

\begin{enumerate}
    \item Separability indices~\cite{greene2001feature,mthembu2008note} such as $\text{TSI}(\scalbb{X}, \scalbb{Y}) = 1 - \dfrac{1}{n}\sum_{i=1}^n \dfrac{1}{k} \sum_{j \in A_k(\scalbb{x}_i)} \scalbb{1}_{j \not\in B_k(\scalbb{y}_i) },$
    \item ``Trustworthiness" and ``continuity" measures proposed by Kaski et al.~\cite{kaski2003trustworthiness}, given by 
    \begin{align*}
        \text{Trustworthiness} & = 1 - \dfrac{2}{nk(2n-3k-1)} \sum_{i=1}^n \sum_{j \in B_k(\scalbb{y}_i) \backslash A_k(\scalbb{x}_i) } (r_{k, i}(\scalbb{y}_j) - k)\\
        \text{Continuity} & = 1 - \dfrac{2}{nk(2n - 3k-1)} \sum_{i=1}^n \sum_{j \in A_k(\scalbb{x}_i) \backslash B_k(\scalbb{y}_i) } (\widetilde{r}_{k, i}(\scalbb{x}_j) - k)
    \end{align*}
    These two quantities are created based on the intuitive idea that to measure trustworthiness of the output of an algorithm, if the low dimensional visualization shows that two points are close together, then they should also remain close together in the original dataset. The continuity is the other side of the coin, points in a vicinity in the original space should be projected close together, so that the visualization of a new test datapoint can be obtained based on the proximity calculations. 
\end{enumerate}

\subsection{Contributions of this study}\label{sec:aims}

Despite having a large number of different metrics for evaluating different dimensionality reduction algorithms, there is no consensus about a single metric. Also, most of these metrics, using some form of $k$-nearest neighbours, only focus on retaining the local structure, then they are only meaningful when the reduced data is used for clustering and classification using nearest neighbour-based algorithms. Our contribution would be to formally define a dimensionality reduction framework, under which we will provide a precise definition for an algorithm to qualify as a dimensionality reduction algorithm such that almost all of the existing dimensionality reduction algorithms will obey the definition. Then, we will consider two metrics under that general setup based on two important properties of the ideal dimensionality reduction algorithm, named as ``Trustability index'' and ``Consistency index''. 

While these two metrics will allow us to compare performances of different algorithms, we shall try to explore how these metrics can be reduced and what are the essential components of an algorithm that naturally correlates with these metrics. These, we shall try to build a dimensionality reduction algorithm by integrating those components which should guarantee better performances than the existing algorithms.

For brevity of presentation, the proofs of all the theorems are given in the Appendix~\ref{appendix:proofs}.

\section{A Generalized Framework}

\subsection{The Framework for Dimensionality Reduction}\label{sec:framework}

We start with a generic setup where $\rvbb{x}$ is a $\R^p$-valued random variable where $p$ is large. The support of the distribution of $\rvbb{x}$ is approximately a lower dimensional manifold. Denoting this intrinsic dimension as $d$, there should exist $\phi_1, \dots \phi_d$ functions from $\R^p \rightarrow \R$ such that each of them represents a coordinate system along the manifold. When one takes account of all these coordinates, the part left out must be a random variation that the coordinate system cannot take an account for. With this interpretation, we define the following.

\begin{definition}\label{defn:reduction}
A $d$-dimensional reduction for the $\R^p$ valued random variable $\rvbb{x}$ is a specification of the functions $\phi_1, \phi_2, \dots \phi_d$ with each $\phi_i : \R^p \rightarrow \R$ for $i=1,2,\dots d$ and $f : \R^d \rightarrow \R^p$ such that,

\begin{enumerate}
    \item $\rvbb{x} = f\left( \phi_1(\rvbb{x}), \dots \phi_d(\rvbb{x}) \right) + \epsilon(\rvbb{x})$ almost surely.
    \item $\phi_1, \phi_2, \dots \phi_d$ are measurable (may be continuous and smooth) functions. Each of these embeds the original data into the lower dimensional space, hence are called embedding functions.
    \item $f : \R^d \rightarrow \R^p$ is a measurable reconstruction function that approximately retrieves $\rvbb{x}$.
    \item $\epsilon(\rvbb{x})$ is the error in reconstruction, which is independent of the $\sigma$-field generated by $\{ \phi_1(\rvbb{x}), \dots \phi_d(\rvbb{x}) \}$.
    \item $\E(\epsilon(\rvbb{x})) = 0$.
\end{enumerate}
\end{definition}

Following are some of the examples to show that the above definition is a very general one.

\begin{example}
The first and trivial example of a dimensional reduction is the set of coordinate functions (or any permutation of that). Formally, a $p$-dimensional reduction for any $\R^p$-valued random variable $\rvbb{x}$ is given by $\phi_i(\scalbb{x}) = x_{\pi(i)}$, where $x_i$ is the $i$-th coordinate of $\scalbb{x}$ and $\pi : \{1, 2, \dots p \} \rightarrow \{1, 2, \dots p \}$ is a permutation function. Clearly, $\epsilon(\rvbb{x}) = 0$ and $f(x) = (x_{\pi^{-1}(1)}, \dots x_{\pi^{-1}(p)})$. 
\end{example}

\begin{example}
Let, $\rvbb{x}$ has a multivariate Gaussian distribution. Then, starting with any orthonormal basis $B = \{ \scalbb{v}_1, \dots, \scalbb{v}_d, \dots \scalbb{v}_p \}$ (for $d < p$), one may decompose, $\rvbb{x} = \sum_{i=1}^d \phi_i(\rvbb{x}) \scalbb{v}_i + \sum_{i = (d+1)}^p \psi_i(\rvbb{x}) \scalbb{v}_i$. Here, $\phi_i(\rvbb{x}) = \scalbb{v}_i\transpose \rvbb{x}$ are the embedding functions and $f(x_1, \dots x_d) = \sum_{i=1}^d x_i \scalbb{v}_i$ is the reconstruction function. Due to multivariate normality of $\rvbb{x}$ and orthogonality of $B$, the error $\sum_{i = (d+1)}^p \psi_i(\rvbb{x}) \scalbb{v}_i$ is independent of the $\sigma$-field generated by the embedding functions. 

Note that, the same example also holds if $\rvbb{x}$ follows a mixture of Gaussian distributions. Any linear dimensionality reduction technique that computes the embedding functions with respect to some orthogonal basis (including PCA) falls under this category.
\end{example}

\begin{theorem}\label{thm:uniqueness}
    If there exists two measurable functions $f$ and $g$ such that
    $$
    \rvbb{x} = f(\phi_1(\rvbb{x}), \dots \phi_d(\rvbb{x})) + \epsilon_1(\rvbb{x}) = g(\phi_1(\rvbb{x}), \dots \phi_d(\rvbb{x})) + \epsilon_2(\rvbb{x})
    $$
    \noindent i.e. both can be regarded as reconstruction function corresponding to the reduction $(\phi_1, \dots \phi_d)$, then both reconstructions are same almost surely.
\end{theorem}

The implication of this result is that to determine such a dimensional reduction, it is enough to specify the embedding functions $\phi_1, \phi_2, \dots \phi_d$ only, since in view of Theorem~\ref{thm:uniqueness}, any reconstruction function will output the same decoded version almost surely.

\subsection{Dimensionality reduction algorithms}

Now, turning to the real world datasets, we do not have the random variable $\rvbb{x}$, but instead we have i.i.d. data $\rvbb{x}_1, \dots \rvbb{x}_n$, each being $p$-dimensional. Therefore, any dimensionality reduction algorithm basically defines a loss function (convex or nonconvex)~\cite{van2009dimensionality} which is a function of the original datapoints, the embedding outputs $\rv{y}_{ij} = \phi_j(\rvbb{x}_i)$ for $j = 1, 2, \dots d$ and $i = 1, 2, \dots n$, the reconstruction function $f$, and the errors $\epsilon$. However, in view of Theorem~\ref{thm:uniqueness}, the specification of $f$ is not necessary when $\rv{y}_{ij}$ is known, and also when the reconstruction and the errors are known, specification of the original dataset is not needed. Therefore, one may consider a loss function as $L\left( \{ \phi_j(\rvbb{x}_i) : i = 1, \dots n; j = 1, \dots d \}, \epsilon(\rvbb{x}_1), \dots \epsilon(\rvbb{x}_n) \right)$ for $n$ observations. However, such specification requires different loss functions to be specified for each choice of the intrinsic dimension $d$ (which is not known a-priori), as the loss function above is a function from $\R^{nd} \times \R^{np}$ to $\R$. This requires special analysis for each of these loss functions, hence we extend this notion and consider only a single loss function $L : \R^{2np} \rightarrow \R$ given by 

$$
L\left( \{ \phi_j(\rvbb{x}_i) : i = 1, \dots n; j = 1, \dots p \}, \epsilon(\rvbb{x}_1), \dots \epsilon(\rvbb{x}_n) \right)
$$

\noindent while we perform the minimization of this loss subject to the constraints that $\phi_{j}(\rvbb{x}) = 0$ for all $j = (d+1),\dots p$ and for all $\rvbb{x}$, in order to estimate the embedding functions.

\begin{definition}\label{defn:dr-algorithm}
A dimensionality reduction algorithm $\mathcal{A}_n$ is a function from $\{ 1, 2, \dots p \} \times \R^{np}$ to a restricted class $\mathcal{C}$ of functions (say smooth, linear, piecewise linear etc.) along with a specification of the loss function $L : \R^{2np} \rightarrow \R$. This means, the algorithm takes the first argument as the number of embedding functions (i.e the lower dimension $d$) and the second argument as the $n$ datapoints each of dimension $p$. The algorithm then chooses the best possible reduction by minimizing the associated loss function for any $d = 1, 2, \dots p$,

\begin{multline}
(\widehat{\phi}_1, \dots \widehat{\phi}_d) = \arg\min_{\phi_1, \dots \phi_d \in \mathcal{C}} L\left( \{ \phi_j(\rvbb{x}_i) : i = 1, \dots n; j = 1, \dots p \}, \epsilon(\rvbb{x}_1), \dots \epsilon(\rvbb{x}_n) \right),\\
\text{subject to the constraints } \phi_j(\rvbb{x}) = 0 \ \forall j = (d+1), \dots p, \ \forall \rvbb{x} \in \R^p
\label{eqn:minimize-loss}
\end{multline}

\noindent The embedded lower dimensional representation is denoted by $\rvbb{Y} = \mathcal{A}_n(d, \rvbb{X})(\rvbb{X})$, where $\rvbb{Y}$ is a $n \times d$ matrix, whose rows are $\rvbb{y}_i = (\widehat{\phi}_1(\rvbb{x}_i), \dots \widehat{\phi}_d(\rvbb{x}_i))$.
\end{definition}

The loss function should not be any arbitrary one, there are some obvious restrictions.

\begin{enumerate}
    \item $L\left( \{ \phi_j(\rvbb{x}_i)\}_{i,j=1}^{n,p}, \{\epsilon(\rvbb{x}_i)\}_{i=1}^n \right) = L\left( \{ a_j + \phi_j(\rvbb{x}_i)\}_{i,j=1}^{n,p}, \{\epsilon(\rvbb{x}_i)\}_{i=1}^n \right)$, i.e. a translation to the embedding functions preserves the shape and the structure of the data, hence the loss function should not change. Note that, the translation amount may be different in different coordinates.
    \item $L\left( \{ \phi_j(\rvbb{x}_i)\}_{i,j=1}^{n,p}, \{\epsilon(\rvbb{x}_i)\}_{i=1}^n \right) = L\left( \{ (T\circ \phi)_j(\rvbb{x}_i)\}_{i,j=1}^{n,p}, \{\epsilon(\rvbb{x}_i)\}_{i=1}^n \right)$, where $T : \R^p \rightarrow \R^p$ is an orthonormal transformation. Since, orthonormal transformations simply rotate the original data, the structure and shape of the data is preserved.
    \item $L\left( \{ \phi_j(\rvbb{x}_i)\}_{i,j=1}^{n,p}, \{\epsilon(\rvbb{x}_i)\}_{i=1}^n \right) = L\left( \{ a\phi_j(\rvbb{x}_i)\}_{i,j=1}^{n,p}, \{\epsilon(\rvbb{x}_i)\}_{i=1}^n \right)$. If all the axis are scaled by the same amount, the loss function should not change. If the scaling factors are different, the shape of the function may distort.
    \item For any other transformation that distorts the shape of the data, the loss function should output something different.
    \item $L\left( \{ \phi_j(\rvbb{x}_i)\}_{i,j=1}^{n,p}, \{\epsilon(\rvbb{x}_i)\}_{i=1}^n \right) \geq 0$ and is equal to $0$ if and only if $\epsilon(\rvbb{x}_i) = 0$ for all $i = 1, \dots n$.
\end{enumerate}

\noindent Note that, the rotational invariance imply permutation invariance, hence the optimization problem given in Eq.~\eqref{eqn:minimize-loss} can also be re-framed by constraining $\phi_j(\rvbb{x}) = 0$ for any $(p-d)$ functions, while the optimization is done on the rest.

\subsection{Nonparametric estimator of out-of-the-sample reduction}\label{sec:out-sample-estimate}

Many dimensionality reduction algorithms (which specifically uses non-convex loss functions), only outputs the embedded lower dimensional representations $\rvbb{y}_1, \dots \rvbb{y}_n$ (each being $d$ dimensional) of the training dataset $\rvbb{x}_1, \dots \rvbb{x}_n$ (each being $p$ dimensional) without explicitly specifying the embedding functions $\phi_1,\dots \phi_d$, in accordance to Defn.~\ref{defn:dr-algorithm}. However, if one wants to find the embedding of a test datapoint $\rvbb{x}$, one needs to know either a nonparametric or parametric form of $\phi_j$'s. The following theorem precisely provides that under the general setup considered so far assuming the search space $\mathcal{C}$ for an algorithm $\mathcal{A}_n$ is subset of some Reproducible Kernel Hilbert Space (RKHS) (see Appendix~\ref{appendix:RKHS}).

\begin{theorem}\label{thm:find-embedding}
    For any dimensionality reduction algorithm $\mathcal{A}_n$, given that the restricted class of functions in the search space $\mathcal{C}$ is a subset of some RKHS for some kernel $k$, then the embedding functions $\phi_j$ for any $j = 1, 2, \dots d$ can be evaluated as,
    
    \begin{equation}
    \phi_j(\rvbb{x}) = \rvbb{y}_j\transpose \scalbb{K}^{-1} \scalbb{K}(\rvbb{x})
    \label{eqn:nonparametric-estimator}
    \end{equation}
    
    \noindent where, 

    $$
    \rvbb{y}_j = \begin{pmatrix}
    \rv{y}_{1j}\\
    \dots \\
    \rv{y}_{nj}
    \end{pmatrix}, \ 
    \scalbb{K} = \begin{pmatrix}
    k(\rvbb{x}_1, \rvbb{x}_1) & k(\rvbb{x}_1, \rvbb{x}_2) & \dots & k(\rvbb{x}_1, \rvbb{x}_n)\\
    k(\rvbb{x}_2, \rvbb{x}_1) & k(\rvbb{x}_2, \rvbb{x}_2) & \dots & k(\rvbb{x}_2, \rvbb{x}_n)\\
    \vdots & \vdots & \ddots & \vdots\\
    k(\rvbb{x}_n, \rvbb{x}_1) & k(\rvbb{x}_n, \rvbb{x}_2) & \dots & k(\rvbb{x}_n, \rvbb{x}_n)
    \end{pmatrix},
    \ 
    \text{ and } 
    \scalbb{K}(\rvbb{x}) = \begin{pmatrix}
    k(\rvbb{x}, \rvbb{x}_1)\\
    \dots \\
    k(\rvbb{x}, \rvbb{x}_n)
    \end{pmatrix},
    $$
    
    \noindent and $\rv{y}_{ij} = \phi_j(\rvbb{x}_i)$ for any $i = 1, 2, \dots n; j = 1, 2, \dots d$. This gives us a way to obtain nonparametric estimator of the low dimensional embedding on the basis of observed embedding for training samples, for any dimensionality reduction algorithm. 
\end{theorem}

\subsection{Building the reconstruction function}\label{sec:reconstruction}

This section concerns with the derivation of the reconstruction function $f$. Theorem~\ref{thm:uniqueness} says that given the specific embedding functions, the reconstruction function is unique almost surely, it does not provide a constructive way to produce it. This is difficult to obtain in its generality, because the independence of $\epsilon$ and $\sigma\left(\{ \phi_1, \dots \phi_d \}\right)$ is hard to verify. Therefore, we shall assume that each coordinate function $f_l$ of the reconstruction function $f$ belongs to a RKHS. This simplification actually allows one to reduce the independence condition to zero-covariance operator in RKHS domain (see Theorem~\ref{thm:independence-covariance} for details).

\begin{theorem}\label{thm:reconstruction}
    For any dimensionality reduction algorithm $\mathcal{A}_n$ with embedding functions $\phi_1(\rvbb{x}), \dots \phi_d(\rvbb{x})$ and the training dataset $(\rvbb{x}_1, \dots \rvbb{x}_n)$, if
    \begin{enumerate}
        \item Each $\phi_j(\cdot)$ is a member of some RKHS $\mathcal{H}_x$ generated by a universal kernel function $k_x$ given by $\phi_j(\rvbb{x}) = \sum_{i=1}^n \alpha_{ij} k_x(\rvbb{x}, \rvbb{x}_i)$ with $\E(k_x(\rvbb{x}, \rvbb{x})) < \infty$.
        \item Each coordinate function of $f$ i.e. $f_l$ is a member of some RKHS $\mathcal{H}_y$ generated by a universal kernel function $k_y$ with $\E(k_y(\rvbb{y}, \rvbb{y})) < \infty$.
        \item The probability measure $P_{x,y}$ defining the joint distribution of $(\rvbb{x}, \rvbb{y})$ has the support as a compact subset of $\R^p \times \R^d$.
    \end{enumerate}
    
    \noindent Then the coordinate functions of the reconstruction function $f$ can be expressed as,
    
    $$
    f_l(\rvbb{y}) = n^{-1}\sum_{i=1}^{n}\rvbb{x}_{il} + \sum_{i=1}^n \beta_{il} \left( k_y(\rvbb{y}, \rvbb{y}_i) - n^{-1}\sum_{j=1}^n k_y(\rvbb{y}_j,\rvbb{y}_i) \right), \ l = 1, 2, \dots p
    $$
    
    \noindent where $\rvbb{y}_i = (\rv{y}_{i1}, \dots \rv{y}_{id})\transpose$ and $\rv{y}_{ij} = \phi_j(\rvbb{x}_i)$ and $
    \beta_{il} = \scalbb{A} \left( \scalbb{A}\transpose \scalbb{A} \right)^{-1} \scalbb{c}_l$, and the matrix $\scalbb{A}_{n \times d}$ has entries $a_{rj} = (\scalbb{G}_y \scalbb{G}_x \scalbb{\alpha})_{rj}$, 
    
    $$
    \scalbb{\alpha} = \begin{bmatrix}
        \alpha_{11} & \dots & \alpha_{1d}\\
        \vdots & \ddots & \vdots \\
        \alpha_{n1} & \dots & \alpha_{nd}
    \end{bmatrix}_{n \times d}
    $$
    
    \noindent and $\scalbb{G}_x, \scalbb{G}_y$ are gram matrices corresponding to RKHS $\mathcal{H}_x$ and $\mathcal{H}_y$ as given in Eq.~\eqref{eqn:Gx-1}-Eq.~\eqref{eqn:Gx-2}, and $\scalbb{c}_l = (c_{1l}, \dots c_{dl})$ where 
    
    $$
    c_{jl} = n^{-1}\sum_{i=1}^n \rv{x}_{il}\rv{y}_{ij} - \left( n^{-1} \sum_{i=1}^n \rv{x}_{il}\right)\left( n^{-1} \sum_{i=1}^n \rv{y}_{ij}\right)
    $$
\end{theorem}

\subsection{Trustability and Consistency index}\label{sec:new-indices}

A dimensionality reduction algorithm $\mathcal{A}_n$ is characterized by the loss function $L(\phi_1, \dots \phi_p, \epsilon)$ according to Definition~\ref{defn:dr-algorithm}, hence one may measure the trustability of the algorithm based on the loss function itself. 

Consider the following situation: starting with any algorithm $\mathcal{A}_n$ and a $p$-dimensional dataset $\rvbb{X}_{n\times p}$, we ask the algorithm to reduce the data into a $p$-dimensional output $\rvbb{Y}_{n \times p}$. Clearly, the original data itself becomes the ``best'' possible reduction achieving $0$ loss as $\epsilon(\rvbb{x}_i) = 0$ for all $i=1,\dots n$. Since the algorithm minimizes the loss function given in Eq.~\eqref{eqn:minimize-loss}, which is invariant under translation, uniform scaling and rotation, we should have $\rvbb{Y} \approx \scalbb{1}_n\scalbb{\mu}\transpose + \lambda \scalbb{P}\rvbb{X}$ for some mean vector $\scalbb{\mu}$, real number $\lambda$ and orthogonal matrix $\scalbb{P}$. Therefore, the quantity $\min_{\scalbb{\mu}, \lambda, \scalbb{P}} \Vert \rvbb{Y} - \scalbb{1}_n\scalbb{\mu}\transpose - \lambda \scalbb{P}\rvbb{X} \Vert_2^2$ could be regarded a measure of how trustworthy the algorithm is in preserving shape and structure of the original dataset. The above minimization problem can be exactly solved by the Procrustes problem (see Appendix~\ref{appendix:procrustes}), which results in the construction of a \textbf{Trustability Index} for any dimensionality reduction algorithm.

\begin{definition}
The \textbf{Trustability Index} of any dimensionality reduction algorithm $\mathcal{A}_n$ for any dataset $\rvbb{X}_{n \times p}$ is given by,

$$
\text{TI}(\mathcal{A}_n, \rvbb{X}) := \sum \text{sing}(\scalbb{\Sigma}_{\rvbb{YY}}) - \left( \sum \text{sing}(\scalbb{\Sigma}_{\rvbb{XX}}) \right)^{-1} \left(\sum \text{sing}\scalbb{\Sigma}_{\rvbb{XY}} \right)^2
$$

\noindent where $\sum \text{sing}(\scalbb{A})$ denotes the sum of the singular values of the matrix $\scalbb{A}$, and
\begin{equation*}
    \scalbb{\Sigma}_{\rvbb{XX}} = (\rvbb{X} - \overline{\rvbb{X}})\transpose (\rvbb{X} - \overline{\rvbb{X}}),\
    \scalbb{\Sigma}_{\rvbb{XY}} = (\rvbb{X} - \overline{\rvbb{X}})\transpose (\rvbb{Y} - \overline{\rvbb{Y}}),\
    \scalbb{\Sigma}_{\rvbb{YY}} = (\rvbb{Y} - \overline{\rvbb{Y}})\transpose (\rvbb{Y} - \overline{\rvbb{Y}})
\end{equation*}

\noindent where $\rvbb{Y} = \mathcal{A}_n(p, \rvbb{X})(\rvbb{X})$ and $\overline{\rvbb{X}}, \overline{\rvbb{Y}}$ are the $n \times p$ matrices whose each row is $n^{-1}\scalbb{1}_n\transpose \rvbb{X}$ and $n^{-1}\scalbb{1}_n\transpose \rvbb{Y}$ respectively.
\end{definition}

\begin{remark}
    The more trustable an algorithm is, the less is the value of the index. This means, the most trustworthy algorithm attains the index value equal to $0$.
\end{remark}

\begin{example}
    The algorithm that has a $0$-$1$ loss function, outputs a random subset of size $d$ of the available $p$ variables. Thus, with $d = p$, it outputs a permutation of columns of $\rvbb{X}$, hence attains TI equal to 0. Similarly, for Principal Component Analysis, since $\rvbb{Y}$ is simply a orthogonal transformation of $\rvbb{X}$, it is also most trustable algorithm. 
\end{example}

Naturally, the above algorithms are not reasonably well if $d < p$. Therefore, along with Trustability Index, we need another index which is competing in nature. The problem in the aforementioned algorithms is that they have a large variability in its output under different transformations of the original data.

\begin{theorem}\label{thm:consistency-index}
    Let, $\scalbb{z} = T(\scalbb{x})$ be some invertible transformation $T: \R^p \rightarrow \R^p$. For a dimensionality reduction algorithm $\mathcal{A}_n$, with loss function $L$ and restricted class of functions $\mathcal{C}$, if 
    \begin{enumerate}
        \item For any $\phi \in \mathcal{C}$, both $\phi\circ T $ and $\phi \circ T^{-1}$ belong to $\mathcal{C}$.
        \item $\epsilon(\rvbb{x}_i) = \epsilon(\rvbb{z}_i) = \rvbb{\epsilon}_i$ for all samples $\rvbb{x}_1, \dots \rvbb{x}_n$ and transformed samples $\rvbb{z}_i = T(\rvbb{x}_i)$.
    \end{enumerate}
    
    \noindent Then,
    $$
    \min_{\phi_1, \dots \phi_d \in \mathcal{C}} L\left( \{ \phi_j(\rvbb{x}_i) \}_{i,j=1}^{n,p}, \{ \rvbb{\epsilon}_i \}_{i=1}^n \right) = \min_{\psi_1, \dots \psi_d \in \mathcal{C}} L\left( \{ \psi_j(\rvbb{z}_i) \}_{i,j=1}^{n,p}, \{ \rvbb{\epsilon}_i \}_{i=1}^n \right)
    $$
    
    \noindent where both the above minimization are performed subject to the conditions that $\phi_j(\rvbb{x}) = \psi_j(\rvbb{z}) = 0$ for all $j = (d+1), \dots p$.
\end{theorem}

\begin{figure}[ht]
    \centering
    \begin{subfigure}{0.49\linewidth}
        \centering
        \includegraphics[width = \textwidth]{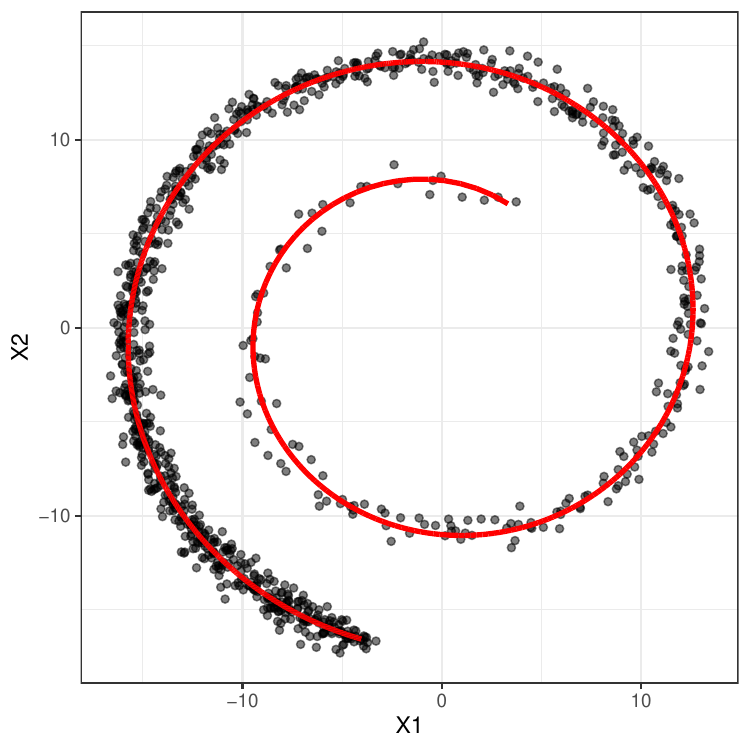}
    \end{subfigure}
    \hfill
    \begin{subfigure}{0.49\linewidth}
        \centering
        \includegraphics[width = \textwidth]{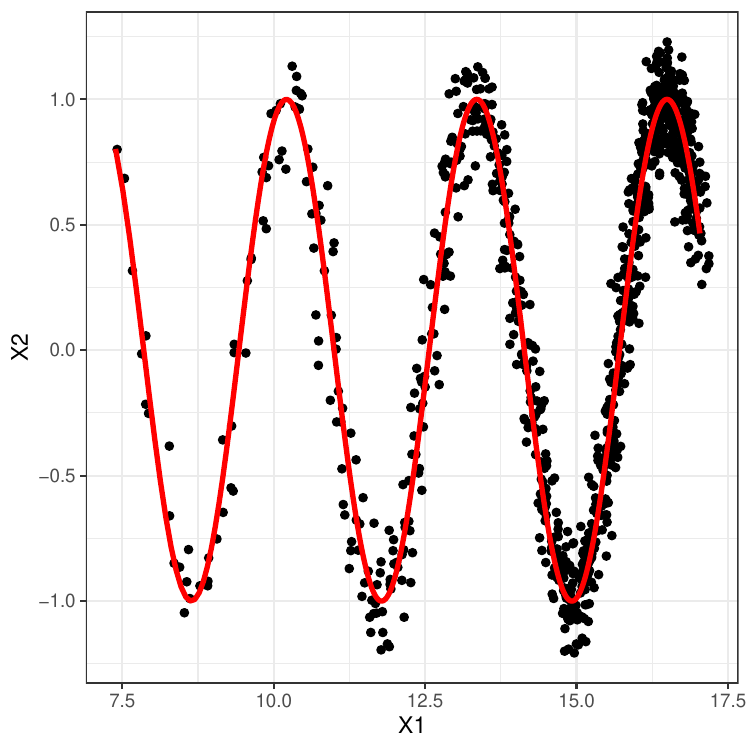}
    \end{subfigure}
    \caption{Two high dimensional data which have equivalent low dimensional underlying manifold}
    \label{fig:consistency-demo}
\end{figure}

Theorem~\ref{thm:consistency-index} opens up the possibility that even if the underlying intrinsic manifold is changed, as long as the errors in reconstruction remain the same, the loss also remains the same. For example, in Figure~\ref{fig:consistency-demo}, we demonstrate two different 2D datasets, both of which have an underlying one-dimensional manifold. Clearly, the underlying one-dimensional manifolds can be transformed into each other employing a simple one-to-one transformation, hence a desirable algorithm should output the same one-dimensional coordinate representation for both these datasets. In Theorem~\ref{thm:consistency-index}, the assumption that the errors remain the same even after the transformation is crucial. Therefore, even if $T(\scalbb{x}) = \lambda \scalbb{x}$ for any $\lambda > 0$ is a one-to-one transformation, it cannot be readily applied to the data as it may shrink (or inflate) both the errors and the extent of the low dimensional manifold simultaneously. We, on the other hand, want to apply this transformation only to the underlying manifold and keeping the reconstruction error fixed. Therefore, we may employ the following scheme.

\begin{enumerate}
    \item From $\rvbb{Y} = \mathcal{A}_n(d, \rvbb{X})(\rvbb{X})$, obtain the reconstruction function $f$ as shown in Section~\ref{sec:reconstruction}, and let the reconstructed version of $\rvbb{X}$ is denoted by $\rvbb{X}_{\mathcal{A}, d}$. 
    \item Fix any one-to-one transformation $T : \R^p \rightarrow \R^p$, and obtain $\rvbb{Z} = (\rvbb{z}_1, \dots \rvbb{z}_n)$, (the $n$ transformed datapoints) such that $T((\rvbb{X}_{\mathcal{A}, d})_i) = \rvbb{z}_i$ for any $i = 1, 2, \dots n$.
    \item Create new dataset $\tilde{\rvbb{X}} = \rvbb{Z} + (\rvbb{X} - \rvbb{X}_{\mathcal{A}, d})$.
    \item By means of Theorem~\ref{thm:consistency-index}, the output of the algorithm $\mathcal{A}_n$ applied on this new dataset i.e. $\mathcal{A}_n(d, \tilde{\rvbb{X}})(\tilde{\rvbb{X}})$ should be simply a translated, uniformly scaled and rotated version of $\rvbb{Y}$.
\end{enumerate}

\noindent Again, one can use Procrustes's problem (see Appendix~\ref{appendix:procrustes}) to measure the discrepancy and obtain the \textbf{consistency index}, which measures the variability of the output.

\begin{definition}[Consistency Index]\label{defn:ci-index}
     The \textbf{Consistency Index} for any algorithm $\mathcal{A}_n$ for any dataset $\rvbb{X}$ at intrinsic dimension $d$, is given by,
     $$
     \text{CI}(\mathcal{A}_n, \rvbb{X}, d) = \max_{T \text{ is invertible} }\min_{\scalbb{\mu}, \lambda, \scalbb{P}} \Vert \mathcal{A}_n(d, \tilde{\rvbb{X}})(\tilde{\rvbb{X}}) - \scalbb{1}_n\scalbb{\mu}\transpose - \lambda \scalbb{P}\mathcal{A}_n(d, \rvbb{X})(\rvbb{X}) \Vert_2^2
     $$
     \noindent where $\tilde{\rvbb{X}} = T(\rvbb{X}_{\mathcal{A}, d}) + (\rvbb{X} - \rvbb{X}_{\mathcal{A}, d})$. Similar to the trustability index, lower values of consistency index implies that algorithm is more consistent in its outputs.
\end{definition}

\begin{example}
    Let us consider another simple dimensionality reduction algorithm, which always outputs the same constant, i.e. $\phi_j(\rvbb{x}) = \text{constant}$ for any $\rvbb{x}$. Then, it is possible to achieve the highest consistency in output (i.e. the consistency index is equal to $0$), however, the algorithm is not at all trustable, since at $d = p$, the algorithm does not output a rotated, uniformly scaled and translated version of the original dataset.
\end{example}

In its generality, it is not possible to compute the consistency Index for an algorithm, since the maximization happens over all invertible functions $T$. Rather, we shall look at the maximization subject to the constraint that the invertible function $T$ lie in some RKHS $\mathcal{H}$ containing $\R^p$-valued functions. Note that, it is not true that all functions of the RKHS is invertible (for example, if $f \in \mathcal{H}$, then $(-f) \in \mathcal{H}$ and hence $f + (-f) \equiv 0 \in \mathcal{H}$, which is non-invertible). Also, we would like the invertible functions to be such that it does not scale the dataset too small or too large. Therefore, we restrict our attention to the subclass of functions (i.e. a subset of the RKHS $\mathcal{H}$ generated by a matrix-valued kernel $k$),

$$\mathcal{H}^\ast = \left\{ f : f \in \mathcal{H}, f(\cdot) = \sum_{i=1}^\infty \sum_{j=1}^p \beta_{ij}k(\cdot, \rvbb{x}_i)\scalbb{e}_j, 0 \leq \beta_{ij} \leq 1, \sum_{i=1}^\infty \sum_{j=1}^p \beta_{ij} = 1, \text{ and } f \text{ is invertible} \right\}.
$$

\noindent Even then, since the index itself depends on the particular algorithm $\mathcal{A}_n$ we choose, unless we have an understanding of how the algorithm behaves, it is hard to evaluate the maximum value analytically. Since $\mathcal{H}^\ast$ is an infinite set, it is also not computationally possible to maximize by searching over the whole space $\mathcal{H}^\ast$. Under certain simple conditions, this maximization over all $T \in \mathcal{H}^\ast$ can be reduced to maximization over only a finite set of functions, which make this problem computationally tractable. The following theorem proves this fact.

\begin{theorem}\label{thm:ci-index-boundary}
    If $S(T) = \min_{\scalbb{\mu}, \lambda, \scalbb{P}} \Vert \mathcal{A}_n(d, \tilde{\rvbb{X}})(\tilde{\rvbb{X}}) - \scalbb{1}_n\scalbb{\mu}\transpose - \lambda \scalbb{P}\mathcal{A}_n(d, \rvbb{X})(\rvbb{X}) \Vert_2^2$, where $\tilde{\rvbb{X}}$ is as given in Definition~\ref{defn:ci-index}, is a convex function in $T$, then $\max_{T \in \mathcal{H}^\ast } S(T) = \max_{T \in \partial\mathcal{H}^\ast } S(T)$, where
    $$
    \partial\mathcal{H}^\ast = \left\{ k(\cdot, \rvbb{x}_i)\scalbb{e}_j : i = 1, \dots n;\ j = 1, \dots p \right\}.
    $$
\end{theorem}

\noindent Since, $\partial\mathcal{H}^\ast$ is now a finite set, therefore, we can use it to compute the consistency Index for any algorithm $\mathcal{A}_n$ approximately. Therefore, we have the following approximate version of consistency Index, called Tractable Consistency Index.

\begin{definition}
    \textbf{Tractable Consistency Index} for any algorithm $\mathcal{A}_n$ for any dataset $\rvbb{X}$ at intrinsic dimension $d$ is given by,
    $$
     \text{TCI}(\mathcal{A}_n, \rvbb{X}, d) = \max_{T \in \partial\mathcal{H}^\ast }\min_{\scalbb{\mu}, \lambda, \scalbb{P}} \Vert \mathcal{A}_n(d, \tilde{\rvbb{X}})(\tilde{\rvbb{X}}) - \scalbb{1}_n\scalbb{\mu}\transpose - \lambda \scalbb{P}\mathcal{A}_n(d, \rvbb{X})(\rvbb{X}) \Vert_2^2
     $$
     \noindent where $\tilde{\rvbb{X}} = T(\rvbb{X}_{\mathcal{A}, d}) + (\rvbb{X} - \rvbb{X}_{\mathcal{A}, d})$, and $
     \partial\mathcal{H}^\ast = \left\{ k(\cdot, \rvbb{x}_i)\scalbb{e}_j : i = 1, \dots n;\ j = 1, \dots p \right\}
     $ where $k(\cdot, \cdot)$ is a positive definite matrix-valued kernel function from $\R^p \times \R^p$ to $\R^p \times \R^p$. 
\end{definition}

\section{Localized Skeletonization and Dimensionality Reduction (LSDR)}

\subsection{Manifold Approximation Graph}\label{sec:G-alpha}

\begin{figure}[ht]
    \centering
    \begin{subfigure}{0.45\textwidth}
        \includegraphics[width = \textwidth]{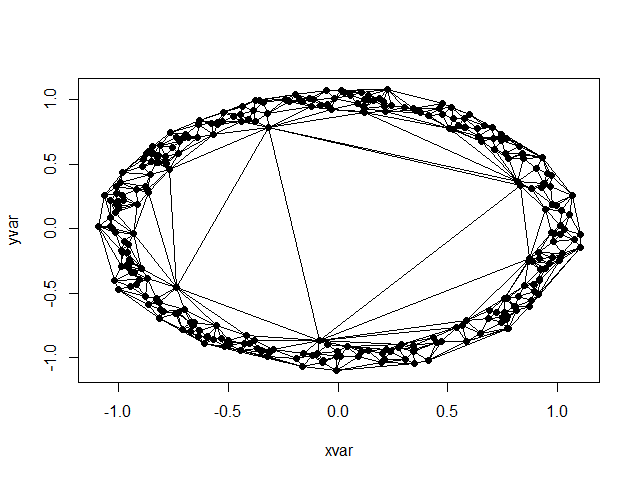}
        \caption{Delaunay triangulation for points on a circle}
        \label{fig:delaunay-1}
    \end{subfigure}
    \begin{subfigure}{0.54\textwidth}
        \includegraphics[width = \textwidth]{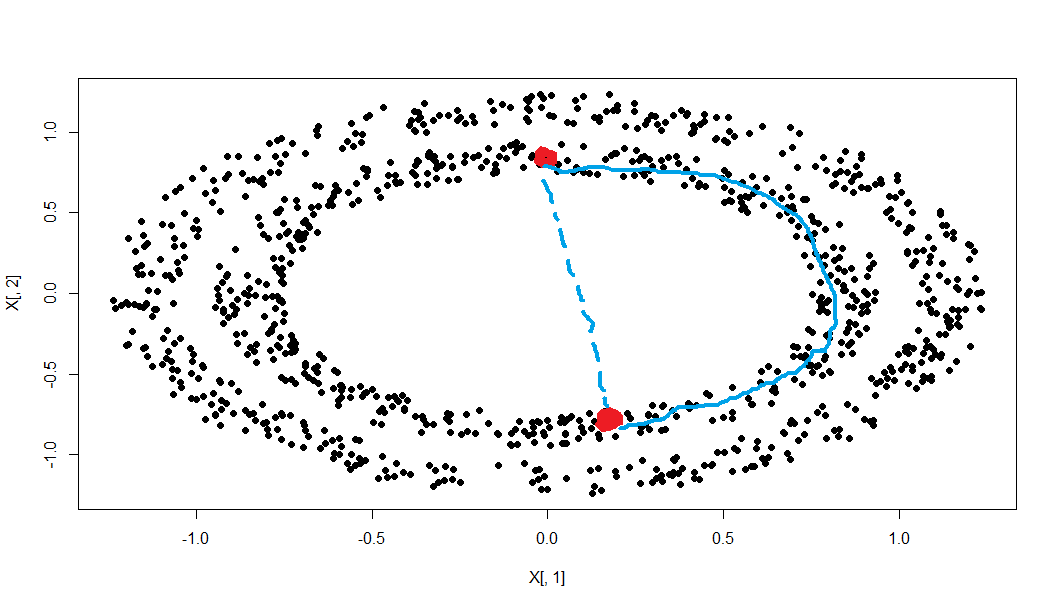}
        \caption{Euclidean distance vs Manifold distance}
        \label{fig:delaunay-2}
    \end{subfigure}
    \begin{subfigure}{0.49\textwidth}
        \includegraphics[width = \textwidth]{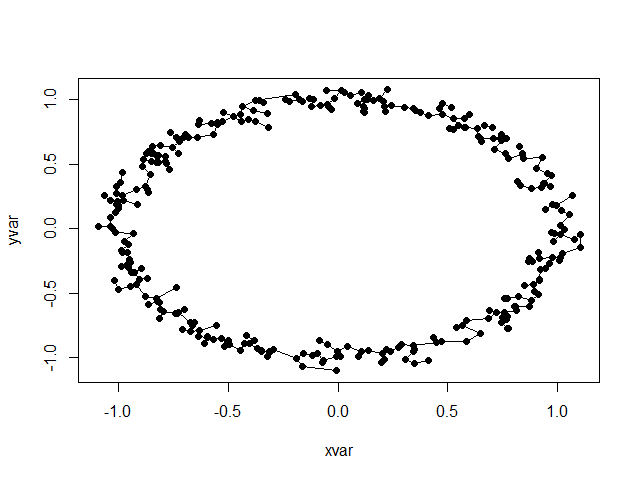}
        \caption{Euclidean MCST}
        \label{fig:delaunay-3}
    \end{subfigure}
    \begin{subfigure}{0.49\textwidth}
        \includegraphics[width = \textwidth]{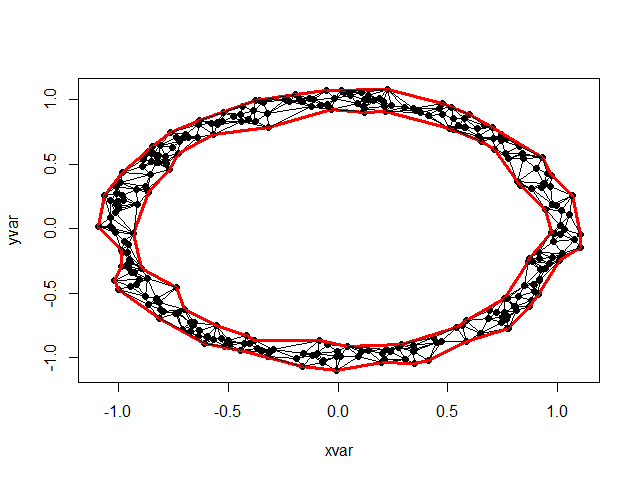}
        \caption{Manifold approximating graph $G_\alpha(P)$}
        \label{fig:delaunay-4}
    \end{subfigure}
    \caption{Effect of Delaunay triangulation, Minimum cost spanning tree, and the manifold approximating graph $G_\alpha(P)$}
\end{figure}

As discussed by Maaten and Hinton~\cite{van2008visualizing}, one cannot hope to preserve all pairwise distances in the lower dimensional space, since the same $\epsilon$-ball in $\R^p$ can hold more equidistant points than in $\R^d$ (with $d < p$). This phenomenon is popularly referred to as the ``Crowding problem". Therefore, the authors argue that only preserving the local distances can be achieved. A visualisation of this argument is given in Figure~\ref{fig:delaunay-2}, where the two red points shown have very different distances in Euclidean sense and in the distance through the manifold sense. As we notice, this distance through the manifold can be measured by connecting several smaller localized distances, where a point is connected to its local neighbourhood and only these small distances are summed up to calculate the larger distances. Hence, if one can obtain a graph $G$ connecting these datapoints by edges whenever two points are in close proximity, then the corresponding graph would be an approximation of the underlying manifold. Consequently, the distance between any two points through the manifold can be computed by the length of the shortest path between those two points through the graph. Delaunay tessellation and Minimum Cost Spanning Tree (MCST) are two integral components of creation of such approximation graphs.

Delaunay triangulation (also known as a Delone triangulation, in short DT) for a given set $P$ of discrete points in a general position on a plane, is a triangulation $DT(P)$ such that no point in $P$ is inside the circumcircle of any triangle in $DT(P)$. It can be shown that this triangulation avoids narrow triangles and maximizes the sum of the minimum angles of all triangles~\cite{de2008computational}, hence provide a good triangular approximation of the underlying mesh. Delaunay~\cite{delaunay1934sphere} extended the idea of triangulation into tessellation which can be obtained for any set of points $P$ lying in general position in any dimensional Euclidean vector space. Figure~\ref{fig:delaunay-1} shows one such example of DT for points distributed on a circle on $\R^2$. One could simply use DT as the required graph $G$, but there are some longer edges that connect points lying on the opposite parts of the circle as DT outputs a graph with only triangles. However, if Delaunay triangulation is modified to have polygonal tessellation as well, then such tessellation would allow us to create the required graph.

Let, $P = \{ \scalbb{x}_1, \dots \scalbb{x}_n \}$ be a set of points lying in the Euclidean space $\R^p$, and its Delaunay triangulation is denoted as $DT(P) = (G_{DT}(P), E_{DT}(P))$. Consider a point $\scalbb{x}_i$ and let $e_1, \dots e_k \in E_{DT}(P)$ be the associated edges with one endpoint equal to $\scalbb{x}_i$. Then, considering a multivariate normal distribution centered at $\scalbb{x}_i$ with variance $\sigma I_p$, one can show that $\dfrac{\Vert e_j \Vert^2}{\sigma^2}$ has a $\chi^2_p$ distribution. Similarly, $\dfrac{1}{\sigma^2}\sum_{j=1}^k \Vert e_j \Vert^2 \sim \Gamma(kp/2, 1/2)$ distribution. Therefore, one can show that 

$$
\dfrac{\Vert e_j\Vert^2}{\sum_{j' = 1}^k \Vert e_{j'}\Vert^2} \sim \text{Beta}\left( \dfrac{p}{2}, \dfrac{(k-1)p}{2} \right)
$$

\noindent Hence, one can choose some $\alpha \approx 0.95$ and use the above statistic and compare it with the threshold $\text{Beta}\left( \frac{p}{2}, \frac{(k-1)p}{2}, \alpha \right)$ which denotes the $\alpha$-th quantile of $\text{Beta}\left( \frac{p}{2}, \frac{(k-1)p}{2} \right)$ distribution, and remove the particular edge $e_j$ if the statistic exceeds this threshold. This is similar to a statistical hypothesis testing setup where the null hypothesis assumes that the points in the locality of $\scalbb{x}_i$ which are joined by DT follows a multivariate normal distribution. When the null hypothesis is false, the high value of the test statistic would indicate the presence of a longer edge $e_j$ in DT, hence the removal of this edge make the graph $DT(P)$ a better approximation to the manifold.

The edge removal process will only keep the small edges in $DT(P)$. However, if there are two or more clusters present in the data, then some global structure or characteristic need to be retained (Otherwise, similar problems as described in Section~\ref{sec:flaws} and Figure~\ref{fig:umap-problem} may occur). This global structure can only be preserved through some large edges, which may be deleted in the aforementioned edge removal process. In this case, the Euclidean minimum cost spanning tree (MCST) provides us with useful insights. Euclidean MCST is a subgraph of the Delaunay tessellation $DT(P)$ such that it is a spanning tree with the sum of its edge lengths equal to the minimum possible value. The Euclidean MCST for the same circular data is shown in Figure~\ref{fig:delaunay-3}. The following result provides a theoretical justification of why MCST is a reasonable choice to preserve global structure between several clusters.

\begin{theorem}\label{thm:mcst-1}
    Let us assume we have two groups of points given by,
    $$
    \scalbb{x} = \scalbb{\mu}_1 \scalbb{1}_{C_1} + \scalbb{\mu}_2 \scalbb{1}_{C_2} + \epsilon
    $$
    such that $|C_1| = m, |C_2| = n$ and the errors $\epsilon$ are independently distributed. Assume that for any $\delta > 0$, $$P(\Vert \scalbb{x}_i - \scalbb{x}_j \Vert_2 > \delta) < 1$$ for any $\scalbb{x}_i, \scalbb{x}_j$ lying in the same class. Then, the probability that the euclidean MCST of these set of points has two edges connecting different classes tends to zero as $m, n \rightarrow \infty$, namely the MCST would have two subtrees, each of them being a spanning tree for each of the class, and there will be only one edge that connects two classes.    
\end{theorem}

\begin{remark}
    The condition required for Theorem~\ref{thm:mcst-1} is satisfied if $\epsilon$ has a continuous distribution with support containing $0$. It simply requires that the points within the same cluster can be probabilistically arbitrarily close, which if not met, indicates that the cluster itself can be broken into more than one sub-clusters. Also, Theorem~\ref{thm:mcst-1} holds for more than two clusters as well, where between any two clusters there is at most a single edge of MCST.
\end{remark}

Theorem~\ref{thm:mcst-1} shows that the edges which are part of the MCST are extremely crucial to approximate the underlying manifold of a point set $P$. Therefore, we start with the Delaunay triangulation $DT(P)$ of the point set $P$, then perform the edge removal process by comparing the edge length based test statistic with the cut-off from beta distribution, and only remove the edge it is not a part of the MCST. This whole procedure yields us a graph $G_\alpha(P)$ for the point set. An example of such $G_\alpha(P)$ is shown in Figure~\ref{fig:delaunay-4}. We call such a graph as $\alpha$-approximation of a manifold generated by the point set $P$.

\subsection{Skeletonization}\label{sec:skeletonization}

The next step in our algorithm is to obtain a skeletal representation of the manifold from its graph approximation. This skeletal representation would allow one to obtain points lying on the underlying low-dimensional manifold (i.e. without any errors). In mathematical terms, a skeletal representation of a point set $P = \{ \scalbb{x}_1, \scalbb{x}_2, \dots \scalbb{x}_n \}$ is given by a subset of points $Sk(P) = \{ \scalbb{x}_{i_1}, \dots \scalbb{x}_{i_k} \}$ such that $\scalbb{x}_{i_j} = f(\phi_1(\scalbb{x}_{i_j}), \dots, \phi_d(\scalbb{x}_{i_j}))
$ i.e. $\epsilon(\scalbb{x}_{i_j}) = 0$ for all $j = 1, 2, \dots k$ (see Definition~\ref{defn:reduction}). Let us now consider a single point $\scalbb{x} = f(\phi_1(\scalbb{x}), \dots \phi_d(\scalbb{x})) + \epsilon(\scalbb{x})$ where $\scalbb{x} \in \R^p$. Let $\scalbb{v} \in \R^p$ be a vector such that $\scalbb{v}\transpose \nabla \phi_j(\scalbb{x}) = 0$ for all $j = 1, 2, \dots d$, i.e. $v$ is locally orthogonal to each of the embedding functions. Applying Taylor's theorem, we then get $\phi_j(\scalbb{x} + \scalbb{v}) \approx \phi_j(\scalbb{x}) + \scalbb{v}\transpose \nabla \phi_j(\scalbb{x}) = \phi_j(\scalbb{x})$, i.e. the embedding remains approximately constant on the line $\scalbb{x} + \lambda \scalbb{v}$ where $\lambda \in \R$ is a parameter. Observe that, in this case, $\epsilon(\scalbb{x} + \lambda\scalbb{v}) \approx \epsilon(\scalbb{x}) + \lambda\scalbb{v}$, which shows that the error is locally linear on that orthogonal line. This realization then allows us to define boundaries of the manifold in mathematical terms, a point $\scalbb{x}_b$ is a boundary point if either $\scalbb{x}_b = \displaystyle\max_{\scalbb{x} + \lambda \scalbb{v} \in P} (\scalbb{x} + \lambda \scalbb{v})$ or $\scalbb{x}_b = \displaystyle\min_{\scalbb{x} + \lambda \scalbb{v} \in P} (\scalbb{x} + \lambda \scalbb{v})$ for some $\scalbb{x} \in Sk(P)$. The set of all boundary points are denoted as $B(P)$.

Now consider a point $\scalbb{x} \in Sk(P)$ and let $\scalbb{x}_b \in B(P)$ be a corresponding boundary point. Assume that $\scalbb{x}_b$ is a part of a $(p-1)$-dimensional hyperplane $\mathcal{P}$ constituted of some edges in the graph $G_\alpha(P)$ obtained from Delaunay tessellation. If this hyperplane is part of two $(p+1)$-simplices $S_1$ and $S_2$, they must lie in the opposite side of $\mathcal{P}$, therefore, the ray parallel to the vector $(\scalbb{x}_b - \scalbb{x})$ starting from $\scalbb{x}$ must intersect $S_1$ and $S_2$ both. Clearly, for one of this intersection point, the distance from the intersection to $\scalbb{x}$ must be larger than $\Vert \scalbb{x}_b - \scalbb{x}\Vert$ which contradicts the extremality of the boundary point. Therefore, we have the following result.

\begin{theorem}\label{thm:boundary}
    Any boundary point of the manifold in $\R^p$ must lie on a $(p-1)$ dimensional hyperplane which is a part of atmost one $(p+1)$-simplex.
\end{theorem}

\begin{figure}[ht]
    \centering
    \begin{subfigure}{0.54\textwidth}
        \includegraphics[width = \textwidth]{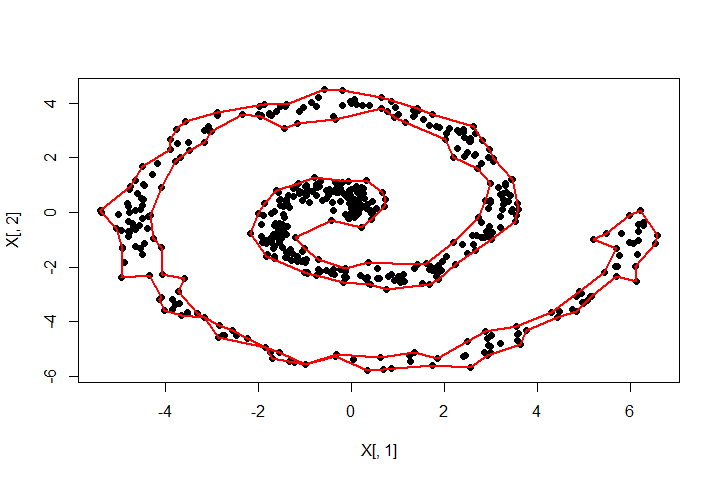}
        \caption{$G_\alpha(P)$ and its boundary}
        \label{fig:skeleton-1}
    \end{subfigure}
    \begin{subfigure}{0.44\textwidth}
        \includegraphics[width = \textwidth]{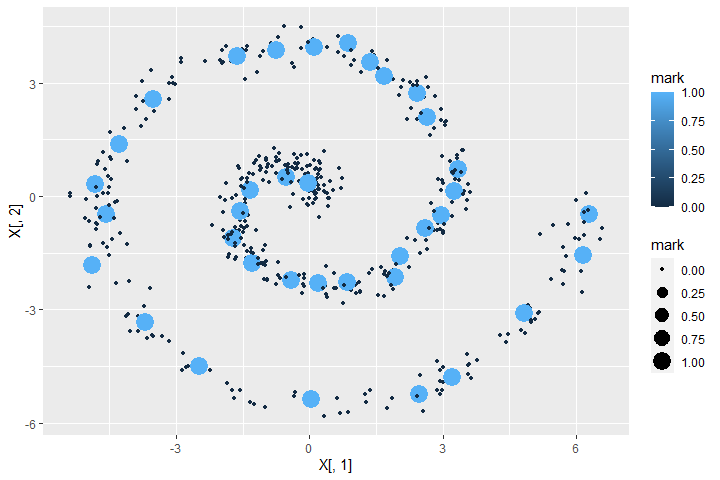}
        \caption{$G_\alpha(P)$ and its skeletal representation}
        \label{fig:skeleton-2}
    \end{subfigure}
    \caption{A point set $P$ in the form of a one dimensional manifold in a 2D plane, its approximation graph $G_\alpha(P)$ with its boundary and skeletal representation.}
\end{figure}

\noindent Due to Theorem~\ref{thm:boundary}, it is possible to obtain the boundary points by simply verifying the condition for each vertex in $G_\alpha(P)$. Figure~\ref{fig:skeleton-1} shows the boundary of $G_\alpha(P)$ computed in this way for a point set arranged in a spiral. However, the significance of this procedure is that it enables us to detect the boundary without any reference to the skeletal representation, hence one can obtain skeletal representation from the boundary itself. To see this, note that the skeletal points are given by those where the error is equal to $0$, and the boundary points are essentially where the error attains its maximum value in magnitude. Since the error is locally linear as we have shown, one can move away from the boundary inside the point cloud which should necessarily decrease the error. Therefore, the point that is the most distance away from the boundaries should have the smallest error in magnitude, which may serve our purpose as a skeletal point.

In order to put these ideas into a working algorithm, for every point $\scalbb{x} \in P$, we compute its distance from the boundary as 

$$
d_G(\scalbb{x}, B(P)) = \min_{\scalbb{x}_b \in B(P)} d_G(\scalbb{x}, \scalbb{x}_b)
$$

\noindent where $d_G(\scalbb{x}_i, \scalbb{x}_j)$ is the sum of the length of edges in the shortest path between $\scalbb{x}_i$ and $\scalbb{x}_j$ through the graph $G_\alpha(P)$. Then, we iterate over all points $\scalbb{x} \in P$ again and mark them as a skeletal point if 

$$
d_G(\scalbb{x}, B(P)) = \max_{\scalbb{w} \in NN_k(\scalbb{x}) } d_G(\scalbb{w}, B(P)) 
$$

\noindent where $NN_k(\scalbb{x})$ denotes the set of $k$-nearest neighbours of $\scalbb{x}$. When $k$ is sufficiently small, the above criterion assures that the identified skeletal points lie in the more inner region than their neighbours. This idea of the skeleton as the most distant points from its boundary dates back to Blum, 1967~\cite{blum1967transformation} and is later proved to have many optimality criterion~\cite{siddiqi2008medial}. Such a skeletal representation is shown in Figure~\ref{fig:skeleton-2} where the skeletal points are identified with a larger size and different colour. As seen from Figure~\ref{fig:skeleton-2}, the skeleton points lie approximately in the middle of the one-dimensional spiral manifold.

\subsection{Multidimensional Scaling and Kernel embedding}

Multidimensional scaling (MDS) attempts to find a low dimensional representation $\{ \scalbb{y}_1, \dots \scalbb{y}_n \}$ with each $\scalbb{y}_i \in \R^d$ such that the usual Euclidean distance between $\scalbb{y}_i$ and $\scalbb{y}_j$ approximately equals $q_{ij}$ for all $1\leq i < j\leq n$, for pre-specified choice of $q_{ij}$~\cite{borg2005modern}. Therefore, for pre-specified values of the distances, it attempts to find points in a $d$-dimensional space that preserves the distances as much as possible, by minimizing the stress function

$$
\text{Stress}_D(\scalbb{y}_1, \dots \scalbb{y}_n) = \left(\sum_{1 \leq i < j \leq n} (q_{ij} - \Vert \scalbb{y}_i - \scalbb{y}_j \Vert)^2 \right)^{1/2}.
$$

\noindent This variation is known as the metric MDS. Since the distances are invariant under translation and rotational transformation of $\scalbb{y}_i$'s, the solution to metric MDS is not unique. However, one can show that a condition similar to the desired values of trustability index mentioned in Section~\ref{sec:new-indices} can be achieved by metric MDS.

\begin{theorem}\label{thm:mds-1}
    Let us assume for a dataset $\{ \scalbb{x}_1, \dots \scalbb{x}_n\}$, the metric MDS is applied on $q_{ij} = \Vert \scalbb{x}_i - \scalbb{x}_j \Vert$ to obtain an embedding $\scalbb{y}_i \in \R^p$. Then, $\scalbb{y}_i = \scalbb{v} + \scalbb{P}\scalbb{x}_i$ for all $i = 1, 2, \dots n$ for some $\scalbb{a} \in \R^p$ and some orthogonal matrix $\scalbb{P}$.
\end{theorem}

Theorem~\ref{thm:mds-1} shows the connection between metric MDS and the trustability index. In other words, it says that as long as the original distances can be recovered in some proposed DR algorithm, a metric MDS step then ensures that the DR algorithm remains optimal with respect to the trustability index. However, the distances which are needed to pass into the metric MDS step is the Euclidean distance. In contrast, if we pass an approximation to the geodesic distance in the metric MDS setup, then one can show under certain reasonable assumptions that the corresponding DR algorithm will be optimal with respect to the consistency index defined earlier. 

\begin{theorem}\label{thm:mds-2}
    Let $f : \R^d \rightarrow \R^p$ be a smooth co-ordinate function of a Riemannian manifold $(\mathcal{M}, g)$ such that a unit-velocity curve in $\R^d$ induces a constant velocity curve in $\mathcal{M}$. Let, $\phi: \R^p \rightarrow \R^p$ be a local diffeomorphism such that $\phi \circ f: \R^d \rightarrow \R^p$ be the coordinate function of another Riemannian manifold $(\mathcal{M}', g')$ with 
    $$
    g(v, w) = c g'(\phi^\ast v, \phi^\ast w), \qquad \forall v, w \in \mathcal{M}
    $$  
    \noindent where $\phi^\ast$ is the push-forward and $c$ is some positive real constant. Then, for any $\{ \scalbb{\theta}_1, \dots \scalbb{\theta}_n \}$ with each $\scalbb{\theta}_i \in \R^d$, the $d$-dimensional output $\{ \scalbb{y}_1, \dots \scalbb{y}_n \}$ of metric MDS of $\{ f(\scalbb{\theta}_1), \dots f(\scalbb{\theta}_n) \}$ and the $d$-dimensional output $\{ \scalbb{z}_1, \dots \scalbb{z}_n \}$ of metric MDS of $\{ \phi\circ f(\scalbb{\theta}_1), \dots \phi\circ f(\scalbb{\theta}_n) \}$ are related by 

    $$
    \scalbb{z}_i = \scalbb{v} + \lambda\scalbb{Py}_i, \qquad i = 1, 2, \dots n
    $$
    \noindent for some fixed $\scalbb{v} \in \R^p, \lambda \in \R$ and some orthogonal matrix $\scalbb{P}$.
\end{theorem}

From the skeletal points obtained in Section~\ref{sec:skeletonization}, one may assume that these points lie on the low dimensional manifold in $\R^p$. The assumptions made in Theorem~\ref{thm:mds-2} enables one to approximate the geodesic distance on this manifold by summing up local Euclidean distances through the approximation graph $G_\alpha$. Therefore, if one can use the length of the shortest paths between two skeletal points on $G_\alpha$ as an approximation to the geodesic distance, then Theorem~\ref{thm:mds-2} ensures that the output of the metric MDS of these distances would ensure a dimension reduction for the skeletal points and that DR would be optimal with respect to the consistency index. This is precisely what we will do to find the low dimensional embedding of the skeletal points.

However, the metric MDS would be unable to properly find low dimensional embedding for the non-skeletal points, since they do not lie exactly on the manifold. Therefore, given that the skeletal points sufficiently cover the entire structure of the manifold, one may use the technique mentioned in Section~\ref{sec:out-sample-estimate} to obtain a low-dimensional embedding of non-skeletal points. While metric MDS preserves the global structure as much as possible by considering only the skeletal points, the local structure is recovered by the kernel embedding technique, hence the whole procedure is expected to balance the local and global structures found in the high dimensional data nicely.

\begin{algorithm}[ht]
    \SetAlgoLined
    \KwData{$\scalbb{x}_1, \dots \scalbb{x}_n$ where each $\scalbb{x}_i \in \R^p$}
    \KwIn{$\alpha, k$, embedding dimension $d$ and some kernel $K(\cdot, \cdot)$}
    \KwOut{$\scalbb{y}_1, \dots \scalbb{y}_n$ where each $\scalbb{y}_i \in \R^d$ ($d << p$)}
    \Begin(\textbf{Step 1}){
        Let $P = \{ \scalbb{x}_1, \dots \scalbb{x}_n \}$\;
        Compute $p$-dimensional Delaunay tessellation $DT(P)$\;
        Compute Euclidean MCST of the point set $P$\;
        \ForEach{point $\scalbb{x}_i$}{
            Let $e_1, \dots e_{d(\scalbb{x}_i)}$ be the edges with one endpoint equal to $\scalbb{x}_i$ where $d(\scalbb{x}_i)$ is the degree of $\scalbb{x}_i$\;
            \ForEach{edge $e_j$}{
                \If{$e_j \notin E(MCST(P))$}{
                    Compute $T_j = \dfrac{\Vert e_j\Vert^2}{\sum_{j' = 1}^{d(\scalbb{x}_i)}\Vert e_{j'}\Vert^2 }$\;
                    \lIf{$T_j > \text{Beta}(p/2, (d(\scalbb{x}_i) - 1)/2, \alpha)$}{
                        remove $e_j$ from the graph
                    }
                }
            }
        }
        Call the resulting graph $G_\alpha(P)$\;
        Use Dijkstra's algorithm to calculate all graph-based pairwise distances\;
        Let $d_G(\scalbb{x}_i, \scalbb{x}_j)$ be the graph-based distance between $\scalbb{x}_i$ and $\scalbb{x}_j$\;
    }
    \Begin(\textbf{Step 2}){
        Intialize $B(P) = \{\}$\;
        \ForEach{edges $e \in E(G_\alpha(P))$}{
            $Count =$ Number of $(p+1)$-simplices with $e$ as an edge\;
            \lIf{$\text{Count} \leq 1$}{insert edge $e$ into $B(P)$}
        }
        \ForEach{point $\scalbb{x}_i$}{
            Compute $d_B(\scalbb{x}_i) = \min_{\scalbb{x} \in B(P)} d_{G}(\scalbb{x}_i, \scalbb{x})$\;
        }
        Initialize $Sk(P) = \{\}$\;
        \ForEach{point $\scalbb{x}_i$}{
            Calculate the $k$-nearest neighbours $NN_k(\scalbb{x}_i)$ according to $d_G$ metric\;
            \lIf{$d_B(\scalbb{x}_i) = \max_{\scalbb{v} \in NN_k(\scalbb{x}_i)}d_B(\scalbb{v})$}{Mark $\scalbb{x}_i$ as skeletal point and add it to $Sk(P)$}
        }
    }
    \Begin(\textbf{Step 3}){
        Let $Sk(P) = \{ \scalbb{x}_{i_1}, \dots \scalbb{x}_{i_r} \}$\;
        Apply metric MDS on the distance matrix $Q$ with entries $q_{jj'} = d_G(\scalbb{x}_{i_j}, \scalbb{x}_{i_{j'}})$ to obtain $d$-dimensional representation $\scalbb{y}_{i_1}, \dots \scalbb{y}_{i_r}$\;
        \ForEach{$i = 1, 2, \dots n$}{
            $\scalbb{y}_i = \sum_{j=1}^r \scalbb{y}_{i_j}K(\scalbb{x}_i, \scalbb{x}_{i_j}) / \sum_{j=1}^r K(\scalbb{x}_i, \scalbb{x}_{i_j})$\; 
        }
    }
    \caption{Localized Skeletonization and Dimensionality Reduction (LSDR)}
    \label{algo:lsdr}
\end{algorithm}

The final algorithm, which we call ``Localized Skeletonization and Dimensional Reduction'' (LSDR) is described in Algorithm~\ref{algo:lsdr}. While the algorithm in its generality is described by three parameters $\alpha, k$ and the kernel of choice, we found that the standard choice of $\alpha \approx 0.9$ or $0.95$ (the usual choice of statistical significance), $k = 3$ works pretty well. The required kernel could be chosen as the Gaussian kernel with a properly chosen bandwidth in some data-dependent way. The choice of the bandwidth $\sigma$ should be such that if the one considers a $\sigma$-ball around the skeletal points, no point should lie outside of them. In other words, one recommendation could be to take 

$$
\sigma = \max_{\scalbb{x}_i \in Sk(P)} \sqrt{d_B^2(\scalbb{x}_i) + \min_{\scalbb{x}_j \in Sk(P), i \neq j} d_G^2(\scalbb{x}_i, \scalbb{x}_j) }
$$

\section{Applications}

\subsection{Simulation Studies}

\subsubsection{Performance indices}

In this section, we shall compute the trustability index and the Tractable consistency index for various dimensionality reduction algorithms. In particular, we consider the algorithms Principal Component Analysis (PCA), Kernel Principal Component Analysis (kPCA), Locally Linear Embedding (LLE), Hessian Locally Linear Embedding (HLLE), Laplacian Eigenmaps (LE), t-distributed Stochastic Neighbour Embedding (tSNE), Uniform Manifold Approximation and Projection (UMAP) etc. To compute the indices, we shall consider various artificial and real-life datasets as discussed below.

\begin{enumerate}
    \item In simulation setup \textbf{S1}, three clusters of datapoints are generated, with each cluster of datapoints following $10$-dimensional Gaussian distribution with a different location and common covariance matrix.
    \item In simulation setup \textbf{S2} the data has been generated uniformly from the $10$-dimensional hypercube $[0, 1]^{10}$.
    \item In setup \textbf{S3} the datapoints lie approximately on the $2$-D surface of the unit sphere in $\R^3$.
    \item In simulation setup \textbf{S4}, the datapoints lie approximately on a swiss roll manifold in $\R^3$.
    \item \textbf{Sonar} data is a popular benchmark dataset for dimensionality reduction~\cite{gorman1988analysis}. The dataset contains $60$ variables (each between $0$ and $1$) representing the angular deviations of different signals when bouncing off a metal or rock object. 
    \item \textbf{WBCD} or Wisconsin Breast Cancer Dataset from UCI Machine Learning Repository~\cite{Dua2019} contains $32$ attributes of $569$ patients with tumour, including a class variable indicating whether the tumour is malignant or benign.
    \item \textbf{COIL2000} data set~\cite{van2000coil} used in the COIL 2000 Challenge contains information on customers of an insurance company. The data consists of 86 variables and includes product usage data and socio-demographic data.
\end{enumerate}

Since generally the trustability index increases as the number of datapoints $n$, it is meaningful to look at $\text{TI}(\mathcal{A}_n, \rvbb{X})/n$. In each of the simulation setups S1, S2, S3, S4, we consider $n = 200$ datapoints. The normalized trustability and tractable consistency indices have been computed for the aforementioned datasets and for the aforementioned algorithms, which have been summarized in Table~\ref{tbl:trust-index} and Table~\ref{tbl:ci-index} respectively. As seen from these tables, PCA, LLE, Hessian LLE, and Laplacian Eigenmaps provides very trustable outputs, however, they are less consistent. Therefore, their output can be used for subsequent analysis of regression and classification purposes, since these algorithms do not create new patterns in the data. On the other hand, tSNE and UMAP are less trustable but more consistent. Their output should primarily be used for visualisation purposes, but not for subsequent analysis if trustworthiness is a requirement. These algorithms might create new superficial patterns in the data which were not originally present. Also, Laplacian Eigenmaps is generally better than Hessian LLE for both trustability and consistency, hence should be preferred for any kind of data. Finally, our proposed algorithm LSDR is found to be extremely trustable, but without compromising its consistency index to much extent.

\begin{table}[ht]
    \centering
    \begin{tabular}{crrrrrrrr}
        \toprule
        Dataset & PCA & kPCA & LLE & HLLE & LE & tSNE & UMAP & LSDR \\
        \midrule
        S1 & 0 & 138.727 & 10.112 & 0.084 & \textbf{0.083} & 128.366 & 91.807 & 0.995\\
        S2 & 0 & 168.667 & 9.997 & 0.822 & \textbf{0.018} & 32.007 & 4.885 & 1.037 \\
        S3 & 0 & 32.781 & 2.991 & \textbf{0.011} & 0.014 & 4.917 & 11.701 & 0.012 \\
        S4 & 0 & 112.678 & 3 & \textbf{0.021} & 0.034 & 142.131 & 60.023 & 0.159 \\
        Sonar & 0 & 56.897 & 59.984 & 59.983 & \textbf{0.288} & 2.771 & 9.826 & 0.944\\
        WBCD & 0 & 30.867 & 30.021 & 33.309 & \textbf{0.052} & 39.205 & 64.206 & 1.746\\
        COIL2000 & 0 & 118.322 & 86.001 & 92.116 & \textbf{0.172} & 11.819 & 38.311 & 2.193\\
        \bottomrule
    \end{tabular}
    \caption{Normalized Trustability Index ($n^{-1}\text{TI}(\mathcal{A}_n, \rvbb{X})$) of various algorithms on various datasets}
    \label{tbl:trust-index}
\end{table}

\begin{table}[ht]
    \centering
    \begin{tabular}{crrrrrrrr}
        \toprule
        Dataset & PCA & kPCA & LLE & HLLE & LE & tSNE & UMAP & LSDR \\
        \midrule
        S1 & 58434.22 & 9.31 & 232.511 & 203.115 & 153.313 & 23.963 & 11.818 & \textbf{6.93}\\
        S2 & 143.233 & 2.946 & 6.305 & 8.818 & 8.003 & 3.116 & \textbf{2.291} & 2.445\\
        S3 & 67.414 & 10.065 & 15.337 & 29.212 & 29.252 & 13.535 & \textbf{2.904} & 4.196 \\
        S4 & 72.212 & 5.193 & 34.611 & 31.995 & 27.212 & 3.998 & 3.184 & \textbf{1.008} \\
        Sonar & 58.414 & \textbf{2.854} & 18.884 & 21.229 & 18.818 & 3.971 & 4.498 & NA \\
        WBCD & 1.234$\times 10^6$ & 3.214 & 48.913 & 74.227 & 70.212 & \textbf{2.105} & 2.816 & NA \\
        COIL2000 & 109783.7 & 4.009 & 69.913 & 75.200 & 71.200 & \textbf{3.211} & 2.877 & NA \\ 
        \bottomrule
    \end{tabular}
    \caption{Normalized Tractable Consistency Index ($n^{-1}\text{TCI}(\mathcal{A}_n, \rvbb{X})$) of various algorithms (Due to computational complexity limitations, we could not compute consistency index of LSDR for Sonar, WBCD, COIL2000 data; see Section~\ref{sec:conclusions} for details)}
    \label{tbl:ci-index}
\end{table}

\subsubsection{Performance of LSDR}

We have performed various simulation to compare the performance of LSDR with that of tSNE and UMAP, the two state of the art dimensionality reduction algorithms. To compare them on common ground, simulated setups are chosen from the demonstrations by Google engineers~\cite{wattenberg2016how,googlepairumap} who have provided an extensive analysis of tSNE and UMAP over different sorts of simulated data. They have also shown the effect of several hyperparameters of these algorithms and how they behave. Due to limited time and space constraints, we shall discuss only a few here. More discussion of other simulation setups is provided in the Appendix.

\begin{figure}[ht]
    \centering
    \begin{subfigure}{\textwidth}
        \centering
        \includegraphics[width = 0.4\textwidth]{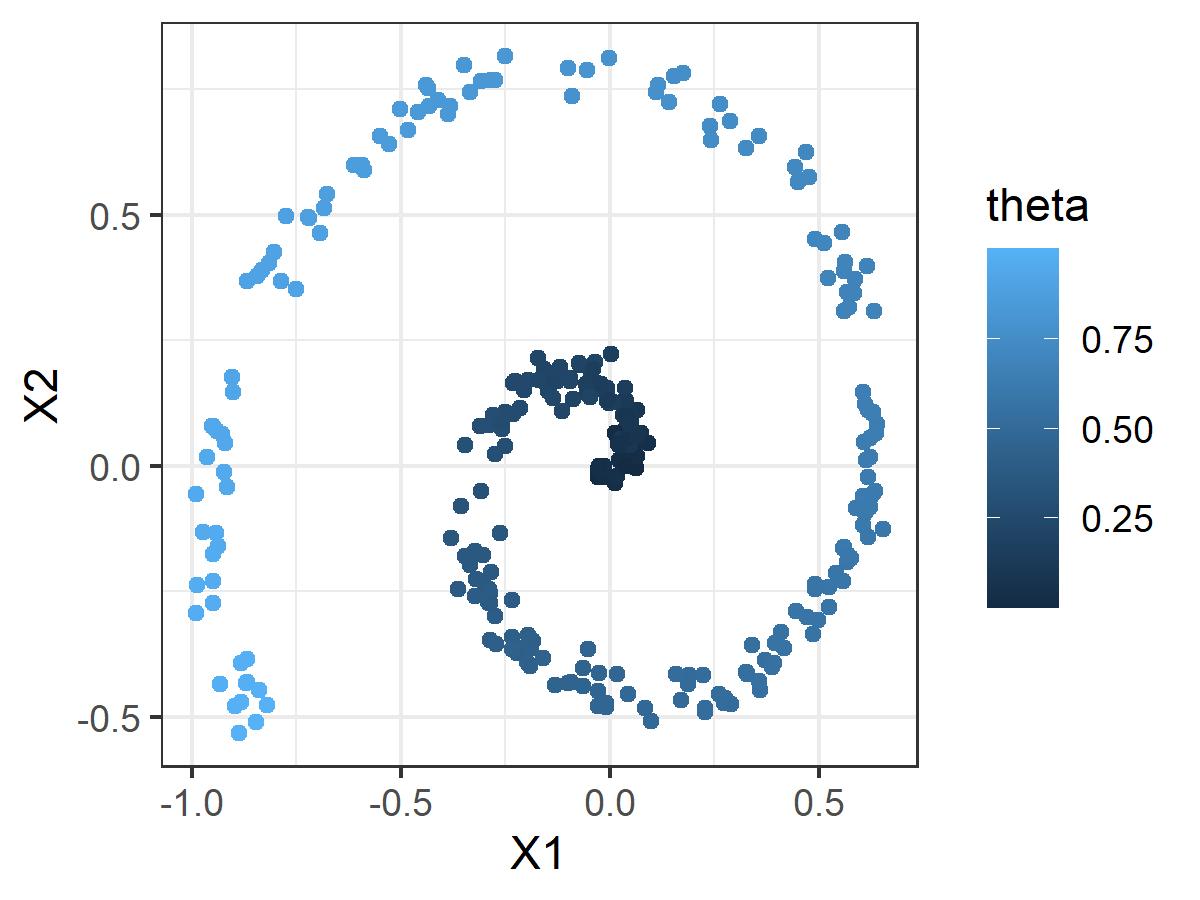}
        \caption{Original Spiral Dataset}
    \end{subfigure}
    \begin{subfigure}{\textwidth}
        \centering
        \includegraphics[width = 0.8\textwidth]{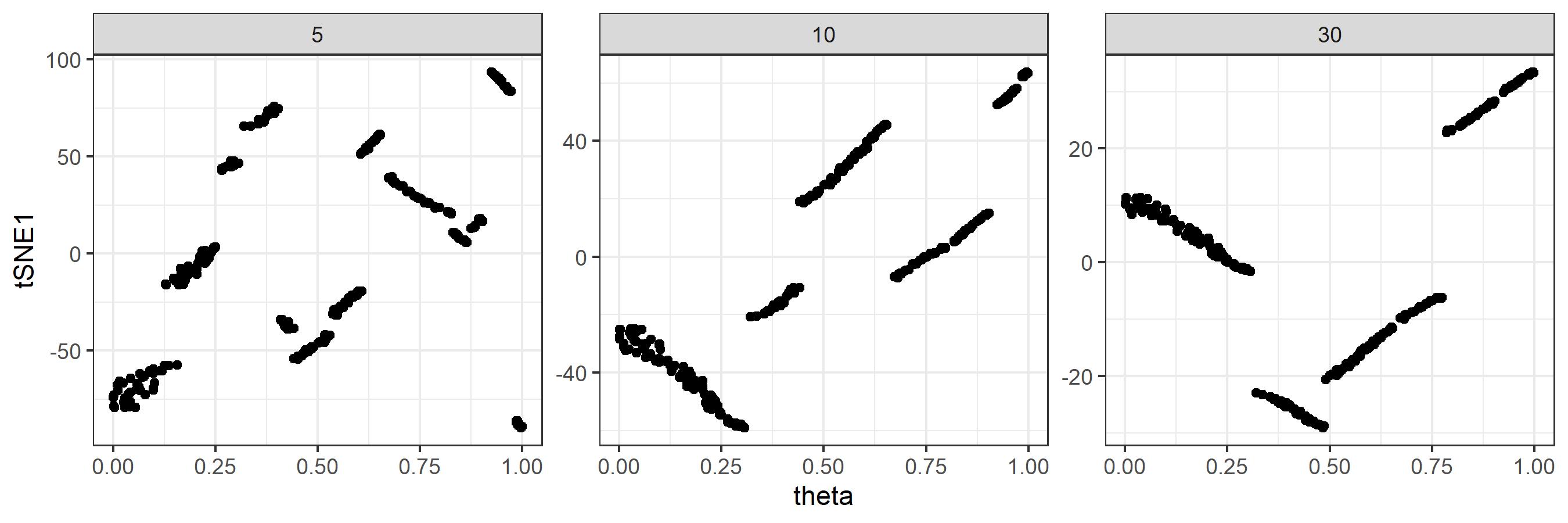}
        \caption{1-dimensional embedding by tSNE vs the true parameter for different perplexity}
    \end{subfigure}
    \begin{subfigure}{\textwidth}
        \centering
        \includegraphics[width = 0.8\textwidth]{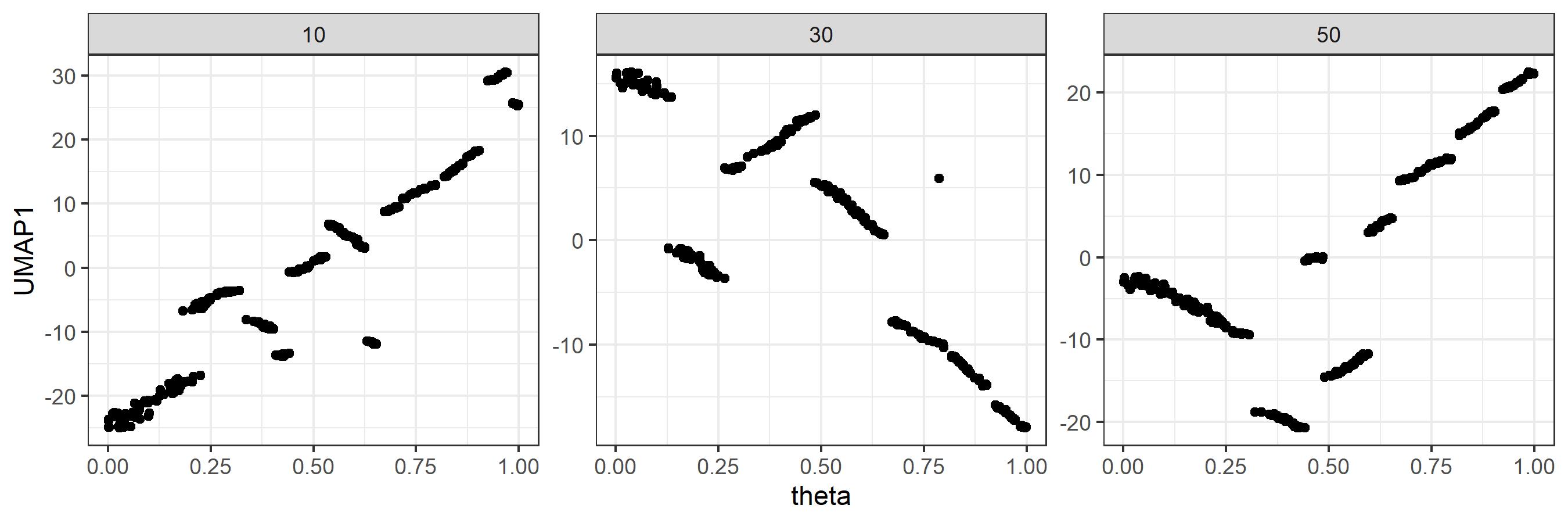}
        \caption{1-dimensional embedding by UMAP vs the true parameter for different neighbours}
    \end{subfigure}
    \begin{subfigure}{\textwidth}
        \centering
        \includegraphics[width = 0.8\textwidth]{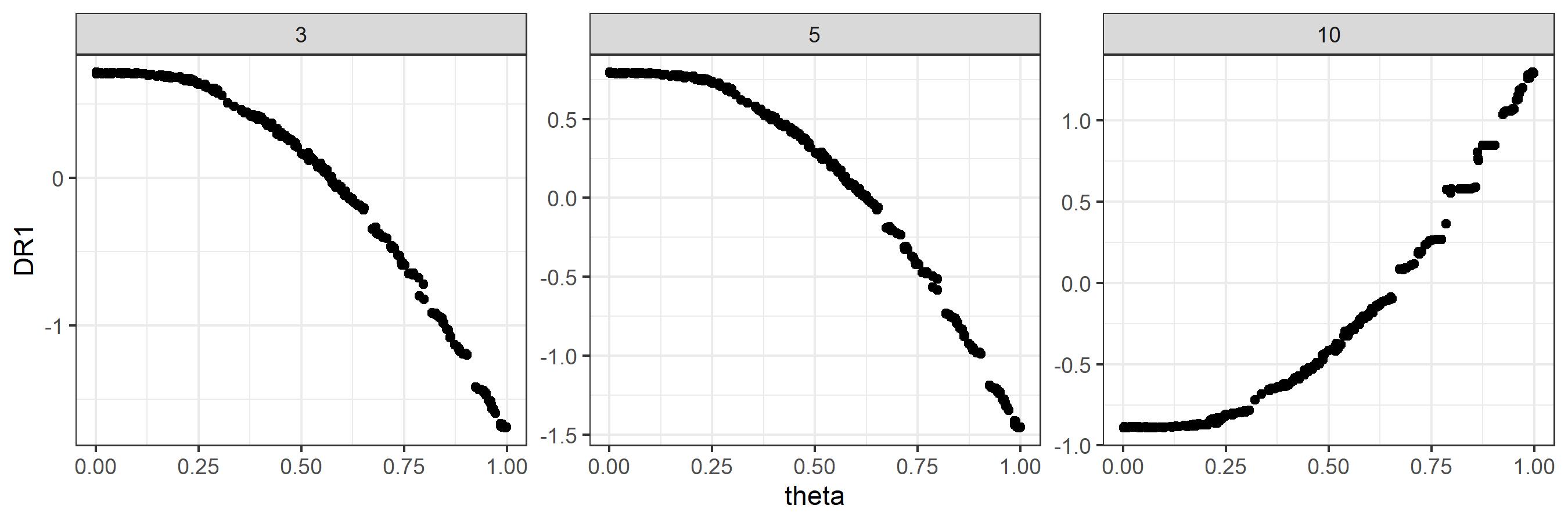}
        \caption{1-dimensional embedding by LSDR vs the true parameter for different NN parameter $k$}
    \end{subfigure}
    \caption{Analysis of Spiral Data}
    \label{fig:spiral-data}
\end{figure}

Our first simulation setup is the popular spiral data in 2-dimension, which is governed by a single parameter $\theta$ denoting the slope of a point with respect to the origin. The original data, its one-dimensional embeddings by tSNE, UMAP and LSDR as a function of the true $\theta$ is shown in Figure~\ref{fig:spiral-data}. As seen from the plots, tSNE and UMAP, even under different parameter values recover $\theta$ only partially, hence results in discontinuous graphs. On the other hand, LSDR outputs a continuous curve even under different choices of its parameters and is extremely well correlated with the true value of $\theta$. However, there is a kernel effect at the tail of the manifold, where the correlation between $\theta$ and the embedding weakens severely. 

\begin{figure}[ht]
    \centering
    \begin{subfigure}{0.49\textwidth}
        \includegraphics[width = \textwidth]{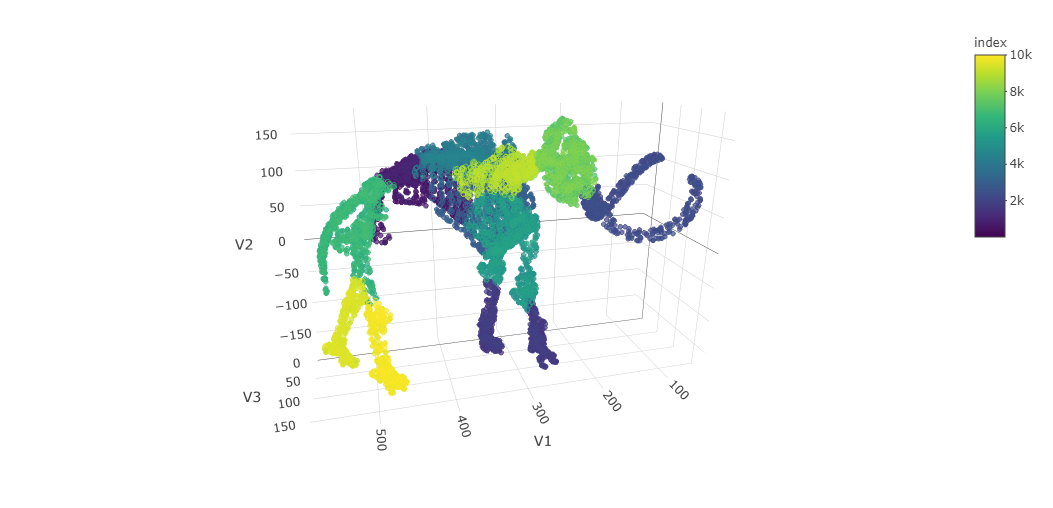}
        \caption{Original 3D Wooly Mammoth Data}
    \end{subfigure}
    \begin{subfigure}{0.49\textwidth}
        \includegraphics[width = \textwidth]{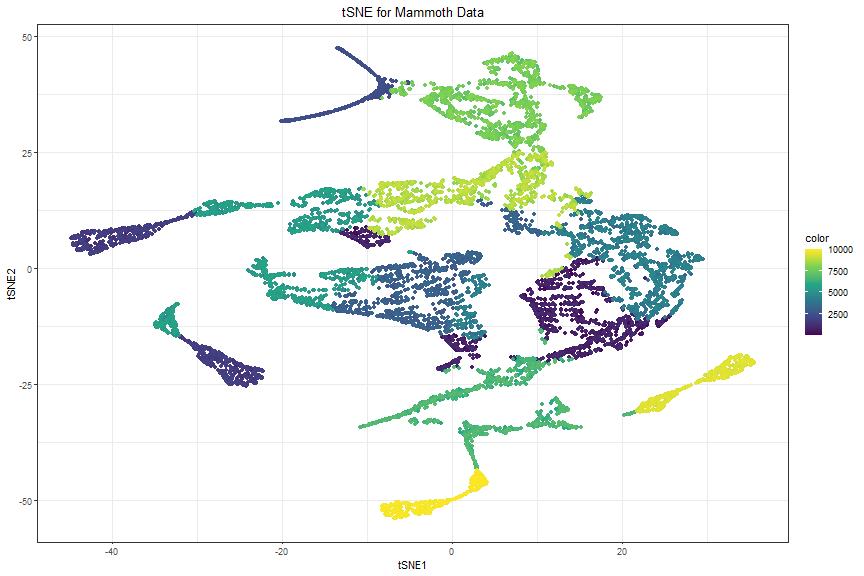}
        \caption{2D Embedding by tSNE of Wooly Mammoth Data}
    \end{subfigure}
    \begin{subfigure}{0.49\textwidth}
        \includegraphics[width = \textwidth]{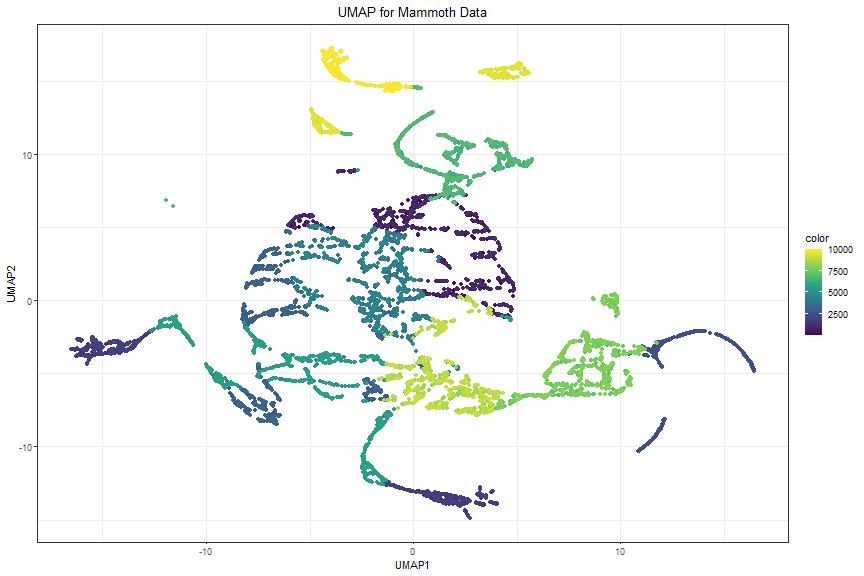}
        \caption{2D Embedding by UMAP of Wooly Mammoth Data}
    \end{subfigure}
    \begin{subfigure}{0.49\textwidth}
        \includegraphics[width = \textwidth]{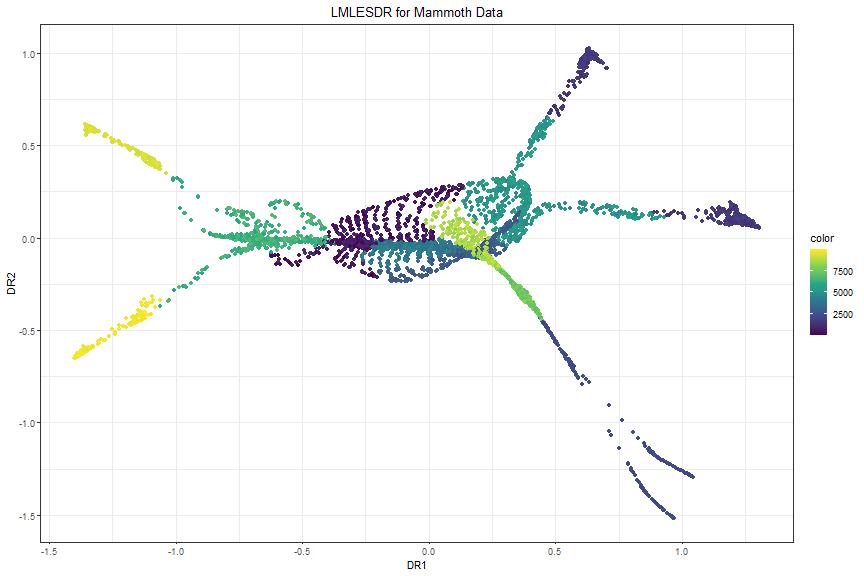}
        \caption{2D Embedding by LSDR of Wooly Mammoth Data}
    \end{subfigure}
    \caption{Analysis of Wooly Mammoth Data}
    \label{fig:wooly-mammoth}
\end{figure}

\begin{figure}[ht]
    \centering
    \begin{subfigure}{\textwidth}
        \centering
        \includegraphics[width = 0.4\textwidth]{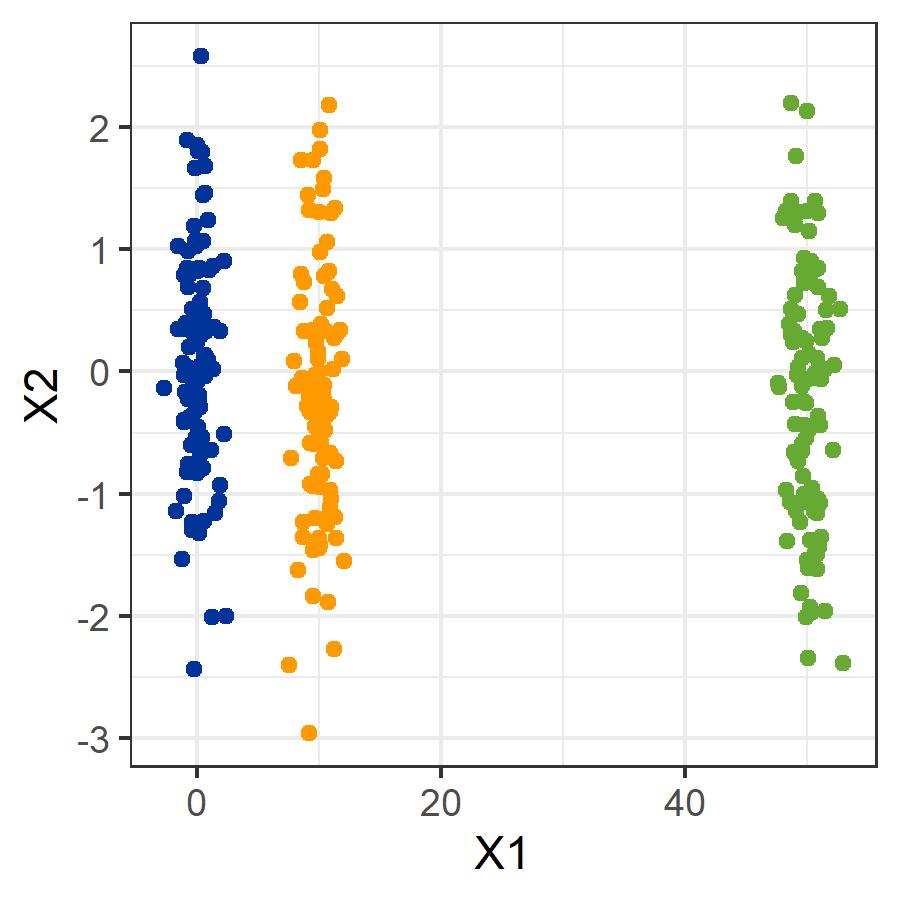}
        \caption{Original Spiral Dataset}
    \end{subfigure}
    \begin{subfigure}{\textwidth}
        \centering
        \includegraphics[width = 0.8\textwidth]{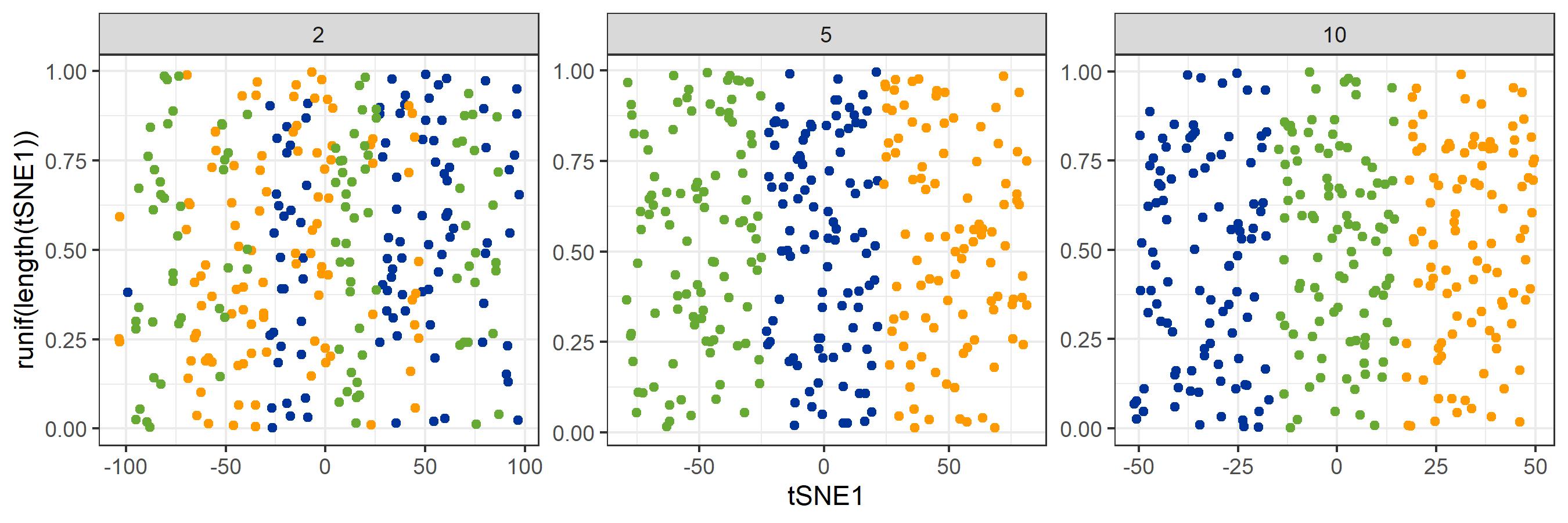}
        \caption{1-dimensional embedding by tSNE for different perplexity}
    \end{subfigure}
    \begin{subfigure}{\textwidth}
        \centering
        \includegraphics[width = 0.8\textwidth]{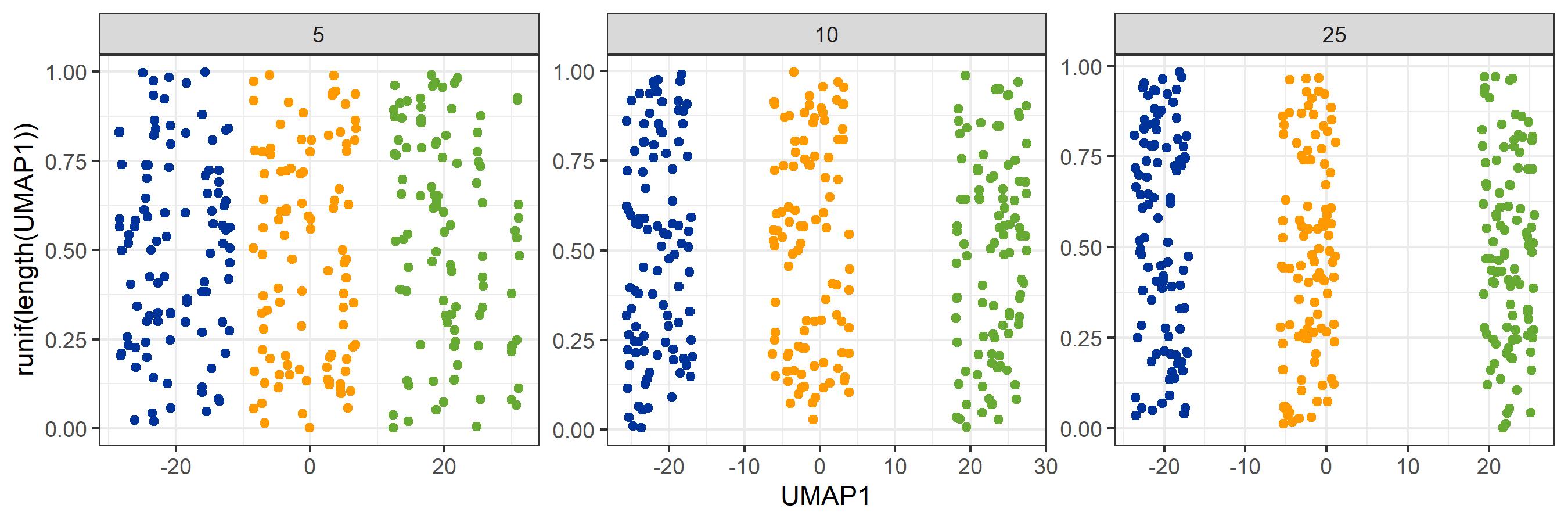}
        \caption{1-dimensional embedding by UMAP for different neighbours}
    \end{subfigure}
    \begin{subfigure}{\textwidth}
        \centering
        \includegraphics[width = 0.8\textwidth]{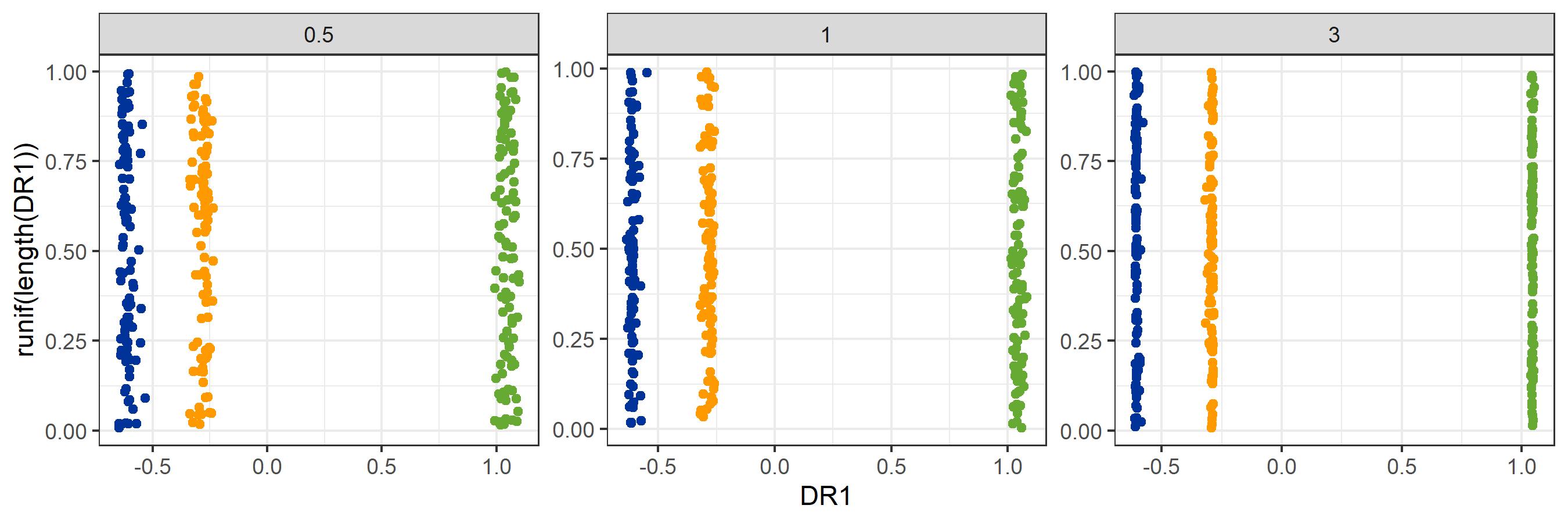}
        \caption{1-dimensional embedding by LSDR for different bandwidths $\sigma$ in radial kernel}
    \end{subfigure}
    \caption{Analysis of Three Cluster Data}
    \label{fig:three-cluster-data}
\end{figure}

Another interesting property of LSDR is that it preserves more global structure than tSNE and UMAP. For example, in the usual Gaussian cluster data~\cite{wattenberg2016how} where there are three clusters with different distances between them, tSNE usually fails to distinguish them in small perplexity parameter. As the perplexity grows, three clusters separate from each other, however, the one-dimensional embedding never shows that there are regions where no data is available. Whereas UMAP overcomes this problem and shows a blank region between clusters, however, the clusters become equidistant despite being different distances away from each other. As shown in Figure~\ref{fig:three-cluster-data}, LSDR preserves this global structure nicely.

Another more close to real-life example is the Wooly Mammoth Data provided by Google PAIR team~\cite{googlepairumap}. The original dataset contains points from the surface of a 3D Wooly mammoth, and a possible best realization of its 2D embedding would be its skin. As shown in Figure~\ref{fig:wooly-mammoth}, both tSNE and UMAP retains the local structure of the data extremely well, however, some parts of the embedding becomes discontinuous. It is quite clear that if the mammoth is skinned and its skin is put flat onto a plane, it would be continuous with the parts corresponding to the legs and trunks sticking out. The 2D embedding obtained by LSDR seems to outperform tSNE and UMAP in this direction by preserving the continuity as well as depicting a very relevant output. LSDR only fails to position the head properly between two frontal legs, which should have been exactly in the middle of those two legs in an ideal dimensionality reduction.

More details about various aspects of comparison between tSNE, UMAP and LSDR are detailed in the Appendix~\ref{appendix:simulation}.

\section{Conclusions and Future Aspects}\label{sec:conclusions}

Analysis of the Trustability index and the Tractable consistency index of various algorithms provide insights into how these algorithms behave in general. In essence, the output of tSNE or UMAP may be used in general for visualisation purposes. However, if the post-reduction analysis requires classification or clustering on the reduced data, then Laplecian Eigenmaps provide a good choice of balance. Also, since the graphical algorithms are generally more trustable, one should use graph-based approximation to the manifold to preserve the global structure, and this primarily gave the intuition of LSDR.

Turning to our proposed algorithm, LSDR outperforms both tSNE and UMAP in terms of its preservation of global structures. Use of metric MDS with the manifold approximation graph $G_\alpha$ provides us with near-optimality properties with respect to both the trustability and consistency indices (see Theorem~\ref{thm:mds-1} and \ref{thm:mds-2}). LSDR preserves the global structure by using skeletal representation, while the local structure is provided by the kernel embedding based usage. In a sense, LSDR is similar to SVM, where the skeletal points act as support points, and any point on the space can be represented in low dimension using the representations of these support points. One of the major problem in the trustability of tSNE and UMAP is that they are iterative in nature, therefore, the final embedding depends a lot on the initialization of the points. In contrast, our proposed LSDR is not an iterative algorithm, hence overcomes this difficulty, but retaining the same level of performance for these setups. The choice of hyperparameters is one of the deciding factors of the performance of LSDR. For example, if $k$ is chosen very large, then there will naturally be fewer skeletal points, and consequently, these points would not be sufficient to describe the manifold's skeleton properly. Therefore, a choice of smaller $k$ (about 3-5) and a higher bandwidth for the kernel is required to ensure that all the points in the dataset are sufficiently represented.

On the other hand, there are some obvious limitations. The most expensive part in LSDR is the Delaunay tessellation which has a worst-case complexity of $O(n^{p/2})$. Therefore, if $p$ is large, it is not computationally feasible in its current form. One could then proceed by making chunks of the whole dataset with respect to close proximities and then perform Delaunay tessellation (this would reduce effective $n$). Another choice could be to use Gabriel graph or Beta skeleton~\cite{siddiqi2008medial,correa2011towards} which are mathematically proven to be a good approximation to the underlying manifold. However, their implementations were not readily available in \texttt{R}, hence with the limited scope of time and space, we could not explore it further. Another problem occurs if there is no error in the data, i.e. the data lies in an exact $d$-dimensional manifold in $\R^p$. In this case, LSDR would detect every point as a point on the boundary $B(P)$, and theoretically, every point would become a skeletal point. While in a realistic scenario, there are always errors, such a case might occur if $n < p$. In this case, an initial PCA (or metric MDS) step is needed to be performed to reduce the dimension while preserving the pairwise distances. Since in a $d$-dimensional plane, only $(d+1)$ distances are needed to be known to define the position of a point, and since there are $(n-1)$ pairwise distances available, it is possible to embed all these $n$ points in $\R^{n-2}$ without changing their pairwise distances. After this initial adjustment, our proposed method LSDR can be used. Finally, LSDR requires its input to become numerical quantities, hence it cannot reduce the dimensionality of the points if only a similarity metric is given. For example, LSDR cannot be used for textual data. However, this problem can again be solved by using an initial metric MDS step to obtain numerical representations in $\R^{n-2}$ by preserving the similarities (or distances).

\section*{Acknowledgements}

I would like to thank my supervisor, Dr. Smarajit Bose, ISRU, Indian Statistical Institute, Kolkata, for his helpful comments and enthusiastic guidance throughout the project.

\appendix

\section{Prerequisites}

\subsection{Reproducible Kernel Hilbert Spaces (RKHS)}\label{appendix:RKHS}

Reproducible Kernel Hilbert Spaces (RKHS) is a Hilbert space of functions which is generated by a basis of kernel functions. If $k : \mathcal{X} \times \mathcal{X} \rightarrow \R$ be a kernel function, then the RKHS generated by its is the set of functions

$$
\mathcal{H} = \left\{ \sum_{i=1}^{\infty} \alpha_i k(\cdot, x_i) : x_i \in \mathcal{X} \text{ and } \sum_{i=1}^{\infty} \alpha_i^2 k(x_i, x_i) < \infty \right\},
$$

\noindent is a Hilbert space equipped with the inner product

$$
\left\langle \sum_{i=1}^{\infty} \alpha_i k(\cdot, x_i), \sum_{j=1}^{\infty} \beta_j k(\cdot, y_j) \right\rangle_{\mathcal{H}} = \sum_{i=1}^\infty \sum_{j=1}^\infty \alpha_i \beta_j k(x_i, y_j) 
$$

Clearly, $\mathcal{H}$ is a set of functions $f$ that maps elements of $\mathcal{X}$ to $\R$. For any choice of positive definite kernel function, the existence of such RKHS is guaranteed by Moore–Aronszajn theorem~\cite{berlinet2011reproducing}. To evaluate any function $f \in \mathcal{H}$ at any $x \in \mathcal{X}$, one can simply compute $\langle f, k(\cdot, x)\rangle_{\mathcal{H}}$ since,

$$
\langle f, k(\cdot, x)\rangle_{\mathcal{H}} = \left\langle \sum_{i=1}^\infty \alpha_i k(\cdot, x_i), k(\cdot, x)\right\rangle_{\mathcal{H}} = \sum_{i=1}^\infty \alpha_i k(x, x_i) = f(x).
$$

\noindent This means, the evaluation functional $T_x : \mathcal{H} \rightarrow \R$ that maps a function $f \rightarrow f(x)$ is a linear functional. This turns out to be a defining property of RKHS in the sense that, any Hilbert space $\mathcal{H}$ on which evaluation functional is a bounded linear functional is generated by some positive definite kernel $k$~\cite{berlinet2011reproducing,li2018sufficient}.

\begin{example}
    Let $\mathcal{H}$ be all functions $f$ on $\R$ such that the support of the Fourier transform of $f$ is contained in $[-a, a]$. Then,

    $$
    k(x, y) = \dfrac{\sin(a(y-x))}{a(y-x)}
    $$

    \noindent and $\langle f, g\rangle = \displaystyle\int fg$, defines a RKHS. 
\end{example}

\begin{example}
    Another popular example of RKHS is the class of functions $f : \R^p \rightarrow \R$ for any $p$, generated by the radial basis kernel 
    $$
    k(\scalbb{x}, \scalbb{y}) = \exp\left( - \lambda \Vert \scalbb{x} - \scalbb{y}\Vert^2 \right), \qquad \scalbb{x}, \scalbb{y} \in \R^p 
    $$
\end{example}

\begin{example}
    Finally, an example of non-continuous kernel is the Bregman kernel defined by,
    $$
    k(x, y) = \begin{cases}
        1 & \text{ if } x = y\\
        0 & \text{ otherwise }
    \end{cases}
    $$
    This kernel could be a good choice if there are categorical variables in the dataset. This kernel reproduces a RKHS $\mathcal{H}$ containing all square integrable real-valued functions.
\end{example}

While the RKHS is introduced in a fixed variable setup, it can generalized to stochastic setup easily by replacing $x_i$'s by $\mathcal{X}$-valued random variables $\rv{x}_i : \Omega \rightarrow \mathcal{X}$. Now, if one considers the linear functional $\mathcal{E} : \mathcal{H} \rightarrow \R$ that maps $f \in \mathcal{H}$ to $\E(f(\rvbb{x}(\omega))$, and assumes that $\E(k(\rvbb{x}(\omega), \rvbb{x}(\omega)) < \infty$, then it follows that $\mathcal{E}$ is a bounded linear functional. Riesz representation theorem~\cite{sunder2008riesz} guarantees that there is a unique member $f_0 \in \mathcal{H}$ such that $\langle f, f_0 \rangle_{\mathcal{H}} = \mathcal{E}(f)$, which one may define as $\E(k(\cdot, \rvbb{x}(\omega))$.

Analogously, if $\E(k(\rvbb{x}(\omega), \rvbb{x}(\omega)) < \infty$, one may define the variance operator of $\rvbb{x}$ in $\mathcal{H}$ as, 

$$
\Gamma_{xx} = \E(k(\cdot, \rvbb{x}(\omega)) \otimes k(\cdot, \rvbb{x}(\omega))) - \E(k(\cdot, \rvbb{x}(\omega))) \otimes \E(k(\cdot, \rvbb{x}(\omega)))
$$

\noindent where $\otimes$ is the symbol for tensor product (see Li~\cite{li2018sufficient} for details). One can further generalize this concept to define covariance operator between functions lying in two different RKHS. Let, $\mathcal{H}_x$ and $\mathcal{H}_y$ be two RKHS generated by two different kernels $k_x$ and $k_y$. Then, the covariance operator is similarly given by 

$$
\Gamma_{xy} = \E(k_x(\cdot, \rvbb{x}(\omega)) \otimes k_y(\cdot, \rvbb{y}(\omega))) - \E(k_x(\cdot, \rvbb{x}(\omega))) \otimes \E(k_y(\cdot, \rvbb{y}(\omega))).
$$

\noindent This means, for any $f(\rvbb{x}) \in \mathcal{H}_x$ and $g(\rvbb{y}) \in \mathcal{H}_y$, $\var(f(\rvbb{x})) = \langle f, \Gamma_{xx} f\rangle_{\mathcal{H}_x}$ and $\cov(f(\rvbb{x}, g(\rvbb{y}) = \langle f, \Gamma_{xy} g\rangle_{\mathcal{H}_x}$. In particular, according to Theorem 12.1 of Li~\cite{li2018sufficient}, when one has i.i.d. sample $\rvbb{x}_1, \dots \rvbb{x}_n$ for $\rvbb{x}$ and $\rvbb{y}_1, \dots \rvbb{y}_n$ for $\rvbb{y}$, these operators are estimated as 

$$
\widehat{\Gamma}_{xx} = \dfrac{1}{n}\scalbb{G}_{x}, \ \widehat{\Gamma}_{xy} = \dfrac{1}{n}\scalbb{G}_{x}\scalbb{G}_{y}, \ \widehat{\Gamma}_{yy} = \dfrac{1}{n}\scalbb{G}_y
$$

\noindent where the matrix $G_x$ and $G_y$ have entries given by,

\begin{align}
    (G_x)_{i,j} & = k_x(\rvbb{x}_i, \rvbb{x}_j) - \dfrac{1}{n}\sum_i k_x(\rvbb{x}_i, \rvbb{x}_j) - \dfrac{1}{n} \sum_j k_x(\rvbb{x}_i, \rvbb{x}_j) + \dfrac{1}{n^2} \sum_i\sum_j k_x(\rvbb{x}_i, \rvbb{x}_j)\label{eqn:Gx-1}\\
    (G_y)_{i,j} & = k_y(\rvbb{y}_i, \rvbb{y}_j) - \dfrac{1}{n}\sum_i k_y(\rvbb{y}_i, \rvbb{y}_j) - \dfrac{1}{n} \sum_j k_y(\rvbb{y}_i, \rvbb{y}_j) + \dfrac{1}{n^2} \sum_i\sum_j k_y(\rvbb{y}_i, \rvbb{y}_j).\label{eqn:Gx-2}
\end{align}

\begin{definition}[Universal Kernels]
    A continuous kernel $k(\cdot,\cdot)$ on a compact metric space $(\mathcal{X},d)$ is called \textbf{universal} if and only if the RKHS $\mathcal{H}$ induced by the kernel is dense in $C(\mathcal{X})$, the space of continuous functions on $\mathcal{X}$, with respect to the infinity norm $\Vert f - g \Vert_{\infty}$.
\end{definition}

\begin{remark}
    Existence of universal kernels allows one to approximate any continuous function by some elements of RKHS. Steinwart~\cite{steinwart2001influence} showed that the radial kernels $k(\scalbb{x}, \scalbb{x}') = \exp(-\lambda \Vert \scalbb{x} - \scalbb{x}'\Vert^2)$ is an example of universal kernel.
\end{remark}

The following theorem provides a easy condition to assure independence of two random variables, by using zero-covariance structure in RKHS reproduced by universal kernels.

\begin{theorem}[Gretton et al.~\cite{gretton2005kernel}]
    \label{thm:independence-covariance}
    If $\mathcal{H}_x$ and $\mathcal{H}_y$ are RKHS generated from universal kernels $k_x, k_y$ on the compact metric spaces $\mathcal{X}$ and $\mathcal{Y}$, then $\rvbb{x}$ and $\rvbb{y}$ are independent with respect to the joint measure $P_{x,y}$ on $(\mathcal{X}\times \mathcal{Y}, \Lambda_x \times \Lambda_y)$ (where $\Lambda_x$ and $\Lambda_y$ are Borel $\sigma$-algebras on $\mathcal{X}$ and $\mathcal{Y}$) if and only if the covariance operator $\Gamma_{xy} = 0$.
\end{theorem}

\begin{remark}
    The theorem can be generalized to non-compact spaces also, without the choice of universal kernel. However, the RKHS should then contain a sufficiently rich class of functions to represent all higher order moments. The theorem for a more lenient assumption about kernels has been proven by Fukumizu~\cite{fukumizu2007kernel} via a notion of characteristic kernels.
\end{remark}

\subsection{Vector Valued RKHS}

As defined in the previous section, RKHS is created to structure the space of functions from $\mathcal{X}$ to $\R$. A generalization to vector valued function (in general operator valued function) is possible~\cite{carmeli2010vector}. Let, $\mathcal{X}$ be the domain space, and $\mathcal{W}$ be the range which is also a Hilbert space. Let, $L(\mathcal{W})$ denotes the space of all bounded linear operators on $\mathcal{W}$. 

\begin{definition}
    A function $k : \mathcal{X} \times \mathcal{X} \rightarrow L(\mathcal{W})$ is said to be an operator-valued positive definite kernel if
    \begin{enumerate}
        \item For each pair $(x, y) \in \mathcal{X} \times \mathcal{X}$, $k(x, y)$ is a self-adjoint operator in $L(\mathcal{W})$, i.e. for any $w_1, w_2 \in \mathcal{W}$, $\langle k(x, y)w_1, w_2 \rangle_{\mathcal{W}} = \langle w_1, k(x,y)w_2 \rangle_{\mathcal{W}}$
        \item For any $x_1, \dots x_n \in \mathcal{X}$ and any $w_1, \dots w_n \in \mathcal{W}$, 
        $$
        \sum_{i=1}^n\sum_{j=1}^n \langle w_i, k(x_i,x_j) w_j\rangle_{\mathcal{W}} \geq 0
        $$
    \end{enumerate}
\end{definition}

\noindent Similar to defining the RKHS, we consider the basis functions given by $k_{x}w(\cdot) = k(\cdot, x)w$ defined by $k_{x}w(y) = k(x, y)w$ for any $x \in \mathcal{X}, w \in \mathcal{W}$. The RKHS $\mathcal{H}$, given by the closure of $\mathcal{H}_0$, where

$$
\mathcal{H}_0 = \text{span}\left\{ k_{x}w(\cdot) : x \in \mathcal{X}, w \in \mathcal{W} \right\}
$$

\noindent equipped with the inner product,

$$
\left\langle \sum_{i=1}^\infty \alpha_i k_{x_i}w_i , \sum_{i=1}^\infty \beta_i k_{y_i}z_i \right\rangle_{\mathcal{H}} = \sum_{i=1}^\infty \sum_{j=1}^\infty \alpha_i\beta_j\langle w_i, k(x_i, y_j)z_j \rangle_{\mathcal{W}}
$$

\noindent The reproducing property here is that $\langle f(x), w \rangle_{\mathcal{W}} = \langle f, k_{x}w \rangle_{\mathcal{H}}$ for any $f \in \mathcal{H}$. In particular, if $\mathcal{W} = \R^p$ the $p$-dimensional Euclidean vector space, then we can choose $w_i$'s as the basis of the vector space, hence, the above infinite summands in expressing the functions in the RKHS can be reduced to a finite sum over choices of $w_i$, as we can simply take $w_i = \scalbb{e}_i$, for $i = 1, 2, \dots p$, where $e_i$ is the $i$-th Euclidean basis vector. Therefore, the functions in the RKHS can be expressed as,

$$
f(\cdot) = \sum_{i=1}^\infty \sum_{j=1}^p \alpha_{ij} k(x_i, \cdot)\scalbb{e}_{j}
$$

\subsection{Procrustes Analysis}\label{appendix:procrustes}

Procrustes Analysis refers to a form of statistical analysis which is used to analyze the shape of the distribution of datasets. The name ``Procrustes" refers to a bandit in greek mythology, who used to cut or stretch the limbs of his victims to fit them into a bed. 

There are many variations of Procrustes problem~\cite{gower2004procrustes}. The simplest variation asks one to find the optimal rotation matrix $\scalbb{P}$ such that the matrix $\scalbb{A}$ can be transformed into $\scalbb{B}$ using the orthogonal transformation $\scalbb{T}$. That means, the problem asks to find $\widehat{\scalbb{P}} = \arg\min_{\scalbb{P}}\Vert \scalbb{A} - \scalbb{PB} \Vert_2$, where $\Vert \cdot \Vert_2$ denotes the usual Frobenius norm subject to $\scalbb{P}\transpose\scalbb{P} = \scalbb{I}$. Here, we consider a general problem instead~\cite{schonemann1966generalized}, that allows translation, uniform scaling and optimal rotation to approximately transform a matrix $\scalbb{A}_{n \times p}$ to $\scalbb{B}_{n \times p}$. Therefore, we consider the generalized problem,

$$
\min_{\scalbb{\mu}, \lambda, \scalbb{P}} \Vert \scalbb{A} - \scalbb{1}_n\scalbb{\mu}\transpose - \lambda \scalbb{PB} \Vert_2^2
$$

\noindent where $\scalbb{\mu}$ is a $(p\times 1)$ column vector, $\lambda \in \R$ is a scaling factor and $\scalbb{P}$ is an orthogonal matrix. Hence, we consider a Lagrangian $L = \Vert \scalbb{A} - \scalbb{1}_n\scalbb{\mu}\transpose - \lambda \scalbb{PB} \Vert_2^2 + \eta \Vert\scalbb{P}\transpose \scalbb{P} - \scalbb{I}\Vert_2^2$, where $\eta$ is the Lagrangian tuning parameter. Now, we compute the derivatives of $L$ with respect to $\scalbb{\mu}, \lambda$ and $\scalbb{P}$ as follows.

\begin{align}
    \dfrac{\partial L}{\partial \scalbb{\mu}} & = -2 ((\scalbb{A} - \lambda \scalbb{PB})\transpose - \scalbb{\mu}\scalbb{1}_n\transpose) \scalbb{1}_n \label{eqn:procrust-deriv-1}\\
    \dfrac{\partial L}{\partial \lambda} & = 2 \tr\left( \scalbb{PB}((\scalbb{A} - \lambda \scalbb{PB})\transpose - \scalbb{\mu}\scalbb{1}_n\transpose) \right) \label{eqn:procrust-deriv-2}\\
    \dfrac{\partial L}{\partial \scalbb{P}} & = -2 \lambda \left( A - \lambda \scalbb{PB} - \scalbb{1}_n\scalbb{\mu}\transpose \right) \scalbb{B}\transpose + 2 \eta \scalbb{P}(\scalbb{P}\transpose\scalbb{P} - \scalbb{I}) \label{eqn:procrust-deriv-3}
\end{align}

Setting these derivatives given in Eq.~\eqref{eqn:procrust-deriv-1}-\eqref{eqn:procrust-deriv-3} equal to zero to obtain the normal equations, we obtain that,

\begin{align}
    \widehat{\mu} & = \dfrac{1}{n} \left(\scalbb{A} - \widehat{\lambda}\scalbb{\widehat{P}B} \right)\transpose \scalbb{1}_n \label{eqn:procrust-normal-1}\\
    \widehat{\lambda} & = \dfrac{\tr \left( \widehat{\scalbb{P}}\scalbb{BA}\transpose (\scalbb{I} - n^{-1}\scalbb{1}_n\scalbb{1}_n\transpose) \right)}{\tr\left( \scalbb{BB}\transpose (\scalbb{I} - n^{-1}\scalbb{1}_n\scalbb{1}_n\transpose) \right)} \label{eqn:procrust-normal-2}
\end{align}

\noindent and $\scalbb{\widehat{P}}$ is the closest orthogonal matrix to $\dfrac{1}{\widehat{\lambda}}\scalbb{\tilde{A}\tilde{B}}\transpose$, where $\tilde{\scalbb{A}}$ and $\tilde{\scalbb{B}}$ are demean versions of $\scalbb{A}$ and $\scalbb{B}$ respectively. Thus, the optimal rotation matrix is given by $\widehat{\scalbb{P}} = \scalbb{UV}\transpose$ such that $\dfrac{1}{\widehat{\lambda}}\scalbb{\tilde{A}\tilde{B}}\transpose$ has a singular value decomposition (SVD) $\scalbb{U\Sigma V}\transpose$. Clearly, since this solution is independent of $\lambda$, one may obtain $\widehat{\scalbb{P}}$ by simply performing SVD of $\scalbb{\tilde{A}\tilde{B}}\transpose$. The optimal value of $\widehat{\lambda}$ and $\widehat{\scalbb{\mu}}$ can then be retrieved by means of Eq.~\eqref{eqn:procrust-normal-1} and Eq.~\eqref{eqn:procrust-normal-2}.

Therefore, the following steps calculates the optimal transformations from $\scalbb{A}$ to $\scalbb{B}$.

\begin{enumerate}
    \item For each column in $\scalbb{A}$, compute its average and subtract it from the corresponding entries. Similarly, demand the columns of $\scalbb{B}$. Denote this demeaned version as $\tilde{\scalbb{A}}$ and $\tilde{\scalbb{B}}$ respectively.
    \item Let $\scalbb{M} = \tilde{\scalbb{A}}\tilde{\scalbb{B}}\transpose$, and let its SVD is denoted as $\scalbb{M} = \scalbb{U\Sigma V}\transpose$. Optimal rotation is $\widehat{\scalbb{P}} = \scalbb{UV}\transpose$.
    \item Finally, the optimal scaling factor is simply,
    $$
    \widehat{\lambda} = \dfrac{\tr\left( \widehat{\scalbb{P}} \tilde{\scalbb{B}}\tilde{\scalbb{A}}\transpose \right)}{\tr\left( \tilde{\scalbb{B}}\tilde{\scalbb{B}}\transpose \right)}
    $$
\end{enumerate}

Note that, in particular, $R = \Vert \scalbb{A} - \scalbb{1}_n\scalbb{\mu}\transpose - \lambda \scalbb{PB} \Vert_2^2 = \Vert \tilde{\scalbb{A}} - \lambda \scalbb{P\tilde{B}} \Vert_2^2$ is simply a quadratic function in $\lambda$, whose minimum is given by 

\begin{align*}
    \min \Vert \tilde{\scalbb{A}} - \lambda \widehat{\scalbb{P}} \tilde{\scalbb{B}}\Vert_2^2
    & = \tr(\tilde{\scalbb{A}}\transpose \tilde{\scalbb{A}}) - \dfrac{\tr(\tilde{\scalbb{B}}\transpose \widehat{\scalbb{P}}\transpose \tilde{\scalbb{A}} )^2}{\tr(\tilde{\scalbb{B}}\transpose \tilde{\scalbb{B}})}\\
    & = \tr(\tilde{\scalbb{A}}\transpose \tilde{\scalbb{A}}) - \dfrac{\tr( \tilde{\scalbb{A}}\tilde{\scalbb{B}}\transpose \scalbb{U}\transpose \scalbb{V} )^2}{\tr(B\transpose B)}\\
    & = \tr(\tilde{\scalbb{A}}\transpose \tilde{\scalbb{A}}) - \dfrac{\tr( \scalbb{U\Sigma V}\transpose \scalbb{U}\transpose \scalbb{V} )^2}{\tr(\tilde{\scalbb{B}}\transpose \tilde{\scalbb{B}})}\\
    & = \tr(\tilde{\scalbb{A}}\transpose \tilde{\scalbb{A}}) - \dfrac{\tr( \scalbb{\Sigma} )^2}{\tr(\tilde{\scalbb{B}}\transpose \tilde{\scalbb{B}})}
\end{align*}

\noindent where $\scalbb{\Sigma}$ is the diagonal matrix consisting of the singular values of the matrix $\tilde{\scalbb{A}}\tilde{\scalbb{B}}\transpose$.

\section{Proofs}\label{appendix:proofs}

\subsection{Proof of Theorem~\ref{thm:uniqueness}}

    Let us denote the errors as $\epsilon_i$ instead of $\epsilon_i(\rvbb{x})$ for $i=1,2$. Note that, $\rvbb{x} = f(\phi_1(\rvbb{x}), \dots \phi_d(\rvbb{x})) + \epsilon_1 = g(\phi_1(\rvbb{x}), \dots \phi_d(\rvbb{x})) + \epsilon_2$ implies that,
    $$f(\phi_1(\rvbb{x}), \dots \phi_d(\rvbb{x})) - g(\phi_1(\rvbb{x}), \dots \phi_d(\rvbb{x})) = \epsilon_2 - \epsilon_1$$ 
    
    \noindent Therefore, $(\epsilon_2 - \epsilon_1)$ is $\mathcal{F} = \sigma(\{ \phi_1(\rvbb{x}), \dots \phi_d(\rvbb{x})\})$-measurable. Now,
    
    \begin{align*}
        \epsilon_2 - \epsilon_1
        & = \E(\epsilon_2 - \epsilon_1 \mid \mathcal{F}), \qquad \text{a.s.}\\
        & = \E(\epsilon_2 \mid \mathcal{F}) - \E(\epsilon_1 \mid \mathcal{F})\\
        & = \E(\epsilon_2) - \E(\epsilon_1), \qquad \text{a.s.}\\
        & \qquad \text{since, both } \epsilon_1 \text{ and } \epsilon_2 \text{ are independent of } \mathcal{F}\\
        & = 0 
    \end{align*}

    Thus, $f(\phi_1(\rvbb{x}), \dots \phi_d(\rvbb{x})) = g(\phi_1(\rvbb{x}), \dots \phi_d(\rvbb{x}))$ almost surely. 

\subsection{Proof of Theorem~\ref{thm:find-embedding}}

For any DR algorithm $\mathcal{A}_n$, the outputted embedding functions $\widehat{\phi}_1, \dots \widehat{\phi}_d$ are implicitly minimizers of the loss function as described in Eq.~\eqref{eqn:minimize-loss}. A representer theorem by Sch{\"o}lkopf, Herbrich and Smola~\cite{scholkopf2001generalized} can then be applied to represent these functions as a finite linear combination of kernel functions, since the search space $\mathcal{C}$ is a subset of some RKHS.

\begin{theorem}[Sch{\"o}lkopf, Herbrich and Smola]
    \label{thm:scholkopf-theorem}
    Suppose we are given a nonempty set $\mathcal{X}$, a positive definite real-valued kernel $k$ on $\mathcal{X} \times \mathcal{X}$, a training sample $(x_1, y_1), \dots, (x_m, y_m) \in \mathcal{X} \times \R$, a strictly monotonically increasing real-valued function $g$ on $[0,\infty]$, an arbitrary cost function $c : (\mathcal{X} \times \R^2)^m \rightarrow \R \cup \{ \infty \}$, and a class of functions
    
    $$
    \mathcal{F} = \left\{ f \in \R^{\mathcal{X}} : f(\cdot) = \sum_{i=1}^{\infty} \beta_i k(\cdot, z_i), \beta_i \in \R, z_i \in \mathcal{X}, \Vert f \Vert < \infty \right\}
    $$
    
    \noindent Then, for any $f \in \mathcal{F}$ minimizing the regularized risk functional 
    
    $$
    c\left( (x_1,y_1,f(x_1)), \dots (x_m, y_m, f(x_m)) \right) + g(\Vert f\Vert)
    $$
    
    \noindent admits a representation of the form 
    
    $$
    f(\cdot) = \sum_{i=1}^m \alpha_i k(\cdot, x_i)
    $$
\end{theorem}

Now, since each of the functions $\phi_j$ belongs to Reproducible Kernel Hilbert Space (RKHS) for some positive definite kernel $k$, one may transform the constrained optimization problem for any algorithm into an unconstrained problem by introducing Lagrangian multipliers as 

$$
\min L\left( \{ \phi_j(\rvbb{x}_i) \}_{i,j=1}^{n,p}, \{ \epsilon(\rvbb{x}_i) \}_{i=1}^n \right) + \sum_{j=(d+1)}^{p} \eta_j \Vert \phi_j \Vert
$$

\noindent In view of a simple extension of Theorem~\ref{thm:scholkopf-theorem} for more than one functions $f$, we see that each $\phi_j$ can be represented as a finite linear combination of kernel basis functions. Mathematically, this means, there exists constants $\alpha_{ij}$ such that,

$$
\phi_j(\rvbb{x}) = \sum_{i=1}^n \alpha_{ij} k(\rvbb{x}, \rvbb{x}_i), \qquad i = 1, \dots n; j = 1, \dots d
$$

\noindent Putting $\rv{y}_{i'j} = \phi_j(\rvbb{x}_{i'})$, we obtain the following system of equations,

\begin{equation}
    \rv{y}_{i'j} = \sum_{i=1}^n \alpha_{ij} k(\rvbb{x}_{i'}, \rvbb{x}_i), \qquad i, i' = 1, \dots n; j = 1, \dots d
    \label{eqn:system-1}
\end{equation}

\noindent For fixed $j$, the system of equations given by Eq.~\eqref{eqn:system-1} is a system of $n$ linear equations in $n$ variables. Also, the system is uniquely solvable since the kernel matrix i.e. the matrix $\scalbb{K}$ with entries $\{ k(\rvbb{x}_i, \rvbb{x}_{i'}) \}_{i,i'=1}^{n,n}$ is a positive definite matrix. Therefore,

$$
\begin{pmatrix}
\alpha_{1j}\\
\alpha_{2j}\\
\dots\\
\alpha_{nj}
\end{pmatrix} = \scalbb{K}^{-1} \begin{pmatrix}
\rv{y}_{1j}\\
\rv{y}_{2j}\\
\dots \\
\rv{y}_{nj}
\end{pmatrix}
$$

\noindent Substituting these values to the form of $\phi_j(\rvbb{x})$ yields the expression given in the theorem.

\subsection{Proof of Theorem~\ref{thm:reconstruction}}

We have that the $l$-th coordinate of the errors is given by $\epsilon_l(\rvbb{x}) = \rv{x}_{\cdot l} - f_l(\rv{y}_{\cdot 1}, \dots \rv{y}_{\cdot d})$, where $\rv{x}_{\cdot l}$ denotes the $l$-th variable in the dataset $\rvbb{X}$. Since we are concerned about the independence of $\epsilon_l(\rvbb{x})$ with the embeddings $\phi_j(\rvbb{x})$, which is usually hard to verify, the assumption on RKHS actually allows one to simplify independence to only zero covariance in RKHS domain. Since $\epsilon_l(\rvbb{x})$ is independent of $\phi_j(\rvbb{x})$ if and only if $\epsilon_l(\rvbb{x}) - c_l$ is independent of $\phi_j(\rvbb{x})$ for any choice of $c_l \in \R$, one may assume without loss of generality that $\epsilon_l(\rvbb{x})$ has mean $0$.

In view of Theorem~\ref{thm:independence-covariance}, to make sure that $\epsilon(\rvbb{x})$ is independent of $\sigma\left(\{ \phi_j(\rvbb{x}): j =1,2\dots d\}\right)$, it is enough to ensure that $\cov\left( \epsilon_l(\rvbb{x}), \phi_j(\rvbb{x}) \right) = 0$ for all $l = 1, 2, \dots p; j = 1, \dots d$. Since, $\epsilon_l(\rvbb{x}) = \rv{x}_{\cdot l} - f_l(\rv{y}_{\cdot 1}, \dots \rv{y}_{\cdot d})$, we require, 

$$
\cov( \rvbb{x}_{\cdot l}, \phi_j(\rvbb{x}) ) = \cov\left( f_l(\rv{y}_{\cdot 1}, \dots \rv{y}_{\cdot d}), \phi_j(\rvbb{x}) \right), \qquad l = 1, 2, \dots p; \ j = 1, \dots d.
$$

\noindent Now, letting $f_l(\cdot) = \sum_{i=1}^n \beta_{il} k_y(\cdot, \rvbb{y}_i)$ where $k_y$ is a kernel on $\R^d \times \R^d$, and $\phi_j(\cdot) = \sum_{i=1}^n \alpha_{ij} k_x(\cdot, \rvbb{x}_i)$, we obtain that,

$$
\cov\left( f_l(\rvbb{y}), \phi_j(\rvbb{x}) \right) = \sum_{r = 1}^n \beta_{rl}  \sum_{s = 1}^n ({\Gamma}_{xy})_{rs} \alpha_{sj} 
$$

\noindent where $\Gamma_{xy}$ is the covariance operator between RKHS $\mathcal{H}_x$ and $\mathcal{H}_y$. Therefore, we can now put the estimates of covariances instead, to obtain the system of equations,

\begin{equation}
    \sum_{r = 1}^n \beta_{rl}  \left[\sum_{s = 1}^n (\widehat{\Gamma}_{xy})_{rs} \alpha_{sj}\right] = \dfrac{1}{n}\sum_{i=1}^n \rv{x}_{il} \phi_j(\rvbb{x}_i) - \left( n^{-1} \sum_{i=1}^n \rv{x}_{il} \right)\left( n^{-1} \sum_{i=1}^n \phi_j(\rvbb{x}_i) \right)
    \label{eqn:system-reconstruction}
\end{equation}

\noindent Note that, in Eq.~\eqref{eqn:system-reconstruction}, the quantities $\alpha_{sj}$'s are known (or can be estimated) using Theorem~\ref{thm:find-embedding}, and $\phi_j(\rvbb{x}_i)$'s are also known for the training samples. For a fixed $l$, the system of linear equations given in Eq.~\eqref{eqn:system-reconstruction} contains $n$ unknown variables $\beta_{1l}, \dots \beta_{rl}$ in $d$ equations. Since generally, $d << n$, the system is underdetermined. Clearly, the system thus has more than one solution, and it is clear that one should find the solution that minimizes the number of nonzero $\beta_{rl}$'s, so that minimum number of parameters are used to describe the reconstruction function. To ease the computation, we seek for a solution to Eq.~\eqref{eqn:system-reconstruction} such that the norm of $\beta = (\beta_{1l}, \dots \beta_{nl})\transpose$ is minimized for each $l = 1, 2, \dots n$. 

Therefore, the problem boils down to finding $\scalbb{\beta}$ with minimum $\scalbb{\beta}\transpose \scalbb{\beta}$ such that $\scalbb{A\beta} = \scalbb{c}$. Introducing the Lagrangian multiplier $\lambda$, we consider the unconstrained minimization problem $\scalbb{\beta}\transpose \scalbb{\beta} + \lambda \left( \scalbb{A\beta} - \scalbb{c} \right)$. Solving it yields the solution, $\scalbb{\beta} = \scalbb{A} (\scalbb{A}\transpose\scalbb{A})^{-1} \scalbb{c}$. This leads us to the result stated in Theorem~\ref{thm:reconstruction}.

\subsection{Proof of Theorem~\ref{thm:consistency-index}}

Note that,
    
    \begin{align*}
        & \min_{\phi_1, \dots \phi_d \in \mathcal{C}} L\left( \{ \phi_j(\rvbb{x}_i) \}_{i,j=1}^{n,p}, \{ \rvbb{\epsilon}_i \}_{i=1}^n \right)\\
        = \quad & \min_{\phi_1, \dots \phi_d \in \mathcal{C}} L\left( \{ \phi_j(T^{-1}(\rvbb{z}_i)) \}_{i,j=1}^{n,p}, \{ \rvbb{\epsilon}_i \}_{i=1}^n \right) \\
        \geq \quad & \min_{\psi_1, \dots \psi_d \in \mathcal{C}} L\left( \{ \psi_j(\rvbb{z}_i) \}_{i,j=1}^{n,p}, \{ \rvbb{\epsilon}_i \}_{i=1}^n \right), \qquad \text{since, } \phi_j\circ T^{-1} \in \mathcal{C} \text{ for any } \phi_j \in \mathcal{C}
    \end{align*}
    
    \noindent Conversely,
    
    \begin{align*}
        & \min_{\psi_1, \dots \psi_d \in \mathcal{C}} L\left( \{ \psi_j(\rvbb{z}_i) \}_{i,j=1}^{n,p}, \{ \rvbb{\epsilon}_i \}_{i=1}^n \right)\\
        = \quad & \min_{\psi_1, \dots \psi_d \in \mathcal{C}} L\left( \{ \phi_j(T(\rvbb{x}_i)) \}_{i,j=1}^{n,p}, \{ \rvbb{\epsilon}_i \}_{i=1}^n \right) \\
        \geq \quad & \min_{\phi_1, \dots \phi_d \in \mathcal{C}} L\left( \{ \phi_j(\rvbb{x}_i) \}_{i,j=1}^{n,p}, \{ \rvbb{\epsilon}_i \}_{i=1}^n \right), \qquad \text{since, } \psi_j\circ T \in \mathcal{C} \text{ for any } \psi_j \in \mathcal{C}
    \end{align*}
    
    \noindent This completes the proof.

\subsection{Proof of Theorem~\ref{thm:ci-index-boundary}}

Since the set of functions in $\mathcal{H}^\ast$ is constrained by choice of $\beta_{ij}$'s such that $0 \leq \beta_{ij} \leq 1$ and their sum $\sum_{i=1}^\infty \sum_{j=1}^p \beta_{ij} = 1$, we may again introduce Lagrangian multipliers to write the problem of obtaining Consistency Index as,

$$
\max_{T \in \mathcal{H}^\ast } \min_{\scalbb{\mu}, \lambda, \scalbb{P}} \Vert \mathcal{A}_n(d, \tilde{\rvbb{X}})(\tilde{\rvbb{X}}) - \scalbb{1}_n\scalbb{\mu}\transpose - \lambda \scalbb{P}\mathcal{A}_n(d, \rvbb{X})(\rvbb{X}) \Vert_2^2 + \eta (\sum_{i,j}\beta_{ij} - 1) + \sum_{i,j}\eta_{ij}\beta_{ij}
$$

Again, we can use a generalized version of representer theorem~\cite{minh2011vector,minh2016unifying} can be used to claim that the maximizer $T$, if belongs to such a space $\mathcal{H}^\ast$, can be represented as linear span of only kernel functions evaluated at $n$ datapoints $\rvbb{x}_1, \dots \rvbb{x}_n$. This means,

\begin{equation}
    \widehat{T}(\rvbb{x}) = \sum_{i=1}^n \sum_{j=1}^p \beta_{ij} k(\rvbb{x}, \rvbb{x}_i)\scalbb{e}_j
    \label{eqn:T-form}    
\end{equation}

\noindent where $k(\cdot, \cdot)$ is the reproducing kernel on $\R^p \times \R^p \rightarrow \R^p \times \R^p$ and $\scalbb{e}_j$'s are the Euclidean basis vector of $\R^p$, and $\widehat{T}$ is the argument of the maximizer in the definition of Consistency Index. However, the number of such functions is uncountable, since there are uncountably many choices for the real numbers $\beta_{ij}$'s. Therefore, although the form of the maximizer function is known by Eq.~\eqref{eqn:T-form}, the maximum value cannot be found by a computer so easily.

So we note that, $\partial\mathcal{H}^\ast \subseteq \mathcal{H}^\ast$, hence clearly, $\max_{T \in \mathcal{H}^\ast} S(T) \geq \max_{T \in \partial\mathcal{H}^\ast} S(T)$.

On the other hand,
\begin{align*}
    \max_{T \in \mathcal{H}^\ast} S(T)
    & \leq \max_{0 \leq \beta_{ij} \leq 1, \sum_{i,j}\beta_{ij} = 1} S\left( \sum_{i=1}^n\sum_{j=1}^p \beta_{ij} k(\cdot, \rvbb{x}_i)\scalbb{e}_j \right), \qquad \text{this is a bigger class}\\
    & \leq \max_{0 \leq \beta_{ij} \leq 1, \sum_{i,j}\beta_{ij} = 1} \left[ \sum_{i=1}^n\sum_{j=1}^p \beta_{ij} S(k(\cdot, \rvbb{x}_i)\scalbb{e}_j) \right], \qquad \text{since S(T) is convex}\\
    & \leq \max_{0 \leq \beta_{ij} \leq 1, \sum_{i,j}\beta_{ij} = 1} \left[ \sum_{i=1}^n\sum_{j=1}^p \beta_{ij} \max_{T \in \partial\mathcal{H}^\ast} S(T) \right], \qquad \text{since, } k(\cdot, \rvbb{x}_i)\scalbb{e}_j \in \partial\mathcal{H}^\ast \\
    & = \max_{T \in \partial\mathcal{H}^\ast} S(T)
\end{align*}

\subsection{Proof of Theorem~\ref{thm:mcst-1}}
\begin{figure}[h]
        \centering
        \includegraphics[width = 0.5\textwidth]{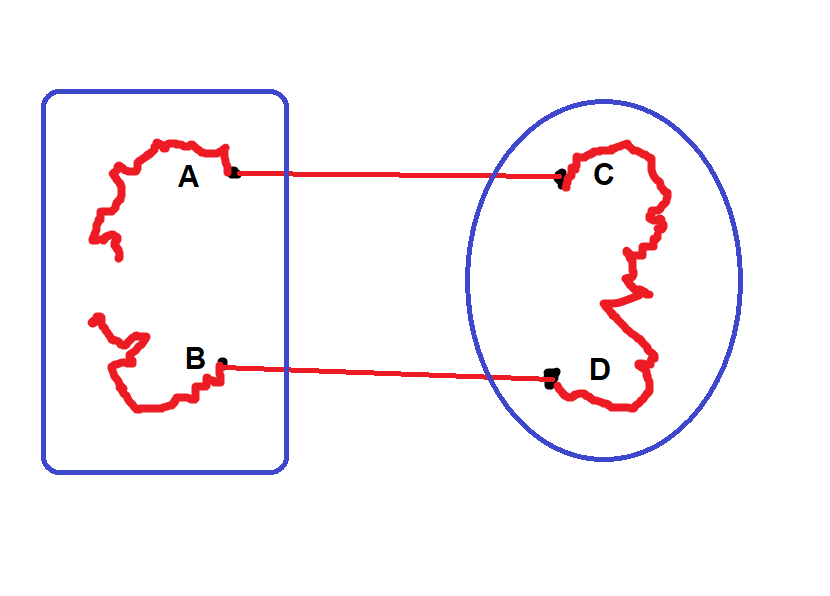}
        \caption{Four arbitrary points with two connections between two distinct clusters in Euclidean MCST}
        \label{fig:mcst-proof-1}
    \end{figure}

    Let, $A, B, C, D$ be 4 points such that $A, B \in C_1$ and $C, D \in C_2$ and lying on the MCST. Figure~\ref{fig:mcst-proof-1} shows one such general configuration.

    Let us compute the conditional probability $P(E | A, B, C, D)$ where $E$ is the required event that the MCST has two edges connecting two classes. Since the MCST is spanning graph and is a tree, if we remove $AC$, then the graph becomes disconnected. In other words, either $A$, $B$ are connected or $C$, $D$ are connected. Otherwise, inclusion of $AC$ would lead to a cycle. 

    Without loss of generality, assume that $A$, $B$ are disconnected. Since the tree is spanning, atleast one of the connected component of $A$ or $B$ (intersected with $C_1$) must be atleast as large as $(m-2)/2$. Say, the connected component of $A$ is larger than $(m-2)/2$. Then, starting with the initial MCST, removing $AC$ and connecting $A$ with any points from the connected component of $B$ will again lead to a spanning tree. The former is a minimum cost spanning tree, if for any such newly formed tree, the cost increases, i.e. the new edge length is more than the length of $AC$. Therefore, we have,

    \begin{align*}
        P(E | A, B, C, D)
        & \leq 4 P(\Vert A - X_i \Vert_2 > |AC|, \forall X_i \in C_1, X_i \rightsquigarrow B \mid A, B, C, D )\\
        & \leq 4 \prod_{i=1}^{(\min(m,n)-2)/2} P(\Vert A - X_i \Vert_2 > |AC| \mid A, B, C, D)\\
        &\qquad \qquad \qquad \text{conditional on A}, \Vert X_i - A\Vert_2 \text{ are independent}\\
        & = 4 \alpha^{(\min(m,n)-2)/2} \qquad (\alpha < 1, \text{ by assumption})\\
        & \rightarrow 0
    \end{align*}

    Now, we can apply DCT to show that the unconditional probability also goes to zero.

\subsection{Proof of Theorem~\ref{thm:mds-1}}

    Let us assume without loss of generality that $\dfrac{1}{n}\sum_i \scalbb{x}_i = \dfrac{1}{n}\sum_i \scalbb{y}_i = 0$. Let, $\scalbb{X}$ and $\scalbb{Y}$ be the $n\times p$ matrices with each row being $\scalbb{x}_i$ and $\scalbb{y}_i$ respectively. Then, the usual double centering mechanism (see Borg and Groenen~\cite{borg2005modern} for details) will indicate that 
    $$
    \scalbb{XX}\transpose = -\dfrac{1}{2} \scalbb{CQC} = \scalbb{YY}\transpose
    $$
    \noindent where $\scalbb{C} = \scalbb{I}_n - \dfrac{1}{n} \scalbb{J}_n$ with $\scalbb{J}_n$ being the $n \times n$ matrix with all entries being equal to $1$, and $\scalbb{Q}$ is the matrix with the entries $q_{ij}^2 = \Vert \scalbb{x}_i - \scalbb{x}_j\Vert^2$. 

    Now, a simultenous singular value decomposition of $\scalbb{X}$ and $\scalbb{Y}$ would yield that 

    $$
    \scalbb{X} = \scalbb{U}\scalbb{D}\scalbb{V}_1\transpose, \qquad 
    \scalbb{Y} = \scalbb{U}\scalbb{D}\scalbb{V}_2\transpose
    $$

    \noindent where the same $\scalbb{U}, \scalbb{V}_1, \scalbb{V}_2$ are unitary matrices and $\scalbb{D}$ is a diagonal matrix. The same $\scalbb{U}$ and $\scalbb{D}$ works since $\scalbb{XX}\transpose = \scalbb{YY}\transpose$. This shows that $\scalbb{X} = \scalbb{Y}\scalbb{V}_2\scalbb{V}_1\transpose$, i.e. the matrix $\scalbb{P} = \scalbb{V}_2\scalbb{V}_1\transpose$ works as the required orthogonal transformation. 

\subsection{Proof of Theorem~\ref{thm:mds-2}}

Since $g$ is a geodesic of $\mathcal{M}$ and a unit-velocity curve in $\R^d$ induces a constant velocity curve in $\mathcal{M}$, it follows that (see Lee~\cite{lee2006riemannian} for details) 
    
    $$
    g(f(\scalbb{\theta}_i), f(\scalbb{\theta}_j)) = c_1 \Vert \scalbb{\theta}_i - \scalbb{\theta}_j\Vert, \qquad \forall 1 \leq i < j \leq n.
    $$

    \noindent Therefore, $\{ \scalbb{y}_1, \dots \scalbb{y}_n \}$ is the output of metric MDS for the distance metric $q_{ij} = c_1\Vert \scalbb{\theta}_i - \scalbb{\theta}_j\Vert$. On the other hand, due to the definition of $g'$, the geodesic on $\mathcal{M}'$, it follows that 

    $$
    g'(\phi\circ f(\scalbb{\theta}_i), \phi \circ f(\scalbb{\theta}_j)) = cc_1 \Vert \scalbb{\theta}_i - \scalbb{\theta}_j\Vert, \qquad \forall 1 \leq i < j \leq n
    $$

    \noindent i.e. $\{ \scalbb{z}_1, \dots \scalbb{z}_n \}$ is the output of metric MDS for the distance metric $q'_{ij} = cc_1\Vert \scalbb{\theta}_i - \scalbb{\theta}_j\Vert = cq_{ij}$. Now, proceeding along the lines of the proof of Theorem~\ref{thm:mds-1}, one can show the required result.

    \section{More Simulated Examples}\label{appendix:simulation}

\begin{figure}[ht]
        \centering
        \begin{subfigure}{\textwidth}
            \centering
            \includegraphics[width = 0.25\textwidth]{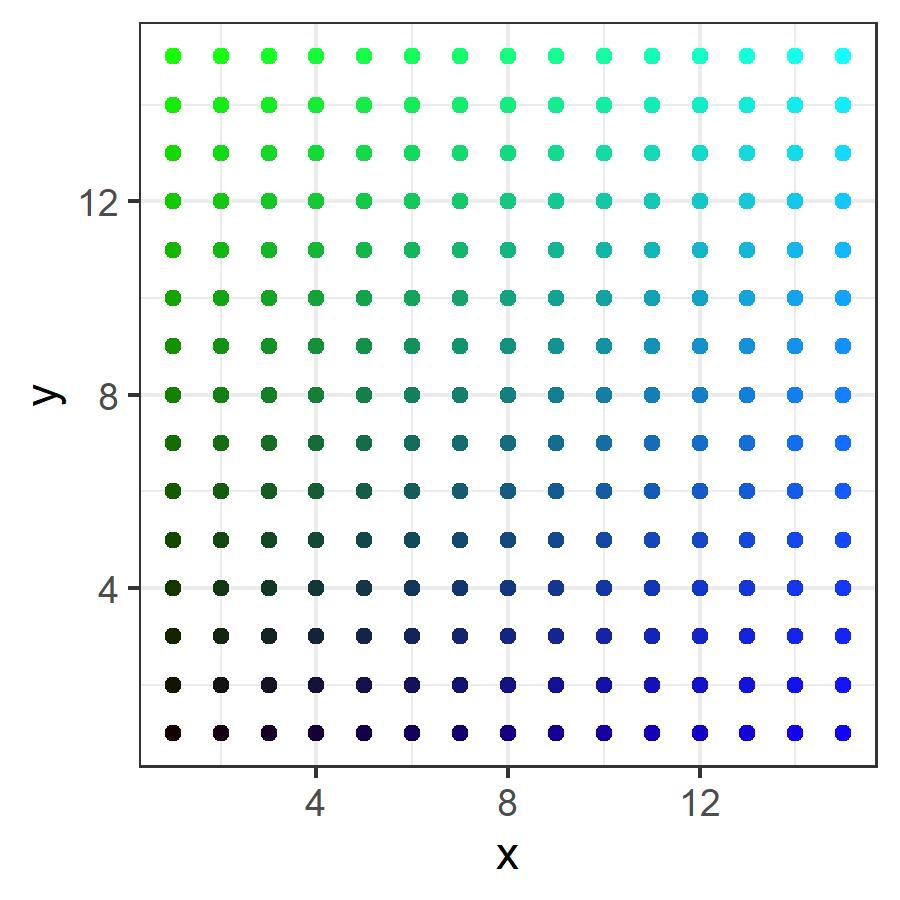}
            \caption{Original Grid Dataset}
        \end{subfigure}
        \begin{subfigure}{\textwidth}
            \centering
            \includegraphics[width = \textwidth]{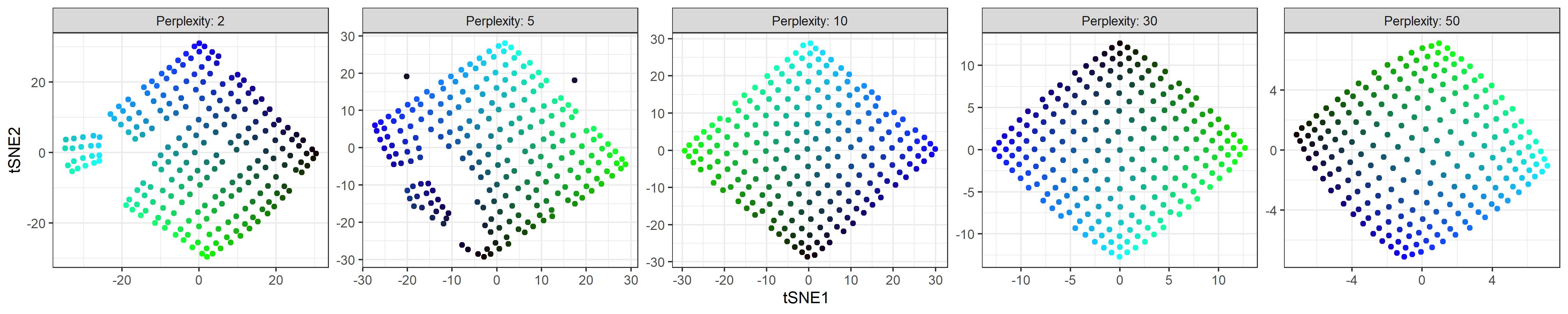}
            \caption{2D embedding of Grid data by tSNE for different perplexities}
        \end{subfigure}
        \begin{subfigure}{\textwidth}
            \centering
            \includegraphics[width = \textwidth]{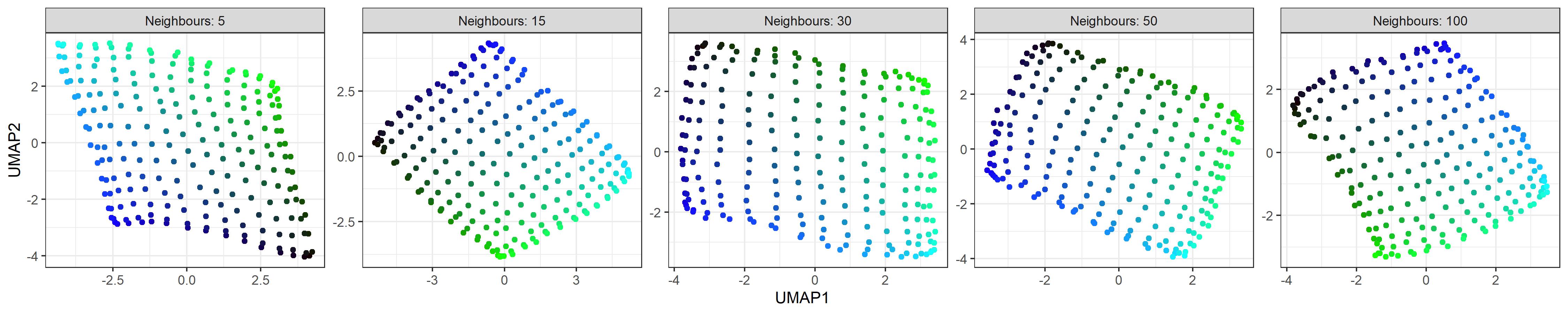}
            \caption{2D embedding of Grid data by UMAP for different n-neighbours}
        \end{subfigure}
        \begin{subfigure}{\textwidth}
            \centering
            \includegraphics[width = \textwidth]{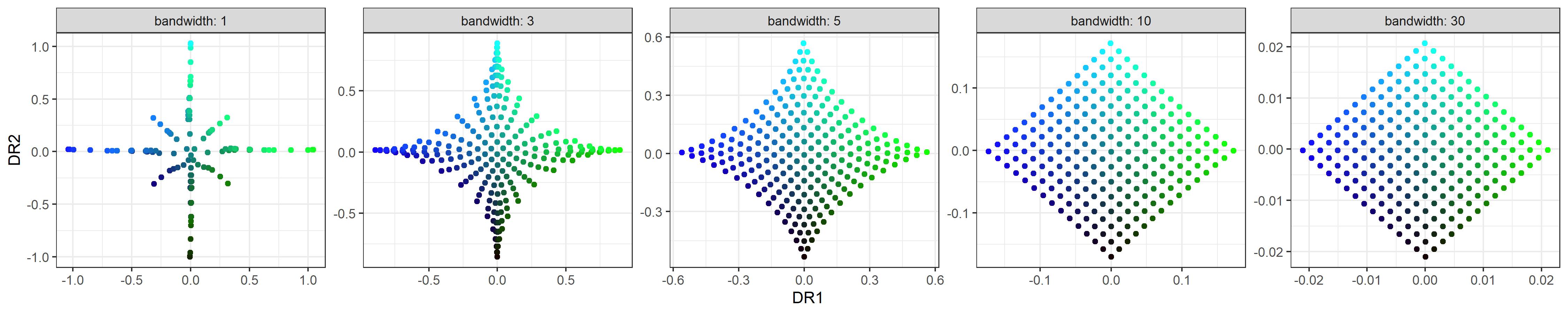}
            \caption{2D embedding of Grid data by tSNE for different kernel bandwidth}
        \end{subfigure}
        \caption{Analysis of Grid Data}
        \label{fig:grid-data}
    \end{figure}

    Several simulated examples, their embeddings using tSNE, UMAP and LSDR, under different hyperparameters such as perplexities, n-neighbours, kernel bandwidths etc. are described in Figure~\ref{fig:grid-data}-\ref{fig:trefoil-knot-data}. These examples are taken from the analysis of Google PAIR engineers~\cite{googlepairumap,wattenberg2016how}, where an extensive comparison of tSNE and UMAP is already available. These are some of the datasets where tSNE and UMAP are found to have better performances, and yet in these datasets, LSDR achieves a similar level of performances compared to them. In particular, in linked circle data (see Figure~\ref{fig:linked-circle-data}) and in trefoil knot data (see Figure~\ref{fig:trefoil-knot-data}), both tSNE and UMAP only sees the local structure, and hence outputs two unlinked circles and a large circle respectively. However, there does not exist a continuous transformation that can map those two linked circles into two unlinked circles, and similarly, the trefoil knot, being a non-trivial knot, is not isomorphic to a single large circle. These notions can only be achieved if we look at a global scale, and LSDR takes care of that.
    
    On the other hand, in the two long linear cluster data (see Figure~\ref{fig:linked-circle-data}), it is possible to separate the two clusters by a single line, hence even a one-dimensional embedding should be able to perform a reduction of higher quality. However, tSNE and UMAP does not necessarily show this importance in their first reduced component. In other words, it is not necessarily true that the first reduced variable obtained by tSNE or UMAP is necessarily better than its second reduced component. But, LSDR turns out to be capable of using such structure in its first reduced output only, while the secondary component (or reduced variable) provides an orthogonal direction to look at. This phenomenon can be explained by the fact that the metric MDS step in LSDR (which provides the low dimensional embeddings for the skeletal points) actually perform an eigendecomposition, which creates several orthogonal eigen-directions in the order of the most explainable variation of distances through the manifold approximation graph $G_\alpha$.

    \begin{figure}[ht]
        \centering
        \begin{subfigure}{\textwidth}
            \centering
            \includegraphics[width = 0.3\textwidth]{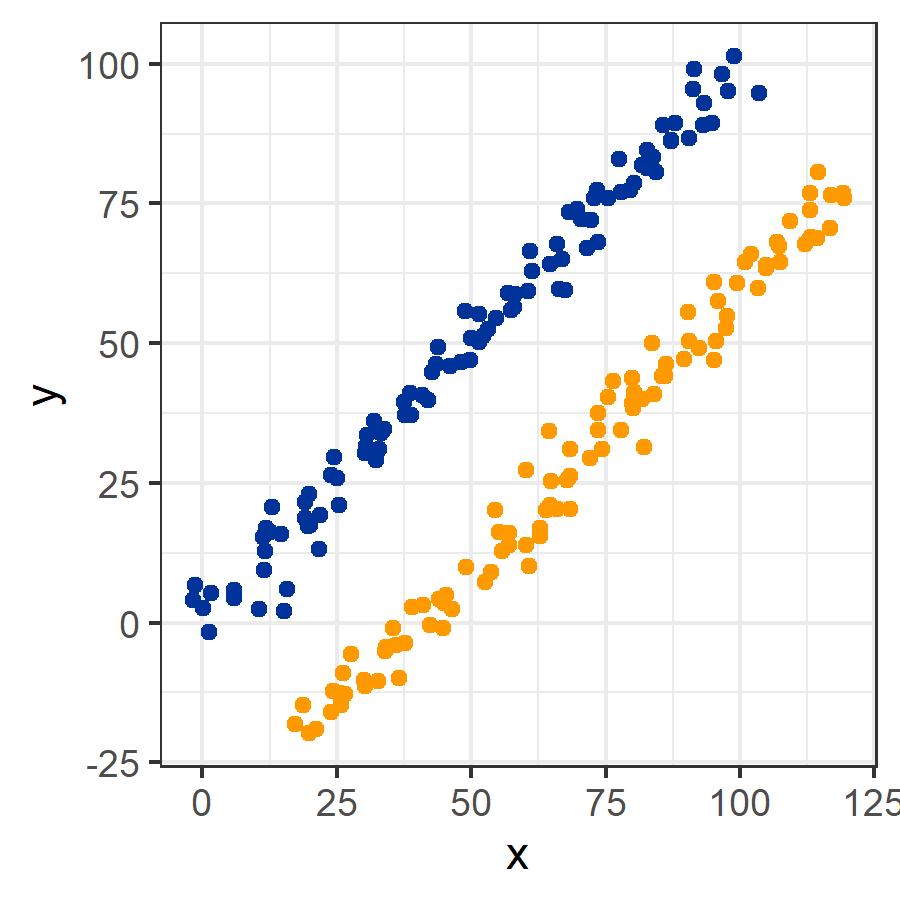}
            \caption{Original data with two long linear clusters}
        \end{subfigure}
        \begin{subfigure}{\textwidth}
            \centering
            \includegraphics[width = \textwidth]{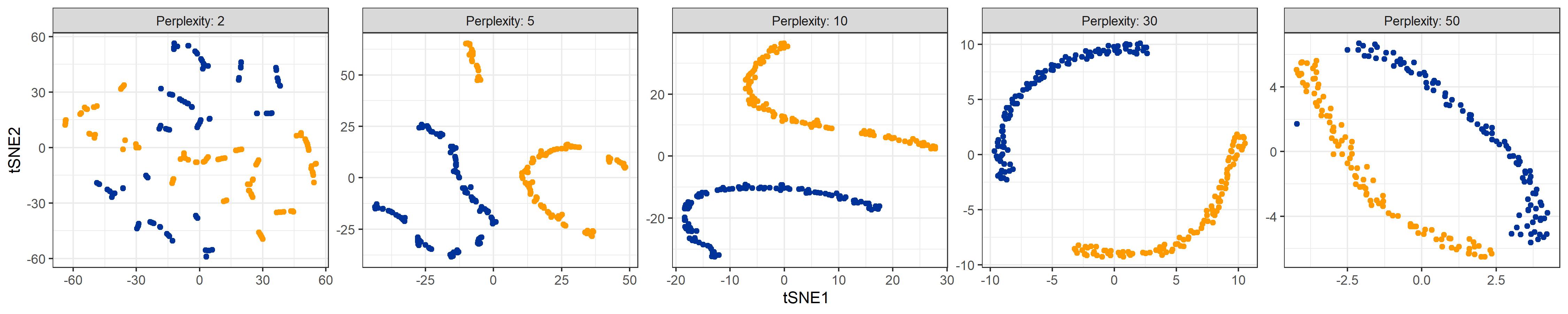}
            \caption{2D Embedding of Two linear clusters data via tSNE for different perplexities}
        \end{subfigure}
        \begin{subfigure}{\textwidth}
            \centering
            \includegraphics[width = \textwidth]{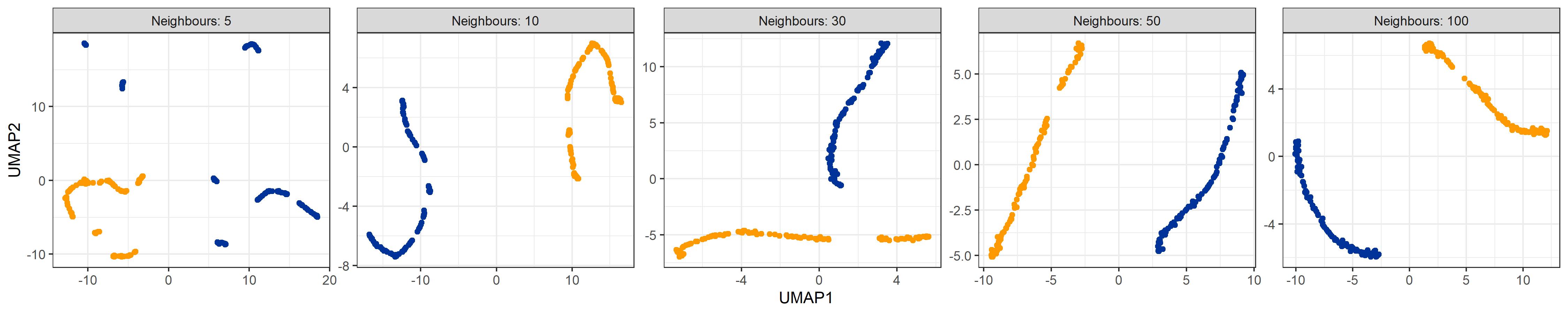}
            \caption{2D Embedding of Two linear clusters data via UMAP for different n-neighbours}
        \end{subfigure}
    \end{figure}
    \begin{figure}[ht]\ContinuedFloat
        \centering
        \begin{subfigure}{\textwidth}
            \centering
            \includegraphics[width = \textwidth]{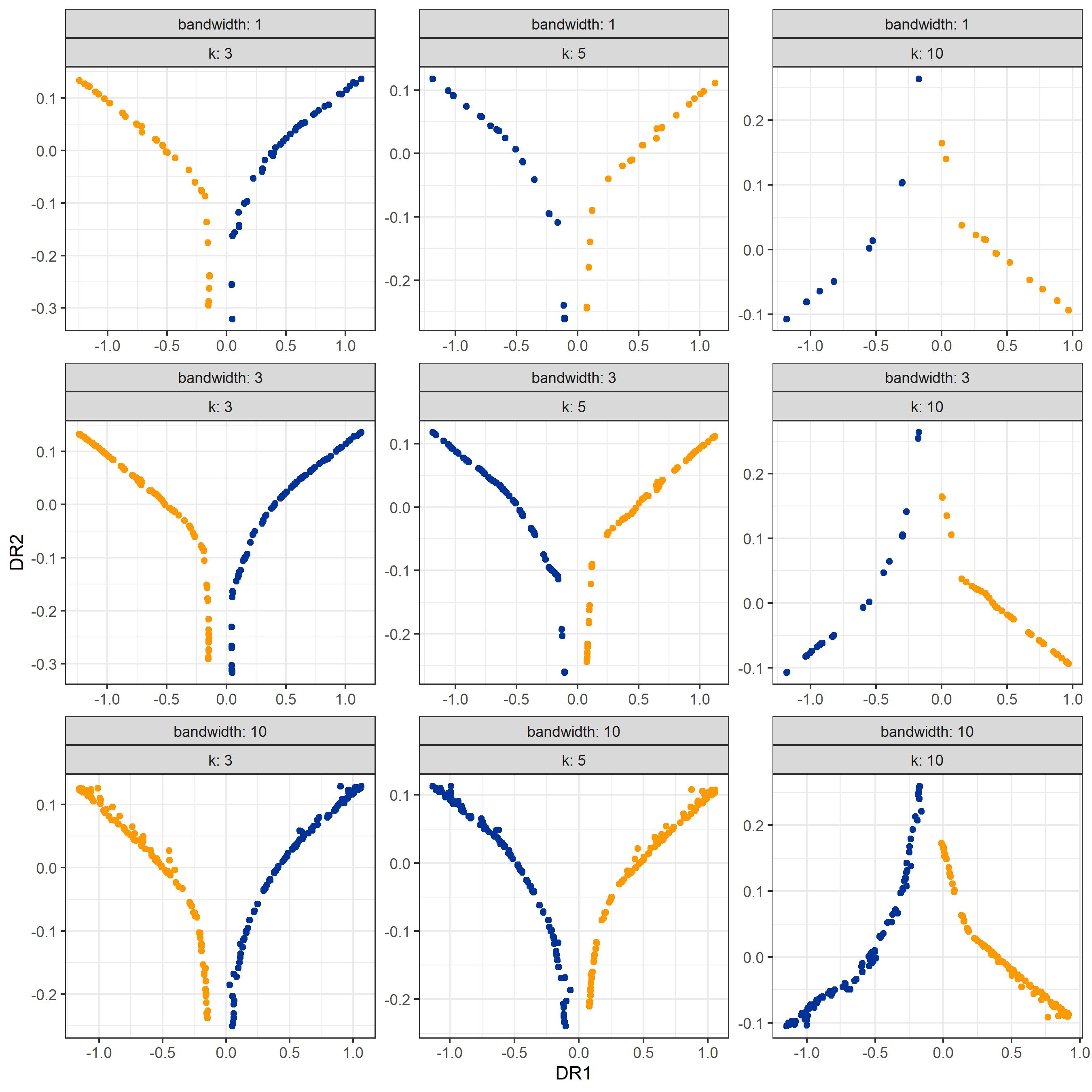}
            \caption{2D Embedding of Two linear clusters data via LSDR for different n-neighbours $k$ and different kernel bandwidth combination}
        \end{subfigure}
        \caption{Analysis of two long linear clusters data}
        \label{fig:long-linear-cluster-data}
    \end{figure}

    \begin{figure}[ht]
        \centering
        \begin{subfigure}{\textwidth}
            \centering
            \includegraphics[width = 0.5\textwidth]{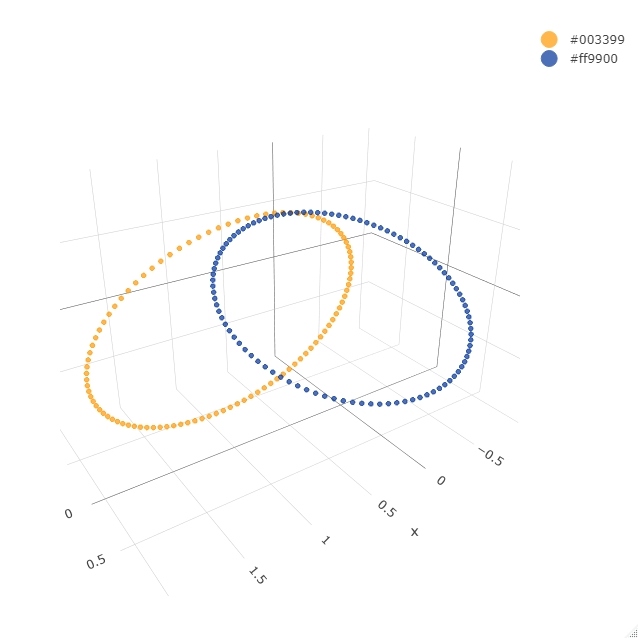}
            \caption{Original data with two linked circle in 3D}
        \end{subfigure}
        \begin{subfigure}{\textwidth}
            \centering
            \includegraphics[width = \textwidth]{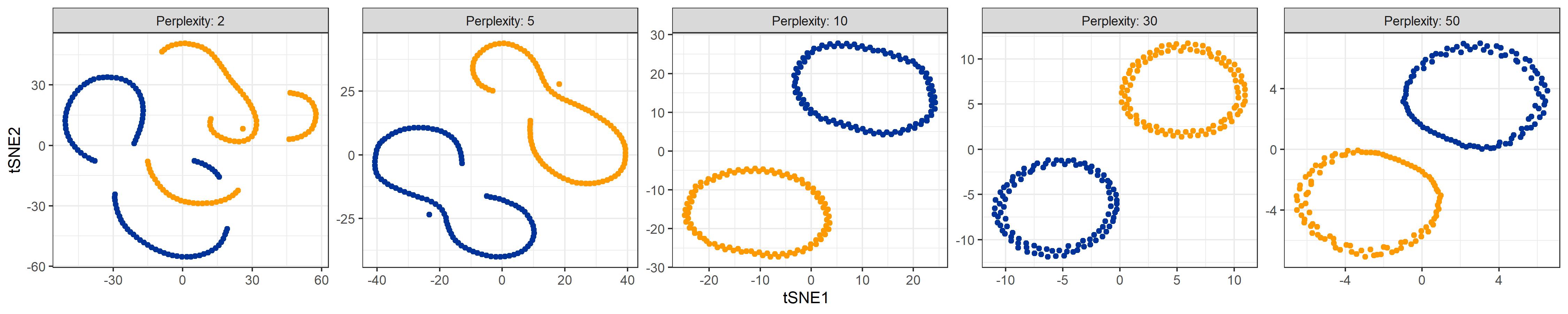}
            \caption{2D Embedding of Two linked circles data via tSNE for different perplexities}
        \end{subfigure}
        \begin{subfigure}{\textwidth}
            \centering
            \includegraphics[width = \textwidth]{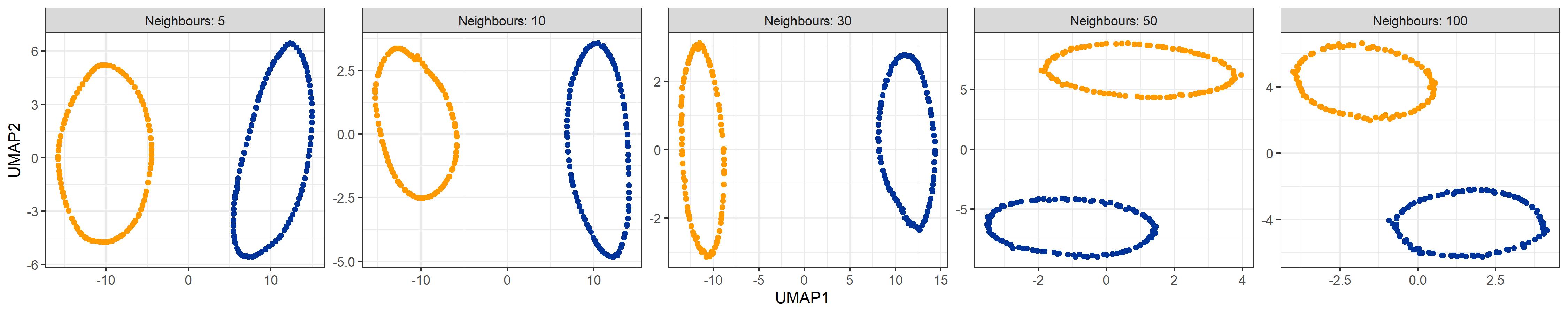}
            \caption{2D Embedding of Two linked circles data via UMAP for different n-neighbours}
        \end{subfigure}
    \end{figure}
    \begin{figure}[ht]\ContinuedFloat
        \centering
        \begin{subfigure}{\textwidth}
            \centering
            \includegraphics[width = \textwidth]{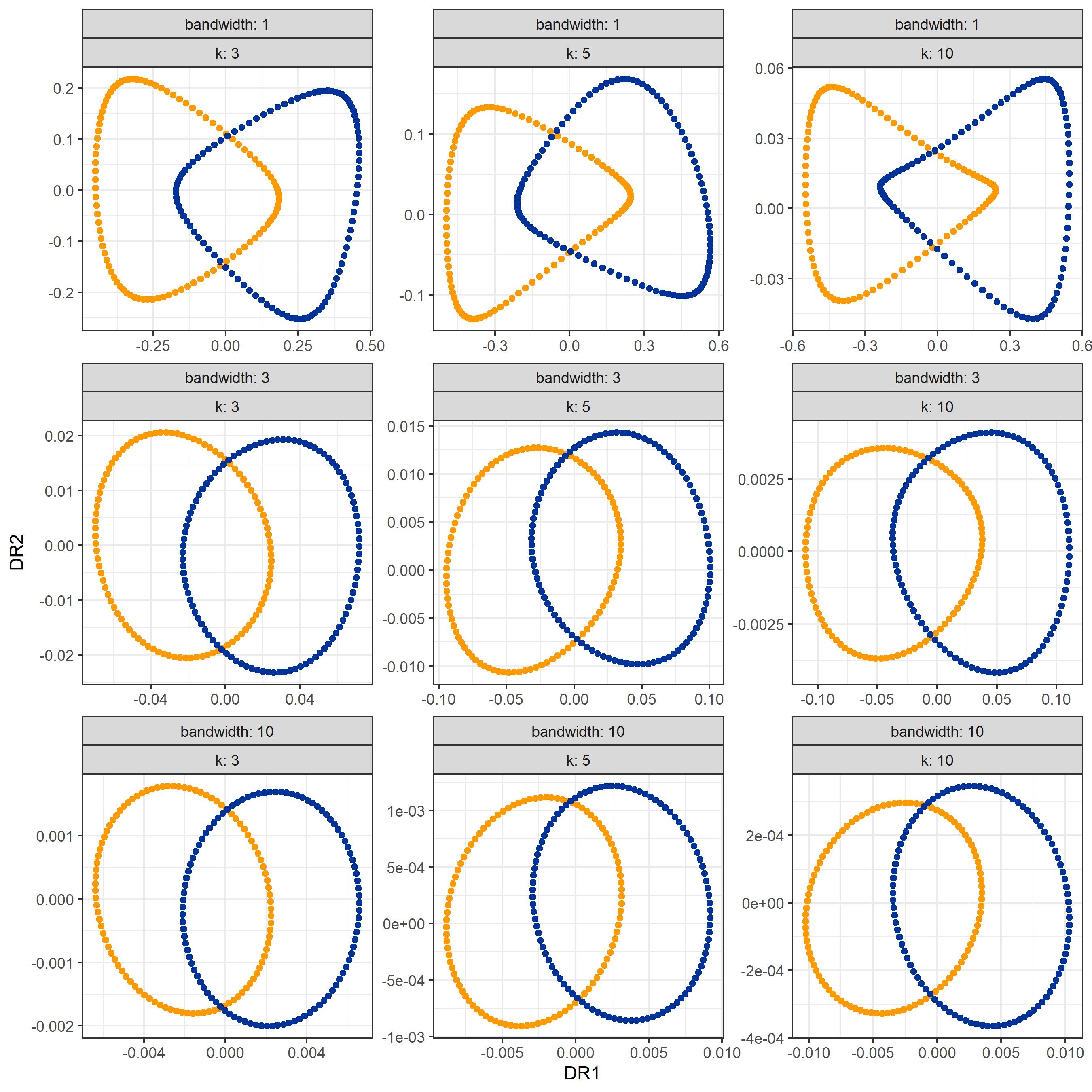}
            \caption{2D Embedding of Two linked circles data via LSDR for different n-neighbours $k$ and different kernel bandwidth combination}
        \end{subfigure}
        \caption{Analysis of two linked circles data}
        \label{fig:linked-circle-data}
    \end{figure}

    \begin{figure}[ht]
        \centering
        \begin{subfigure}{\textwidth}
            \centering
            \includegraphics[width = 0.5\textwidth]{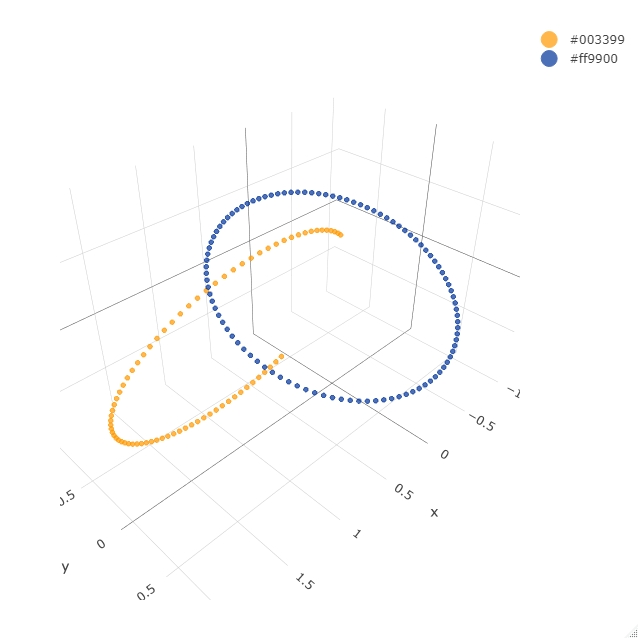}
            \caption{Original data with two unlinked circle in 3D}
        \end{subfigure}
        \begin{subfigure}{\textwidth}
            \centering
            \includegraphics[width = \textwidth]{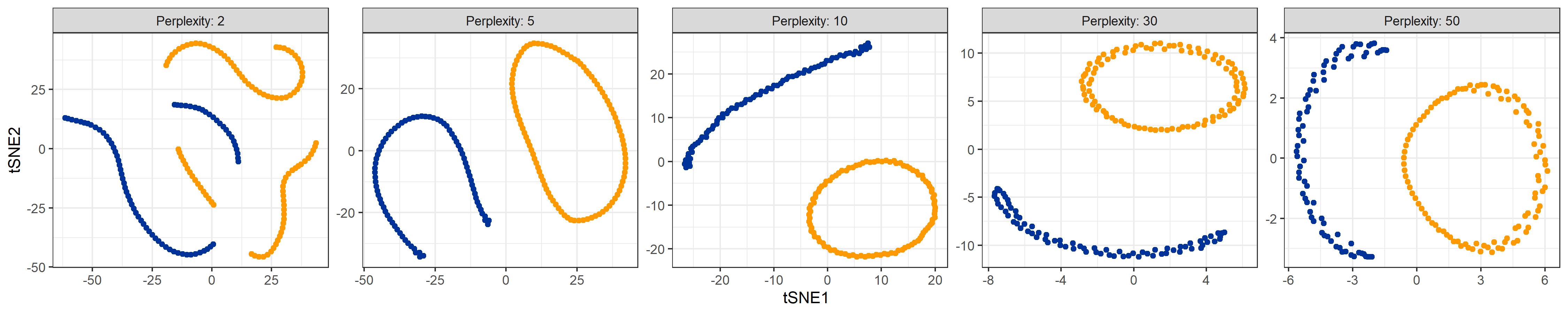}
            \caption{2D Embedding of Two unlinked circles data via tSNE for different perplexities}
        \end{subfigure}
        \begin{subfigure}{\textwidth}
            \centering
            \includegraphics[width = \textwidth]{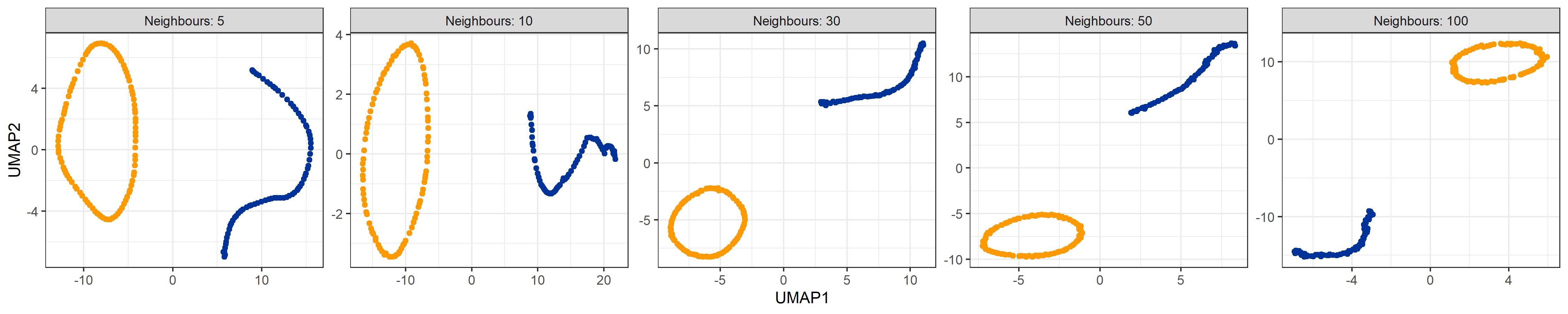}
            \caption{2D Embedding of Two unlinked circles data via UMAP for different n-neighbours}
        \end{subfigure}
    \end{figure}
    \begin{figure}[ht]\ContinuedFloat
        \centering
        \begin{subfigure}{\textwidth}
            \centering
            \includegraphics[width = \textwidth]{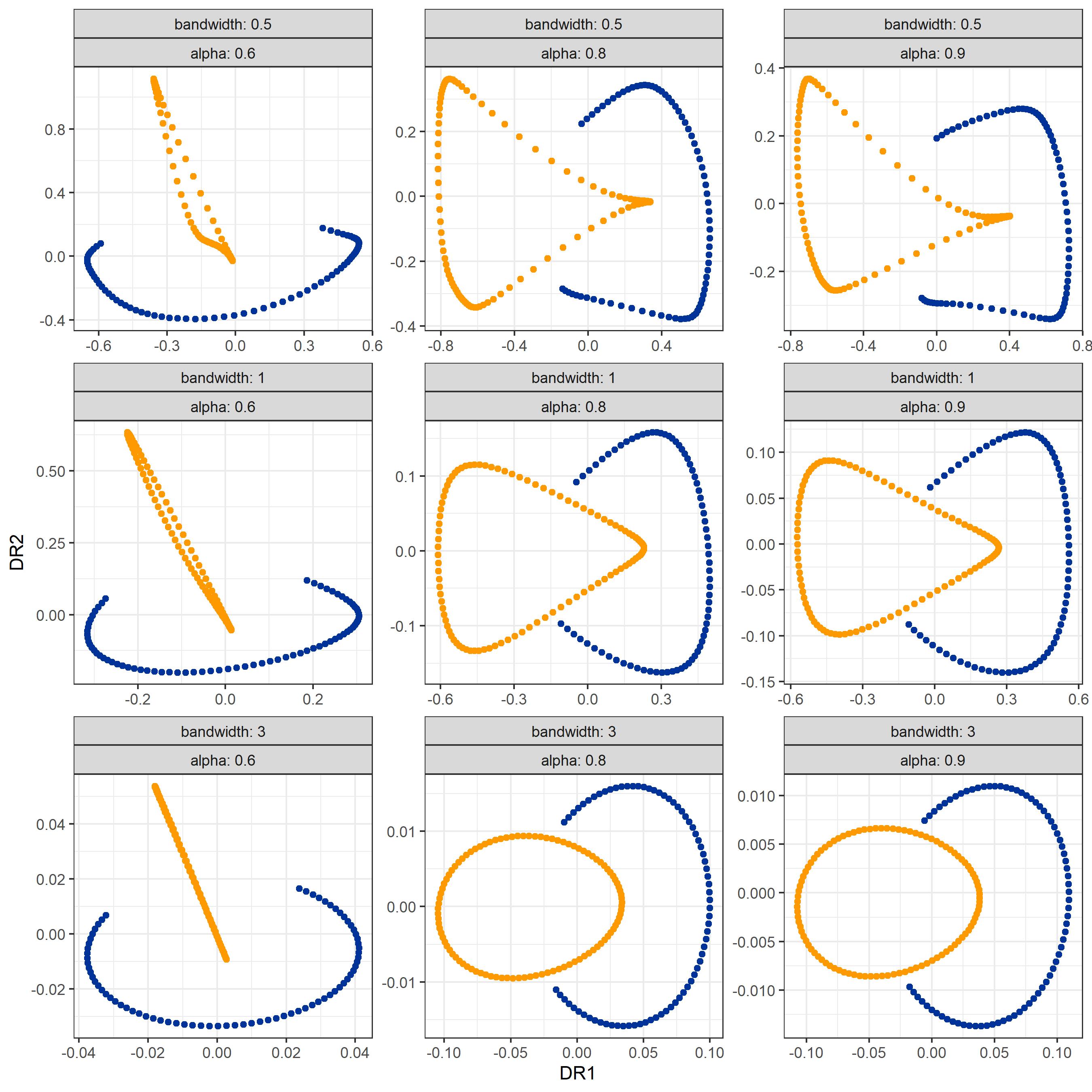}
            \caption{2D Embedding of Two unlinked circles data via LSDR for different choice of $\alpha$-approximation and different kernel bandwidth combination, fixing $k = 5$}
        \end{subfigure}
        \caption{Analysis of two unlinked circles data}
        \label{fig:unlinked-circle-data}
    \end{figure}

    \begin{figure}[ht]
        \centering
        \begin{subfigure}{\textwidth}
            \centering
            \includegraphics[width = 0.3\textwidth]{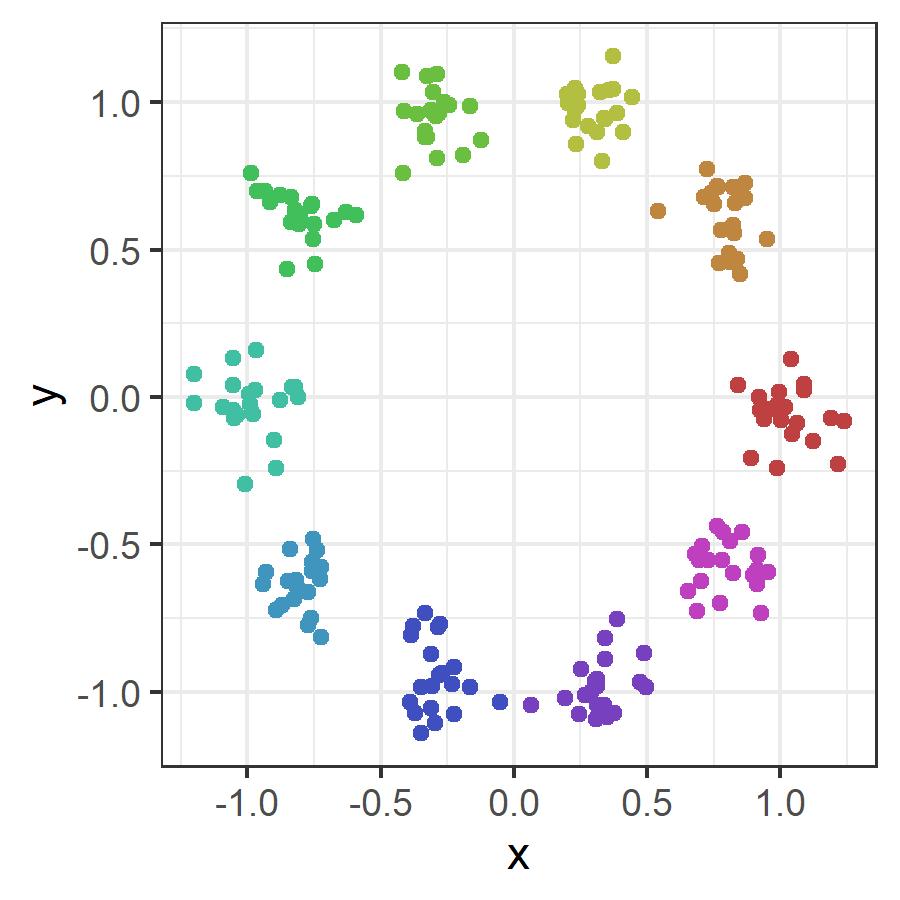}
            \caption{Original data with circular clusters in 3D}
        \end{subfigure}
        \begin{subfigure}{\textwidth}
            \centering
            \includegraphics[width = \textwidth]{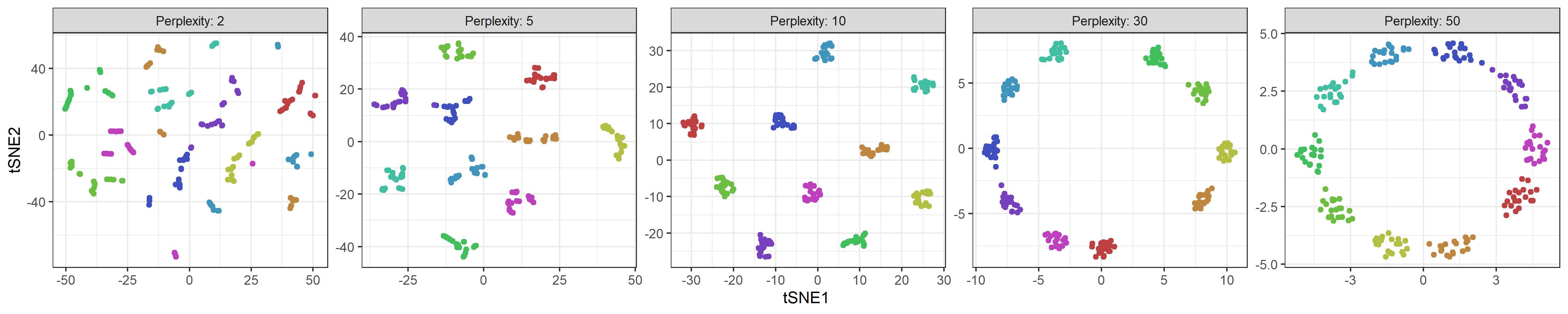}
            \caption{2D Embedding of circular clusters data via tSNE for different perplexities}
        \end{subfigure}
        \begin{subfigure}{\textwidth}
            \centering
            \includegraphics[width = \textwidth]{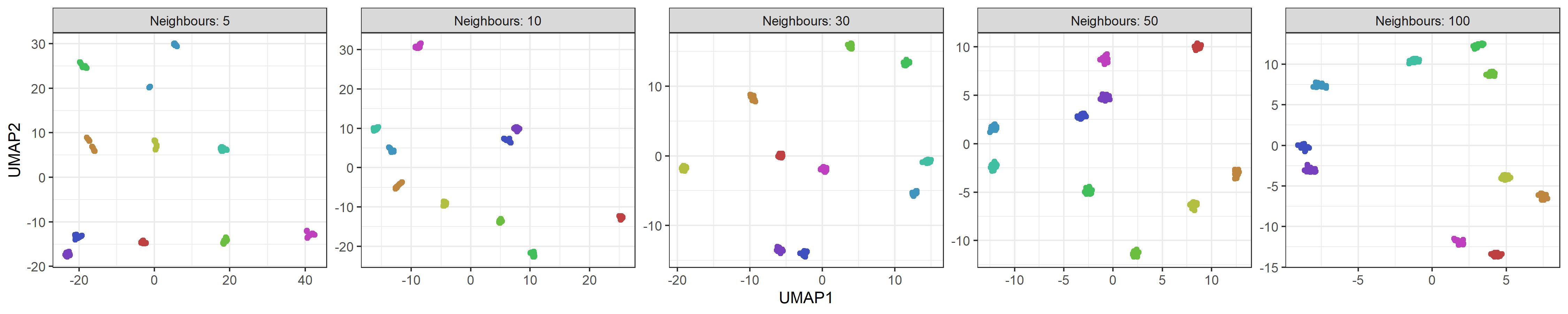}
            \caption{2D Embedding of circular clusters data via UMAP for different n-neighbours}
        \end{subfigure}
    \end{figure}
    \begin{figure}[ht]\ContinuedFloat
        \centering
        \begin{subfigure}{\textwidth}
            \centering
            \includegraphics[width = \textwidth]{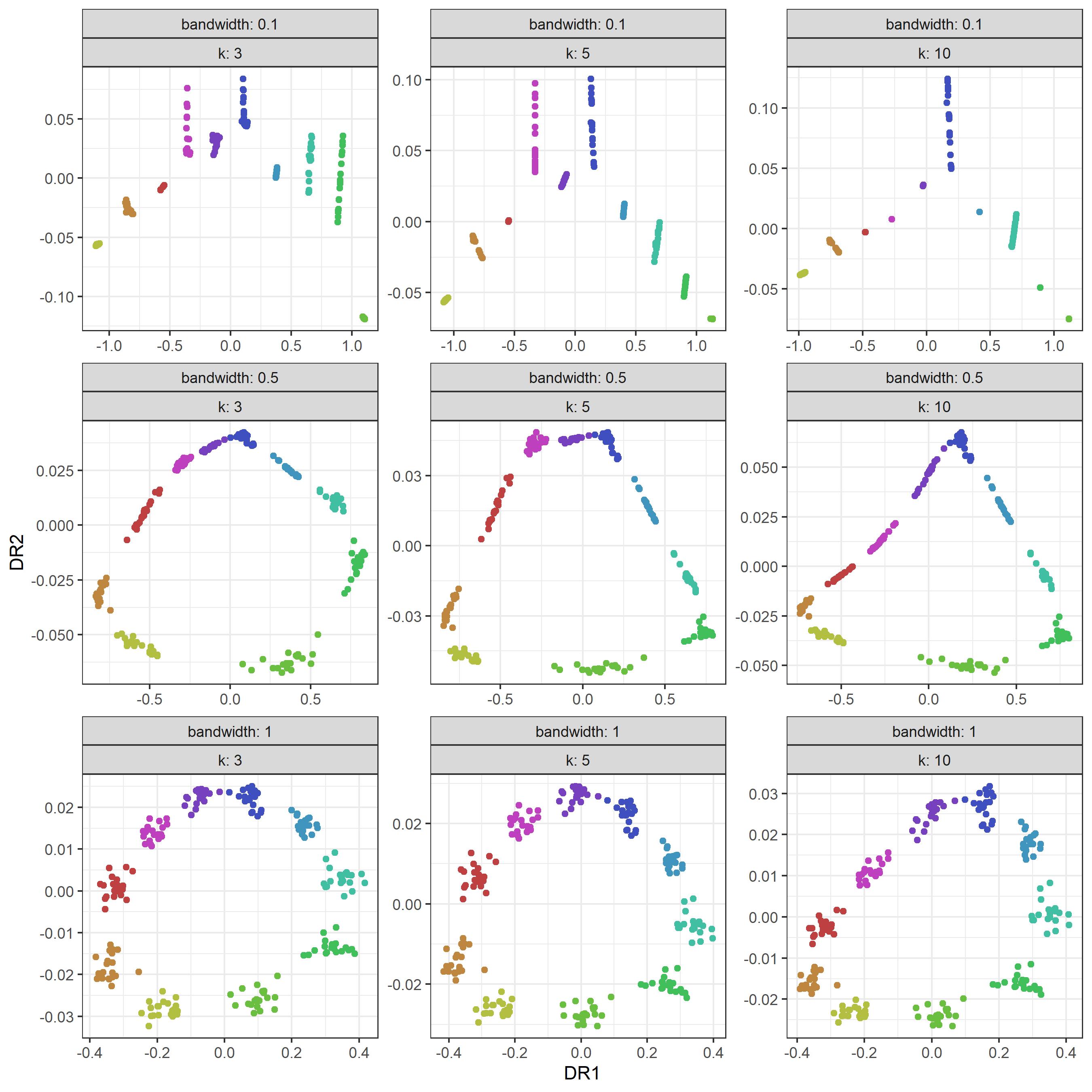}
            \caption{2D Embedding of circular clusters data via LSDR for different choice of n-neighbours $k$ and different kernel bandwidth combination}
        \end{subfigure}
        \caption{Analysis of circular clusters data}
        \label{fig:circular-clusters-data}
    \end{figure}

    \begin{figure}[ht]
        \centering
        \begin{subfigure}{\textwidth}
            \centering
            \includegraphics[width = 0.5\textwidth]{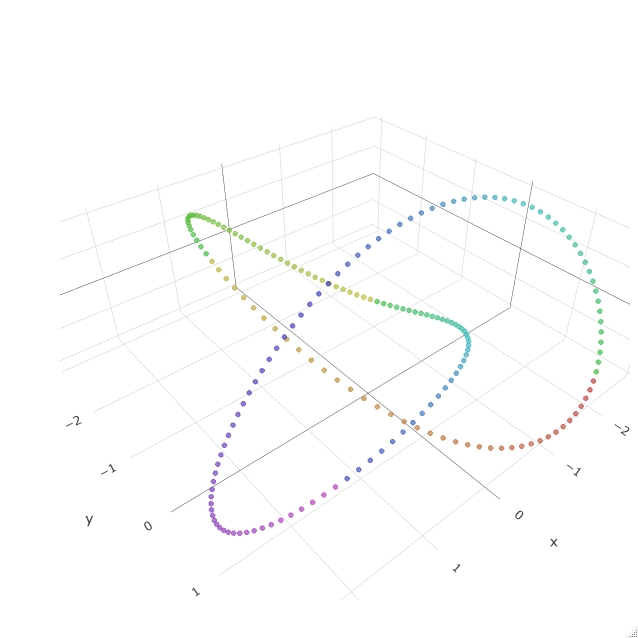}
            \caption{Original trefoil knot data in 3D}
        \end{subfigure}
        \begin{subfigure}{\textwidth}
            \centering
            \includegraphics[width = \textwidth]{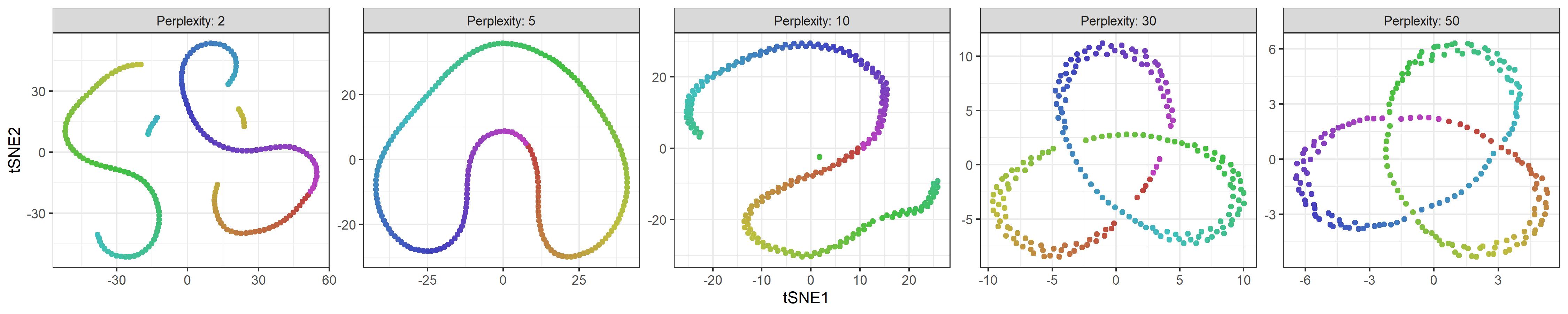}
            \caption{2D Embedding of trefoil knot data via tSNE for different perplexities}
        \end{subfigure}
        \begin{subfigure}{\textwidth}
            \centering
            \includegraphics[width = \textwidth]{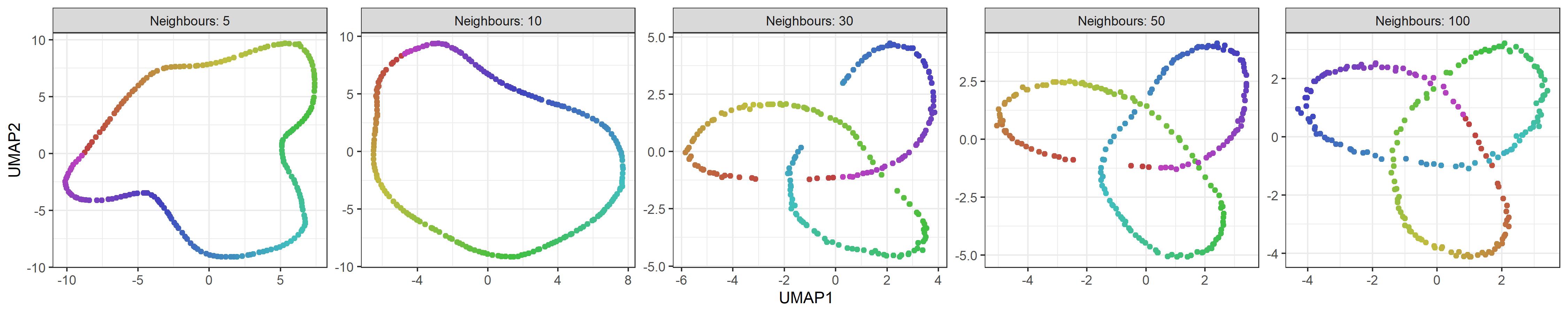}
            \caption{2D Embedding of trefoil knot data via UMAP for different n-neighbours}
        \end{subfigure}
    \end{figure}
    \begin{figure}[ht]\ContinuedFloat
        \centering
        \begin{subfigure}{\textwidth}
            \centering
            \includegraphics[width = \textwidth]{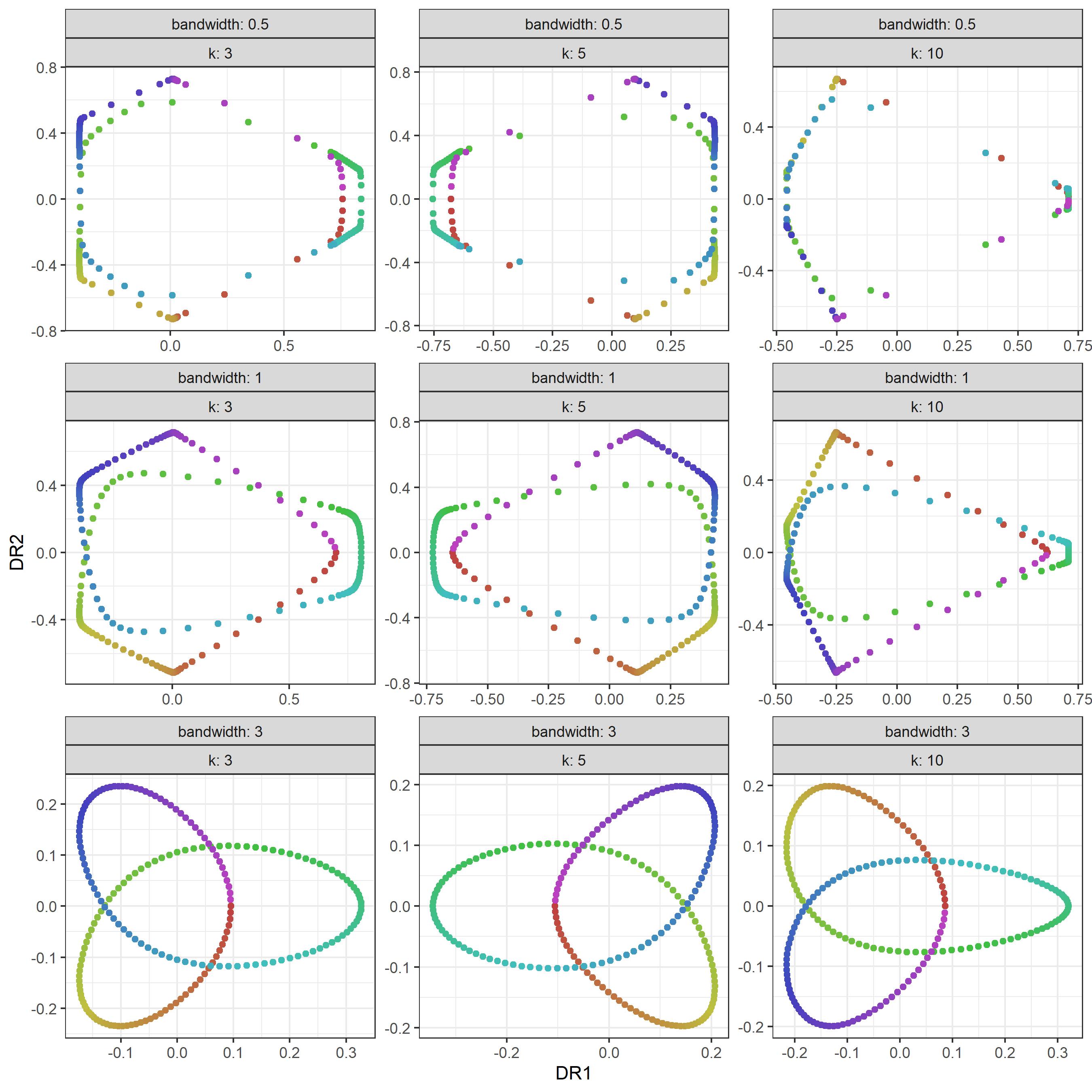}
            \caption{2D Embedding of trefoil knot data via LSDR for different choice of n-neighbours $k$ and different kernel bandwidth combination}
        \end{subfigure}
        \caption{Analysis of trefoil knot data}
        \label{fig:trefoil-knot-data}
    \end{figure}

\bibliographystyle{plain} 
\bibliography{references}

\end{document}